\documentclass[notitlepage,amsmath,preprintnumbers,superscriptaddress,nofootinbib,aps,11pt]{revtex4-1} 

\pdfoutput=1

\usepackage{wasysym}

\usepackage{slashed}

\usepackage{amsmath,amssymb,url,bm}
\usepackage{amssymb,amsfonts}
\usepackage{rotating}
\usepackage{color}

\usepackage{amssymb}
\usepackage{amsmath}
\usepackage{graphicx,subfigure}
\usepackage{longtable}
\usepackage{verbatim}
\usepackage{hyperref}

\usepackage{colortbl}

\usepackage{booktabs}

\usepackage{bm}

\usepackage{xcolor}
\usepackage{array,multirow}

\usepackage{array,mathtools,amssymb,booktabs}
\newcolumntype{C}{>{$}c<{$}}
\AtBeginDocument{
\heavyrulewidth=.16em
\lightrulewidth=.1em
\cmidrulewidth=.03em
\belowrulesep=.4ex
\belowbottomsep=0pt
\aboverulesep=.4ex
\abovetopsep=0pt
\cmidrulesep=\doublerulesep
\cmidrulekern=.5em
\defaultaddspace=.5em
}

\definecolor{Black}{gray}{0}
\definecolor{Gray}{gray}{0.85}
\definecolor{LightGray}{gray}{0.93}
\definecolor{LightGreen}{rgb}{0.88, 1, 0.88}
\definecolor{LightCyan}{rgb}{0.88,1,1}
\definecolor{LightRed}{rgb}{1, 0.25, 0.25}
\definecolor{LightRed}{rgb}{1, 0.85, 0.85}
\definecolor{LightYellow}{rgb}{1, 1, 0.85}
\definecolor{LightYellow}{rgb}{1, 1, 0.85}
\definecolor{LightBlue}{rgb}{0.87, 0.94, 1}
\definecolor{white}{gray}{1}

\newcolumntype{G}{>{\columncolor{LightGray}}c}

\renewcommand{\thesection}{{\bf \Roman{section}}}

\renewcommand{\thesection}{{\bf \Roman{section}}}

\def\di{\displaystyle}

\newcommand*{\fp}[1]{{FP}${}_{#1}$}
\newcommand*{\FP}[1]{$\bm{{\rm FP}_{#1}}$}

\def\beq{\begin{equation}}
\def\eeq{\end{equation}}

\def\bea{\arraycolsep .1em \begin{eqnarray}}
\def\eea{\end{eqnarray}}
\def\Tr{{\rm Tr}}

\def\al#1{\alpha_{\rm {#1}}}
\def\eq#1{(\ref{#1})}

\def\s0#1#2{\mbox{\small{$ \frac{#1}{#2} $}}}
\def\0#1#2{\frac{#1}{#2}}

\def\grgl{\:\hbox to -0.2pt{\lower2.5pt\hbox{$\sim$}\hss}{\raise3pt\hbox{$>$}}\:}
\def\klgl{\:\hbox to -0.2pt{\lower2.5pt\hbox{$\sim$}\hss}{\raise3pt\hbox{$<$}}\:}

\allowdisplaybreaks[1]

\makeatletter
\def\@ssect@ltx#1#2#3#4#5#6[#7]#8{%
  \def\H@svsec{\phantomsection}%
  \@tempskipa #5\relax
  \@ifdim{\@tempskipa>\z@}{%
    \begingroup
      \interlinepenalty \@M
      #6{%
       \@ifundefined{@hangfroms@#1}{\@hang@froms}{\csname @hangfroms@#1\endcsname}%
       {\hskip#3\relax\H@svsec}{#8}%
      }%
      \@@par
    \endgroup
    \@ifundefined{#1smark}{\@gobble}{\csname #1smark\endcsname}{#7}%
  }{%
    \def\@svsechd{%
      #6{%
       \@ifundefined{@runin@tos@#1}{\@runin@tos}{\csname @runin@tos@#1\endcsname}%
       {\hskip#3\relax\H@svsec}{#8}%
      }%
      \@ifundefined{#1smark}{\@gobble}{\csname #1smark\endcsname}{#7}%
      \addcontentsline{toc}{#1}{\protect\numberline{}#8}%
    }%
  }%
  \@xsect{#5}%
}%
\makeatother

\makeatletter
    \def\CT@@do@color{%
      \global\let\CT@do@color\relax
            \@tempdima\wd\z@
            \advance\@tempdima\@tempdimb
            \advance\@tempdima\@tempdimc
    \advance\@tempdimb\tabcolsep
    \advance\@tempdimc\tabcolsep
    \advance\@tempdima2\tabcolsep
            \kern-\@tempdimb
            \leaders\vrule
                    \hskip\@tempdima\@plus  1fill
            \kern-\@tempdimc
            \hskip-\wd\z@ \@plus -1fill }

\usepackage{setspace}
\begin{document}

\begin{flushright}
{\footnotesize DO-TH 17/02, QFET-2017-04}
\end{flushright}

\title{
Directions for model building from asymptotic safety}
\author{Andrew D.~Bond}
\affiliation{Department of Physics and Astronomy, University of Sussex, Brighton,
BN19QH, United Kingdom}
\author{Gudrun Hiller}
\affiliation{Institut f\"ur Physik, Technische Universit\"at Dortmund, D-44221
Dortmund, Germany}
\author{Kamila Kowalska}
\affiliation{Institut f\"ur Physik, Technische Universit\"at Dortmund, D-44221
Dortmund, Germany}
\author{Daniel F.~Litim}
\affiliation{Department of Physics and Astronomy, University of Sussex, Brighton,
BN19QH, United Kingdom}

\begin{abstract}
Building on recent advances in the understanding of gauge-Yukawa theories
we explore possibilities to UV-complete the Standard Model in an asymptotically safe manner.
Minimal extensions are based on a large flavor sector of additional fermions coupled to a scalar singlet matrix field. We find that asymptotic safety requires fermions in higher representations of $SU(3)_C\times SU(2)_L$. Possible signatures at colliders are worked out and include $R$-hadron searches, diboson signatures and
the evolution of the strong and weak coupling constants.
${}$\vskip-1.5cm${}$
\end{abstract}

\maketitle

\begin{spacing}{.9}
\tableofcontents
\end{spacing}

\section{\bf Introduction}\label{sec:intro}

Asymptotic freedom plays a central role in the construction of the Standard Model (SM) of particle physics and extensions thereof \cite{Gross:1973id,Politzer:1973fx}. It predicts that interactions are dynamically switched off at   highest energies due to quantum fluctuations. In the language of the renormalisation group, asymptotic freedom corresponds to a free 
ultraviolet (UV) fixed point.  Asymptotic freedom famously requires the presence of non-abelian gauge fields  \cite{Coleman:1973sx}, together with suitable matter interactions to ensure that Yukawa and scalar couplings reach the free fixed point in the UV alongside the non abelian gauge  coupling \cite{Chang:1974bv}. Identifying viable theories beyond the Standard Model (BSM) with complete asymptotic freedom continues to be an active area of research \cite{Giudice:2014tma,Holdom:2014hla,Pelaggi:2015kna}.

Asymptotic safety states that fundamental quantum fields may very well remain interacting at highest energies \cite{Wilson:1971bg,Weinberg:1980gg}, implying that running couplings reach an interacting (rather than a free) UV fixed point under the renormalisation group evolution. 
If so, theories remain well-behaved and predictive up to highest energies in close analogy to theories with complete asymptotic freedom.   
Asymptotic safety has initially been put forward as a scenario for  quantum gravity \cite{Weinberg:1980gg} where a large amount of evidence has arisen from increasingly sophisticated studies in four dimensions including signatures at colliders  (see \cite{Litim:2011cp} for an overview).  
More recently, necessary and sufficient conditions for asymptotic safety in general weakly coupled gauge theories  (without gravity) have been derived, alongside strict no go theorems \cite{Bond:2016dvk}. 
Most importantly, it was found that Yukawa interactions together with elementary scalar fields such as the Higgs offer a unique key towards asymptotic safety  \cite{Bond:2016dvk}. Moreover, an important proof of existence has been provided in \cite{Litim:2014uca},  and further expanded in \cite{Litim:2015iea}, showing that  exact asymptotic safety with a stable ground state can arise in $SU(N)$ gauge theories under  strict perturbative control in the Veneziano limit.
The feasibility of asymptotic safety is thus well motivated theoretically and opens
intriguing new directions for model building beyond the SM.

In this paper, we make a first step to investigate asymptotically safe extensions of the SM and phenomenological signatures thereof at colliders. Our motivation for doing so is twofold. Firstly, we want to understand whether and how minimal extensions of the SM can be found with weakly interacting UV fixed points. We are particularly interested in the ``phase space'' of such extensions, and in the concrete conditions under which interacting UV fixed points are connected through well-defined trajectories with the SM at low energies. Secondly, we wish to understand how phenomenological constraints  may arise through existing data, and, more generally, the conditions under which asymptotic safety can be tested at colliders.
Our investigation is ``top-down'' in that we begin by requiring conditions under which weakly coupled asymptotic safety can be achieved. Our central  new input are BSM fermions and  scalars, some of which are charged under the gauge symmetries of the SM. Our approach will be minimal in that we add a single BSM Yukawa coupling whose sole task is to negotiate
asymptotically safe UV completions for the SM with $SU(3)_C\otimes SU(2)_L\otimes U(1)_Y$  gauge symmetry.
 
 The paper has the following format. In Sec.~\ref{sec:as} we discuss the basic perturbative mechanism for asymptotic safety in gauge theories including general conditions for existence. In Sec.~\ref{sec:bsm} we investigate minimal extensions of the Standard Model in view of weakly interacting high energy fixed points. In Sec.~\ref{SMmatching}, we explain the conditions under which interacting 
 UV fixed points are connected with the SM at low energies. Phenomenological implications are worked out in Sec.~\ref{Pheno}. We summarize 
in Sec.~\ref{conclusion}. Appendix~\ref{AppA} contains technicalities summarising the perturbative loop coefficients and group theoretical information, and details of  UV-IR connecting separatrices.

\section{\bf Basics of asymptotic safety for gauge theories}\label{sec:as}
In this section, we recall the basic mechanism for asymptotic safety in four-dimensional gauge theories with matter and recall  general theorems for asymptotic safety in weakly coupled gauge theories following \cite{Bond:2016dvk,Litim:2014uca}. We also introduce some notation and conventions.

\subsection{Weakly interacting UV fixed points}\label{simple}
We begin with a discussion of  asymptotic safety in gauge theories and the renormalisation group running of couplings. In the absence of asymptotic freedom, it is well-known that
perturbative couplings would  grow towards higher energies
thereby limiting predictivity to a highest energy scale $\Lambda$. The main feature of asymptotic safety, however, is that the growth of couplings is tamed, dynamically, through a weakly interacting fixed point. An explicit  mechanism which allows quantum fields to avoid the notorious Landau poles of QED-like theories has recently been discovered in \cite{Litim:2014uca}. Strict theorems for asymptotic safety in general weakly coupled gauge theories have been derived in  \cite{Bond:2016dvk}. 

To illustrate the mechanism, and to prepare for our models below, we consider the renormalization group (RG) flow for a simple gauge theory with gauge coupling $\alpha_g=g^2/(4\pi)^2$ interacting with scalars and fermions,  with Yukawa coupling $\alpha_y=y^2/(4\pi)^2$. Within perturbation theory, the RG flow in the gauge-Yukawa system to the leading non-trivial order is given by 
\beq
\label{beta2}
\begin{array}{rcl}
\beta_g&\equiv&
\displaystyle
\frac{d \alpha_g}{d \ln \mu} = 
(- B + C\, \alpha_g  - D\, \alpha_y)\,\alpha_g^2 \,,
\\[1.5ex]
\beta_{y}&\equiv&
\displaystyle
\frac{d \alpha_y}{d \ln \mu} 
= (E\, \alpha_y -F\, \alpha_g)\,\alpha_y\,.
\end{array}
\eeq
Scalar selfcouplings do not impact on interacting fixed points to leading order at weak coupling and can be neglected. The various loop coefficients $B, C, D, E$ and $F$ depend on the matter content  of the theory, which we leave unspecified at this stage. The gauge coupling is asymptotically free (infrared free) provided that the one loop gauge coefficient obeys $B>0$ $(B<0)$. The two loop gauge coefficient $C$ may take either sign depending on the matter content. Provided that asymptotic freedom is absent, $B<0$, it has also been shown that $C>0$ \cite{Bond:2016dvk}.  The other loop coefficients obey $D,E,F >0$ for any quantum field theory, irrespective of the matter content. Also notice that Yukawa couplings always contribute with a negative sign to the running of the gauge coupling, irrespective of the sign of $B$. 

In general, theories with \eq{beta2}  may have  various types of fixed points, depending on the matter content. Equating $\beta_i=0$ for both couplings, three types of fixed points are found. The Gaussian fixed point
\beq\label{Gauss}
(\alpha^*_g,\alpha^*_y)=(0,0)
 \eeq
always exists, and corresponds to the UV (IR) fixed point provided that $B>0$ ($B<0)$. An interacting fixed point where Yukawa interactions are switched-off may also exist, with
\beq\label{CBZ}
(\alpha^*_g,\alpha^*_y)=\left(\frac{B}{C},0\right)\,.
\eeq
This is the well-known Caswell--Banks-Zaks fixed point \cite{Caswell:1974gg,Banks:1981nn} which requires $B\cdot C>0$ to be physical and  $B/C\ll 1$ to be perturbative.  
It is also known that $B\cdot C<0$ as soon as $B<0$ for any quantum field theory \cite{Bond:2016dvk}. This result has the form of a no go theorem: in four dimensions,  weakly coupled gauge theories cannot become asymptotically safe without Yukawa interactions. Hence, Caswell--Banks-Zaks fixed points \eq{CBZ} are invariably IR fixed points.

A fully interacting gauge-Yukawa fixed  point may arise provided that the Yukawa coupling is non-vanishing. Requiring $\beta_y=0$, \eq{beta2} implies that the gauge and Yukawa coupling are proportional to each other, $\alpha_y=\frac{F}{E}\alpha_g$. This nullcline condition modifies the running of the gauge coupling and turns \eq{beta2} into
\beq\label{betag'}
\beta_g=
(- B + C'\, \alpha_g)\,\alpha_g^2 \,,
\eeq
where the  two loop term is effectively shifted $C\to C'$ owing to Yukawa interactions, with
\beq\label{C'}
 C'=C-D\frac{F}{E}<C\,.
\eeq 
This shift term has important implications: Firstly, the  fixed point is now fully interacting, with the gauge coupling taking the form \eq{CBZ} with $C$ shifted as in \eq{C'}, together with the interacting fixed point for the Yukawa coupling,
\beq\label{GY}
(\alpha^*_g\,,\alpha^*_y)=\left(\frac{B}{C'}\,, \frac{B}{C'}\frac{F}{E}\right)\,.
\eeq
Secondly, for theories with asymptotic freedom $(B>0)$, and  provided that $C'>0$, the gauge-Yukawa fixed point \eq{GY} corresponds to an IR fixed point. It can be reached by RG trajectories emanating out of the Gaussian UV fixed point. 
Finally, for theories without asymptotic freedom $(B<0)$ the gauge coupling may now take a viable  interacting fixed point $\alpha_g^*=B/C'>0$ as long as $C'<0$. This is the interacting UV fixed point of asymptotic safety (see Tab.~\ref{tUV} for a summary). The result is in stark contrast to theories without Yukawa interactions, where \eq{CBZ} cannot possibly become an UV fixed point. We conclude that the Yukawa interactions are of crucial importance for asymptotic safety \cite{Bond:2016dvk}.
Moreover, the necessary condition  for asymptotic safety at weak coupling $B, C'<0$, see Tab.~\ref{tUV}~{\bf c)},  now translates into a simple condition relating the one and two loop coefficients appearing in \eq{beta2},
\beq\label{condition}
C'<0\quad \Leftrightarrow\quad DF - CE>0\,.
\eeq
In the remaining part of the paper, we evaluate whether the condition \eq{condition} can be achieved for extensions of the SM.

\subsection{Scaling behaviour}

\begin{table}[t]
\begin{tabular}{ccccc}
     \toprule                     
\rowcolor{LightGreen}
\bf case&
${}\quad$\bf parameter${}\quad$& 
${}\quad$\bf fixed point&
\bf info&
\bf type
\\
\midrule
\rowcolor{LightGray}
\bf a)&
$B>0, C>0$ & \quad{\rm IR}\quad & \quad{\rm asymptotic\ freedom}\quad
&\ \ Caswell--Banks-Zaks (BZ)\ \ \\
\bf b)&
$B>0, C'>0$ & \quad{\rm IR}\quad & \quad{\rm asymptotic\ freedom}\quad
&gauge-Yukawa (GY)\\
\rowcolor{LightGray}
\bf c)&
$B<0, C'<0$&\quad{\rm UV}\quad&\quad{\rm asymptotic\ safety}\quad
&gauge-Yukawa (GY)\\
\bottomrule
\end{tabular}
\caption{\label{tUV}All weakly interacting fixed points $\alpha_*$ of simple gauge theories with \eq{beta2} and their dependence on the matter content expressed through the parameters $B, C$ and $C'$, see \cite{Bond:2016dvk}.}\end{table}

In the vicinity of a free (UV or IR) fixed point the running of couplings is logarithmically slow. In the vicinity of interacting (UV or IR) fixed points, instead, the running of couplings is power law like, characterised by universal scaling exponents $\{\vartheta_i\}$. Linearising the RG flow in the vicinity of a fixed point
\beq\label{linear}
\beta_i=\sum_j M_{ij}\,(\alpha_j-\alpha_j^*)+{\rm subleading}\,,
\eeq
the scaling exponents can be derived as the eigenvalues of the stability matrix $M_{ij}=\partial\beta_i/\partial\alpha_j|_*$. Eigendirections are termed relevant (irrelevant) provided that $\vartheta<0$ ($\vartheta>0$). Marginal couplings have vanishing eigenvalues at linear order \eq{linear}, reflecting logarithmic running. Whether these are marginally relevant ($\vartheta=0^-$) such as in QCD, or marginally irrelevant ($\vartheta=0^+$) such as in QED, is determined beyond leading order. In the vicinity of interacting fixed points couplings scale according to
\beq\label{scaling}
\alpha_i(\mu)=\alpha_i^*+ \sum_n c_n V^n_i \left(\frac{\mu}{\mu_0}\right)^{\vartheta_n}+{\rm subleading}\,,
\eeq
where $V^n$ are the eigenvectors of the stability matrix with eigenvalue $\vartheta_n$, $\mu$ denotes the RG scale, and $c_n$ are free numbers. The significance of \eq{scaling} is as follows \cite{Falls:2013bv,Falls:2014tra}. 
In order to achieve a well-defined UV limit, the parameters $c_n$ related to irrelevant eigenvalues must be set identically to zero, or else the UV fixed point cannot possibly be reached from \eq{scaling} in the limit $\mu\to\infty$. On the other side, the relevant eigendirections are unconstrained and the corresponding numbers $c_n$ are free parameters of the theory. Provided that the number of relevant directions is finite, the theory is predictive with a finite number of free parameters whose values must  be determined by experiment.

Returning to the models at hand, three different types of interacting fixed points arise. At a Caswell--Banks-Zaks fixed point \eq{CBZ}, scaling exponents are given by
\bea
\label{CBZsimple1}
&&\vartheta_1=-BF/C\,,\\
&&\vartheta_2=\ \ B^2/C\,,
\label{CBZsimple2}
\eea
to leading order in $B/C\ll 1$, with $\vartheta_1<0<\vartheta_2$. Consequently, the fixed point has a relevant direction corresponding to the Yukawa interaction, and an irrelevant one, corresponding to the gauge coupling, see Tab.~\ref{tUV}~{\bf a)}.
At the gauge-Yukawa fixed point \eq{GY}, the scaling exponents of the theory are given by
\bea
\label{GYsimple1}
&&\vartheta_1={B^2}/{C'}\,,\\
&&\vartheta_2 ={BF}/{C'}\,,
\label{GYsimple2}
\eea
to leading order in $B/C'\ll 1$.\footnote{Notice that \eq{GYsimple1} and \eq{GYsimple2} do not follow from \eq{CBZsimple1} and \eq{CBZsimple2} by substituting 
$C\to C'$.} For asymptotically free theories, we note that $0<\vartheta_1<\vartheta_2$, meaning that both directions are IR attractive,  see Tab.~\ref{tUV}~{\bf b)}. In the remaining part of the paper we are particularly interested in theories with asymptotic safety where $B<0$. For these, the eigenvalues are of the form $\vartheta_1<0<\vartheta_2$, see Tab.~\ref{tUV}~{\bf c)}. It states that the fixed point has a one dimensional UV critical surface characterised by the relevant direction given through $\vartheta_1$ \cite{Litim:2014uca}. 

\subsection{Theorems for asymptotic safety}
General theorems for asymptotic safety in weakly coupled gauge theories have recently been derived in  \cite{Bond:2016dvk}. In particular, it has been established that Yukawa interactions offer a {\it unique} mechanism towards asymptotic safety. Neither gauge interactions nor scalar self interactions are able to negotiate an interacting UV fixed point at weak coupling. Stated differently, it is impossible to find an asymptotically safe and weakly coupled gauge theory with simple or product gauge groups but without Yukawa interactions. Hence, asymptotic safety in four dimensional gauge theories invariably requires elementary scalars and fermions, besides the gauge fields. Furthermore, fermions must minimally be charged under some or all of the gauge group(s).  For general gauge theories with product gauge group ${\cal G}={\cal G}_1\otimes {\cal G}_2\otimes\cdots\otimes {\cal G}_n$, weakly interacting fixed points arise as  solutions to the linear equations \cite{Bond:2016dvk}
\beq\label{FPij}
B'_i=C_{ij}\,\alpha_j^* \,,
\quad
{\rm subject\ to}\quad \alpha^*_j\ge 0\,,
\eeq
where $C_{ij}$ denotes the matrix of two loop gauge contributions, and $B_i'=B_i+ 2 \,Y_{4,i}^*$ the one-loop coefficient shifted by the Yukawa terms $Y_{4,i}^*=\Tr [{\bf C}^{F_i}_2\, {\bf Y}_*^A\,({\bf Y}_*^A)^\dagger]/d(G_i)\ge 0$  at the interacting fixed point. Here, ${\bf C}^{F_i}_2$ denotes the Casimir of the fermions, ${\bf Y}^A$ the matrix of Yukawa couplings, and $d(G_i)$ the dimension of the group $G_i$ following the conventions of  \cite{Machacek:1983tz,Machacek:1983fi,Machacek:1984zw,Luo:2002ti}.
  It has also been shown in \cite{Bond:2016dvk} that for any infrared free gauge factor $(B_i<0)$, the necessary condition for asymptotic safety amongst the solutions to \eq{FPij} is
\beq\label{Bi'}
B_i'>0\,.
\eeq
It states that Yukawa interactions must effectively change the sign of the one loop coefficient for the infrared free gauge couplings, generalising the necessary condition \eq{condition} to general gauge theories. Sufficiency conditions for asymptotic safety, in addition to  the mandatory presence of Yukawa couplings, have also been detailed in \cite{Bond:2016dvk}. These relate to the specifics of the Yukawa sector as well as to the viability of the scalar sector including the stability of the vacuum. It then remains to investigate whether the mandatory and sufficient conditions for asymptotic safety have viable weakly coupled UV fixed points as their solutions. In the remaining part of the paper, we  investigate in concrete terms the availability of asymptotically safe solutions of \eq{FPij}, \eq{Bi'} for BSM extensions of the $SU(3)_C\times SU(2)_L$ sector of the SM.

\section{\bf Asymptotic safety beyond the Standard Model}
\label{sec:bsm}
In this section, we investigate minimal extensions of the SM and conditions under which asymptotic safety becomes available in the deep UV \cite{Bond:2016dvk}. While designing the structure of the BSM sector, we make use of the 
properties of the gauge-Yukawa theory where asymptotic safety can be achieved by an interplay between 
the gauge and Yukawa interactions of vector-like fermions and a scalar matrix field \cite{Litim:2014uca}.

\subsection{Minimal BSM extensions}
Asymptotic safety in BSM extensions minimally require the presence of new matter fields which carry charges under the SM gauge groups and thereby modify the RG running of couplings. Guided by the findings of \cite{Litim:2014uca,Bond:2016dvk}, we  consider the existence of $N_F$ flavors of BSM vector-like 
fermions $\psi$ which minimally couple to the SM gauge bosons. In general, the BSM fermions may carry
charges under  $SU(3)_C$,
$SU(2)_L$,  or hypercharge $Y$,  meaning 
\beq\label{psi}
\psi_i(R_3,R_2,Y)\,,
\eeq
where $i=1,\cdots,N_F$ denotes the flavor index. Furthermore, the BSM fermions couple via Yukawa interactions to complex scalar fields $S_{ij}$ which we take to be a singlet under the SM. Since the BSM fermions are taken to be vector-like, anomalies are not an issue.  The Yukawa interactions are given by
\beq\label{Yukawa}
L_{\rm BSM,\, Yukawa}=
-y\,\Tr(\overline{\psi}_L\,S\,\psi_R+\overline{\psi}_R\,S^\dagger\,\psi_L)\,.
\eeq
Here, $y$ denotes the BSM Yukawa coupling, the trace $\Tr$ sums over color and flavor indices, and the decomposition $\psi=\psi_L+\psi_R$ with $\psi_{R/L}=\frac 12(1\pm \gamma_5)\psi$ is understood. Yukawa interactions are crucial for asymptotic safety to arise in weakly coupled gauge theories. The BSM sector is invariant under global $U(N_F)\times U(N_F)$ flavor rotations. The full Lagrangean for the BSM extension of the SM is  given by 
\beq\label{L}
L=L_{\rm SM}+L_{\rm BSM,\, kin.} 
+L_{\rm BSM,\, pot.}+L_{\rm BSM,\, Yukawa}\,.
\eeq
Here, $L_{\rm SM}$ denotes the SM Lagrangean and $L_{\rm BSM,\, pot.}$ the interaction  Lagrangean of the BSM scalars. The BSM scalars $S$ can mix with the SM Higgs boson through suitable portal coupling contained in $L_{\rm BSM,\, pot.}$. The BSM kinetic terms are given by
\bea
L_{\rm BSM,\, kin.}&=&\Tr\left(\overline{\psi}\,  i\slashed{D}\, \psi \right)
+\Tr\,(\partial_\mu S^\dagger\, \partial^\mu S) \,.
\eea
 The BSM 
fermions communicate to the SM through the gauge interactions, provided they are charged accordingly. The scalar fields are taken to be singlets under the SM gauge groups. 
We assume that the BSM matter fields develop soft scalar $M_S$ and fermion $M_\psi$ masses for the model to be compatible with data. 

For the sake of this paper we make a few further simplifying assumptions. Firstly, we  limit ourselves to BSM fermions which carry no hypercharge. This assumption can be relaxed without changing the overall picture of results. Secondly, we  neglect the role of quartic self interactions of the BSM scalars as well as  portal couplings to the Higgs.  At weak coupling, neither of these  are relevant for the primary existence of the UV fixed point in the gauge-Yukawa sector.
\footnote{A detailed analysis of the SM Yukawa and scalar sector will be given elsewhere.}  Consequently, the free fundamental parameters of the BSM matter sector are given by their group-theoretical representation under $SU(2)_L$ and $SU(3)_C$, and their flavor multiplicity $N_F$,
\beq\label{parameter}
(R_2,R_3,N_F)\,.
\eeq
A key goal of our study will be to identify viable, weakly coupled UV fixed points for the BSM theory \eq{L} within the parameter space \eq{parameter}.

\subsection{Renormalisation group}

In order to identify interacting fixed points, we must analyse the RG equations for the theory \eq{L}.  Within perturbation theory, weakly interacting fixed points arise for the first time at the two loop level in the gauge sector and at the one loop level in the Yukawa and scalar sectors \cite{Bond:2016dvk}. 
Also, interacting UV fixed points necessarily require the presence of a fixed point in the Yukawa interactions. For these reasons, we consider the RG equations for \eq{L} up to second order in both gauge couplings, and up to first order in the BSM Yukawa coupling. This is the lowest order at which a weakly coupled UV fixed point may arise.

To be concrete, we normalise the gauge and Yukawa couplings with the perturbative loop factor and introduce 
\beq\label{couplings}
\alpha_2=\frac{g_2^2}{(4\pi)^2}\,,\quad\quad
\alpha_3=\frac{g_3^2}{(4\pi)^2}\,,\quad\quad
\alpha_y=\frac{y^2}{(4\pi)^2}\,,
\eeq
to denote the weak,  strong, and  BSM Yukawa coupling, respectively.\footnote{Our definition for the gauge couplings relates to the more standard definition $\alpha_{s}=g_{3}^2/(4\pi)$ as  $\alpha_{s}=4\pi\,\alpha_3$, and similarly for $\alpha_w$.}
Our study will be confined to the perturbative domain where all couplings remain sufficiently small.
For now, we use $\alpha <1$ as a practitioner's criterion for weak coupling. We return to this aspect in Sect.~\ref{synopsis}.
In terms of \eq{couplings}, the RG equations within dimensional regularisation and to the leading non-trivial order are given by \cite{Machacek:1983tz,Machacek:1983fi,Machacek:1984zw,Luo:2002ti}
\bea
\beta_3&\equiv&
\frac{d \alpha_3}{d \ln \mu} = 
 (- B_3 + C_{3}\, \alpha_3  + G_{3}\, \alpha_2-D_3\, \alpha_y)\,\alpha_3^2\,,
\nonumber\\
\label{BSM}
\beta_{2}&\equiv&\frac{d \alpha_2}{d \ln \mu} 
=  (- B_2 + C_{2}\, \alpha_2  + G_{2}\, \alpha_3-D_2\, \alpha_y)\,\alpha_2^2\,,
 \\
 \nonumber
\beta_{y}&\equiv&\frac{d \alpha_y}{d \ln \mu} 
= (E\, \alpha_y  -F_2\, \alpha_2-F_3\, \alpha_3)\,\alpha_y\,.
\eea
A few comments are in order. The one loop gauge coefficients $B_i$ can take either sign, depending on the BSM matter content. The two loop gauge coefficient $C_2$ is positive throughout. The two loop gauge coefficients $C_3$ may  take either sign if $B_3>0$, but is strictly positive as soon as $B_3\le 0$ \cite{Bond:2016dvk}. The two loop gauge mixing terms $G_i$ as well as the two loop Yukawa contribution $D_i$ and the one loop Yukawa terms $E$ and $F_i$ are always positive in any quantum field theory. The Yukawa couplings always contribute with a negative sign to the running of gauge couplings. This is centrally important for interacting UV fixed points to arise at weak coupling. 

Explicit expressions for the various loop coefficients and further details are summarised in the appendix, see \eq{Bs} -- \eq{EFs}. 
In the absence of BSM matter fields, the RG flow \eq{BSM} reduces to the RG flow of the SM with loop parameters given by \eq{SM}. In this limit, the RG flow for the BSM Yukawa coupling becomes obsolete. With \eq{BSM} at hand, we now turn to a systematic fixed point search within the perturbative regime.

\begin{table*}
\begin{center}
\begin{tabular}{cccccc}
\toprule
\rowcolor{LightBlue}
\rowcolor{LightGreen}
&
 \multicolumn{2}{c}{\cellcolor{LightGreen}
 \bf gauge couplings}
&
 \multicolumn{1}{c}{\cellcolor{LightGreen}
 \bf \ Yukawa coupling\ \ }
&\cellcolor{LightGreen}  
&\cellcolor{LightGreen}  
\\[-1mm]
\rowcolor{LightGreen}
\multirow{-2}{*}{\bf \ case\  }
&
${}\quad  {\al 3^*}\quad$
&
${}\quad {\al 2^*}\quad$
&
${}\quad \quad {\al y^*}\quad \quad$
&\cellcolor{LightGreen}  
\multirow{-2}{*}{\bf type}&\cellcolor{LightGreen}  
\multirow{-2}{*}{\cellcolor{LightGreen} \bf \  info\ }
\\
\midrule
\bf FP${}_{\bf 1}$
&0&0&0&{\bf  G $\bm \cdot$ G}&non-interacting\\[1.ex]
\rowcolor{LightGray}
&&&&& \\[-2.5ex]
\rowcolor{LightGray}
\bf FP${}_{\bf 2}$
&0&$\displaystyle\0{B_2}{{C'_2}}$&$\displaystyle\0{F_2}{E}\,\al 2^*$
&
{\bf  G $\bm \cdot$ GY}& partially interacting\\[1.5ex]
&&&&& \\[-2.5ex]
\bf FP${}_{\bf 3}$
&$\displaystyle\0{B_3}{{C'_3}}$&0&$\displaystyle\0{F_3}{E}\,\al 3^*$
&{\bf  GY $\bm \cdot$ G}&partially interacting\\[1.5ex]
\rowcolor{LightGray}
&&&&& \\[-2.5ex]
\rowcolor{LightGray}
\bf FP${}_{\bf 4}$
&$\ \displaystyle\frac{{C'_2}B_3-B_2G'_3}{{C'_2C'_3}-G'_2G'_3}\ $
&$\  \displaystyle\frac{{C'_3}B_2-B_3G'_2}{{C'_2C'_3}-G'_2G'_3}\  $
&$\displaystyle\frac{F_3}{E}\,\al 3^*+\frac{F_2}{E}\,\al 2^*$
&{\bf  GY $\bm \cdot$  GY}& fully interacting\\[1.5ex]
\bottomrule
\end{tabular}
\caption{The four different types of  UV fixed points \fp1 -- \fp4  in minimal BSM extensions of the SM  with \eq{BSM}. The primed and unprimed loop coefficients are defined in App.~\ref{AppA}.
We also indicate how the  fixed points can be interpreted as products of the Gaussian (G) and gauge-Yukawa (GY) fixed points when viewed from the individual gauge group factors (see main text).}
 \label{tFPs}
\end{center}
\end{table*}

\subsection{UV fixed points at weak coupling}

Gauge-Yukawa theories with  \eq{BSM} may display up to four different types of weakly coupled UV fixed points, depending on whether the gauge couplings take  free or interacting values in the UV.  We refer to the different cases as \fp1 -- \fp4, defined as 
\beq\label{FPs}
\begin{array}{rl}
$\fp1$\,:\quad\quad\alpha_2^*=0&\,,\quad\alpha_3^*=0\,,\\
$\fp2$\,:\quad\quad\alpha_2^*> 0&\,,\quad\alpha^*_3 = 0\,,\\
$\fp3$\,:\quad\quad\alpha_2^*=0&\,,\quad\alpha^*_3> 0\,,\\
$\fp4$\,:\quad\quad\alpha_2^*>0&\,,\quad\alpha^*_3> 0\,,
\end{array}
\eeq
see Tab.~\ref{tFPs}. The Gaussian fixed point \fp1, where all couplings vanish, always exists. It qualifies as a candidate for an asymptotically free extension of the SM provided that each gauge sector remains asymptotically free individually. Using the explicit expressions \eq{Bs} this condition translates into bounds
\beq
\label{af_qcd}
\begin{array}{l}
SU(2)_L:\quad N_F < {19}/({8\,S_2(R_2)\,d(R_3))}\,,\\
SU(3)_C:\quad N_F < {21}/({4}\,S_2(R_3)\,d(R_2))\,.
\end{array}
\eeq
In Tabs.~\ref{tFP3} and ~\ref{tFP2} we show the maximal number of BSM vector-like fermions 
$\psi(R_3,R_2)$ compatible with asymptotic freedom, $N_F\le N_{\rm AF}$ for $SU(2)_L$ 
singlets, doublets and triplets, and for different dimensions of the $SU(3)_C$ 
representations. 
We observe a small window for low-dimensional representations where asymptotic freedom persists. Asymptotic freedom is lost as soon as the BSM fermions transform under higher-dimensional representations of the gauge group. See \cite{Giudice:2014tma} for a recent analysis of BSM extensions with complete asymptotic freedom.

\begin{table}[t]
         \begin{center}
         \begin{tabular}{cccccccccc}
\rowcolor{LightGreen}
      \toprule
\rowcolor{LightGreen}
\multicolumn{4}{c|}{\cellcolor{LightBlue}$\bm{$\fp3$}$}
&\multicolumn{2}{c}{\cellcolor{LightGreen}$\ \ \ \ \bm{R_2= 1}\ \ \ \ $}
&\multicolumn{2}{c}{\cellcolor{LightGreen}$\ \ \ \ \bm{R_2= 2}\ \ \ \ $}
&\multicolumn{2}{c}{\cellcolor{LightGreen}$\ \ \ \ \bm{R_2= 3}\ \ \ \ $}\\
\rowcolor{LightYellow}
\cellcolor{LightGreen}
   $\ \ \bm{R_3}\ \ $      & 
   \cellcolor{LightGreen}$\ \ (p,q)\ \ $ &
   \cellcolor{LightGreen} $\ \ C_2(R_3)\ \ $ & 
   \multicolumn{1}{c|}{\cellcolor{LightGreen} $\ \ S_2(R_3)\ 
\ $ }& $\ \ N_{\rm AF}$ & \multicolumn{1}{c}{\cellcolor{LightYellow}$N_{\rm AS}$}& $\ \ N_{\rm AF}$ & $N_{\rm AS}$  & $\ \ 
N_{\rm AF}$ & $N_{\rm AS}$\\ \midrule
\rowcolor{LightGray}
$\bm 3$ & (1,0)& $\s043$  & $\s012$ & 10 & -- & 6 & -- & 3 & --\\ 
$\bm 6$ & (2,0)& $\s0{10}3$ & $\s052$ & 2 & (29) 37 & 1 & (60) 77 & -- & (90) 117 \\
\rowcolor{LightGray}
$\bm 8$ & (1,1)& 3 & 3 & 1 & (62) 96 & -- & (127) 198 & -- & (192) 299 \\
$\bm{10}$ & (3,0) & 6 & $\s0{15}2$ & -- & (16) 18 & -- & (32) 34 & -- & (48) 51\\
\rowcolor{LightGray}
$\bm{15}$ & (2,1) &$\s0{16}3$ & 10 & -- & (28) 30 & -- & (55) 60  & -- & (82) 90 \\
$\bm{15}^\prime$ & (4,0)& $\s0{28}3$ & $\s0{35}2$ & -- & (16) 18 & -- & (32) 33 & -- & (48) 50 \\ 
\bottomrule
         \end{tabular}
         \end{center}
         \caption{Asymptotic freedom versus asymptotic safety at the partially interacting fixed point \fp3: shown are the maximal numbers of BSM  fermion flavors 
compatible with 
asymptotic freedom, $N_{\rm AF}$, and the smallest number of flavors required for an asymptotically safe fixed point \fp3 to exist, $N_{\rm AS}$, both in dependence 
on the fermion representations $R_2$ and $R_3$ under  $SU(2)_L$ and $SU(3)_C$, respectively. $N_{\rm AS}$ values in brackets relate to the absolute lower bound, those without to fixed points with $0<\alpha^*_3,\alpha_y^*<1$. Also indicated  are the weights $(p,q)$, quadratic Casimir, 
and Dynkin index under $SU(3)$.}    
\label{tFP3}
\end{table}

Theories with \eq{BSM} may also display weakly interacting fixed points with $\alpha^*\le 1$. These are either partially or fully interacting.
Conditions for existence  of partially interacting UV fixed points 
such as \fp2 and \fp3 then reduce to those given in Sec.~\ref{simple} for simple gauge theories. Analogous conditions of existence arise for the fully interacting fixed point \fp4.
In either of theses cases, 
for \fp2, \fp3 or \fp4 to qualify as asymptotically safe UV fixed points, 
the Yukawa coupling must take an interacting fixed point by itself. To the leading non-trivial order in perturbation theory, using \eq{BSM}, it follows that the Yukawa coupling at a fixed point is linearly related to the gauge couplings,
\beq\label{Y}
\begin{array}{rl}
$\fp2$\,:\quad\quad\alpha_y^*&=\di\0{F_2}{E}\, \alpha^*_2\,,\\[1.5ex]
$\fp3$\,:\quad\quad\alpha_y^*&=\di\0{F_3}{E}\, \alpha^*_3\,,\\[1.5ex]
$\fp4$\,:\quad\quad\alpha_y^*&=\di\0{F_2}{E}\, \alpha^*_2+\0{F_3}{E}\, \alpha^*_3\,,
\end{array}
\eeq
depending on whether $\alpha_2$, or $\alpha_3$, or both, take interacting fixed points by themselves.
Combining \eq{Y} with the vanishing of the gauge beta functions provides explicit expressions for the different fixed points. An overview of fixed points and their properties is given in Tab.~\ref{tFPs}. Next we analyse minimal conditions that need to be fulfilled in the BSM sector in order to generate partially or fully interacting UV fixed points in the system \eq{BSM}.

\subsection{Partially interacting fixed points}\label{s-partial}

The partially interacting fixed points \fp2 and \fp3  are characterised by one of the gauge couplings, say $\alpha_{\rm AS}$, taking an asymptotically safe  fixed point in the UV whereby the other gauge coupling, say  $\alpha_{\rm AF}$, becomes asymptotically free.  
The Yukawa couplings must take interacting values, $\alpha_y^*\propto\alpha^*_{\rm AS}$, see \eq{Y}. The beta functions \eq{BSM} then take the simplified form
\beq
\begin{array}{rl}
\beta_{\rm AS}
&= 
 (- B_{\rm AS} + C_{\rm AS}\, \alpha_{\rm AS}  -D_{\rm AS}\, \alpha_y)\,\alpha_{\rm AS}^2\,,
\\
\beta_{y}
&= (E\, \alpha_y  -F_{\rm AS}\, \alpha_{\rm AS})\,\alpha_y\,.
\end{array}
\eeq
These expressions  formally agree with \eq{beta2} and therefore offer the same  type of fixed point  solutions.
\begin{table}[t]
         \begin{center}
         \begin{tabular}{cccccccccc}
\rowcolor{LightGreen}
\toprule
\rowcolor{LightGreen}
\multicolumn{4}{c|}{\cellcolor{LightBlue}$\bm{$\fp2$}$}
&\multicolumn{2}{c}{\cellcolor{LightGreen}$\ \ \ \  \bm {R_3=1}\ \ \ \ $}
&\multicolumn{2}{c}{\cellcolor{LightGreen}$\ \ \ \  \bm {R_3= 3}\ \ \ \ $}
&\multicolumn{2}{c}{\cellcolor{LightGreen}$\ \ \ \  \bm {R_3=  6}\ \ \ \ $}\\ 
\rowcolor{LightYellow}
\rowcolor{LightYellow}
\cellcolor{LightGreen}   $\ \ \bm{R_2}\ \ $      
& \cellcolor{LightGreen}$\ \ \ell\ \ $ 
&\cellcolor{LightGreen} $\ \ C_2(R_2)\ \ $ 
&\multicolumn{1}{c|}{\cellcolor{LightGreen}  $\ \ S_2(R_2)\ \ $} 
& $\ \ N_{\rm AF}$ & $N_{\rm AS}$  & $\ \ N_{\rm AF}$ & $N_{\rm AS}$  & $\ \ 
N_{\rm AF}$ & $N_{\rm AS}$\\ 
\midrule
$\bm 2$ & $\s012$& $\s034$  & $\s012$ & 4 & -- & 1 & -- & -- & --\\
\rowcolor{LightGray}
$\bm 3$ & 1 & 2 & 2 & 1 & (26) 53 & -- & (73) 154 & -- & (145) 307 \\
$\bm 4$ & $\s032$ & $\s0{15}4$ & 5 & -- & (7) 9 & -- & (21) 24 & -- & (41) 47\\
\rowcolor{LightGray}
$\bm 5$ & 2 & 6 & 10 & -- & (6) 7 & -- & (17) 18 & -- & (33) 35\\
$\bm 6$ & $\s052$ & $\s0{35}4$ & $\s0{35}2$ & -- & (6) 7 & -- & (16) 17 & -- & (31) 33\\
\bottomrule
         \end{tabular}
         \end{center}
         \caption{Asymptotic freedom versus asymptotic safety at the partially interacting fixed point \fp2: shown are the maximal numbers of BSM  fermion flavors $N_F<N_{\rm AF}$  compatible with 
asymptotic freedom, and the smallest number  $N_F\ge N_{\rm AS}$ required for a weakly-coupled asymptotically safe fixed point, both in dependence 
on the fermion representations $R_2$ and $R_3$ under  $SU(2)_L$ and $SU(3)_C$, respectively. Values for $N_{\rm AS}$ in brackets relate to the absolute lower bound, those without brackets to settings with $0<\alpha^*_2,\alpha_y^*<1$. Also indicated  are the weight $\ell$, the quadratic Casimir, 
and the Dynkin index under $SU(2)_L$.}    
\label{tFP2}
\end{table}
The non-trivial UV fixed point is then of the form \eq{C'}, \eq{GY}, after substituting the appropriate loop coefficients.
A minimal requirement for partially interacting fixed points to be UV fixed points is the loss of asymptotic freedom in the gauge sector $B_{\rm AS}<0$, meaning either 
\beq
\label{as_qcd}
\begin{array}{rl}
$\fp2$\,:&\quad N_F > {19}/({8\,S_2(R_2)\,d(R_3))}\,,\\
{\rm or}\quad 
$\fp3$\,:&\quad N_F > {21}/({4}\,S_2(R_3)\,d(R_2))\,,
\end{array}
\eeq
thus reverting the condition \eq{af_qcd}. 
Associating suitable charges to the BSM fermions, it is then possible to satisfy either of the conditions in \eq{as_qcd}.  Furthermore, the physicality condition \eq{condition} translates into
\beq\label{conditionPI}
\begin{array}{l}
$\fp2$\,:\quad D_2\,F_2-E\,C_2>0\,,\\
$\fp3$\,:\quad D_3\,F_3-E\,C_3>0\,.
\end{array}
\eeq
It  remains to evaluate solutions to the conditions \eq{conditionPI} separately for \fp2 and \fp3, to which we turn next.\\[-1.5ex]

\begin{figure}[t]
\begin{center}
{}\vskip-.3cm
\includegraphics[width=0.9\textwidth]{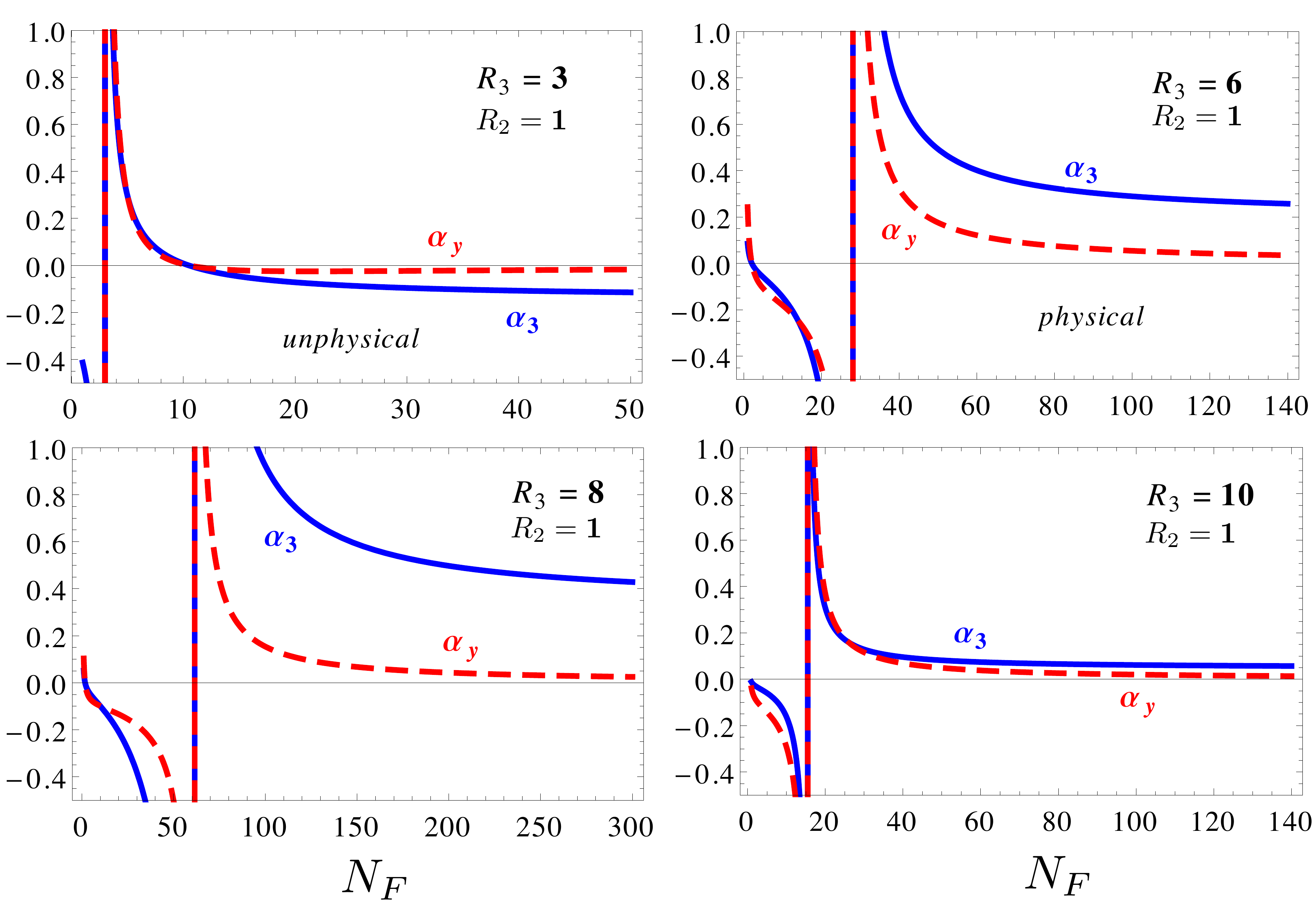}
\end{center}
\vskip-1cm
\caption{Partially interacting fixed point 
\fp3 with $\alpha_2^*=0$, showing the strong coupling 
$\alpha_3^*$ (blue, solid line) and the BSM Yukawa coupling $\alpha_y^*$ (red, dashed) versus the  number $N_F$ of the BSM flavors  for $R_2=\bm{1}$ and different  $SU(3)_C$  representations $R_3=\bm{3}, \bm{6}, \bm{8}$ and $\bm{10}$ 
 (see main text).}
\label{fig:asyA}
\end{figure}

\begin{figure}[t]
\begin{center}
\vskip-.4cm
\includegraphics[width=\textwidth]{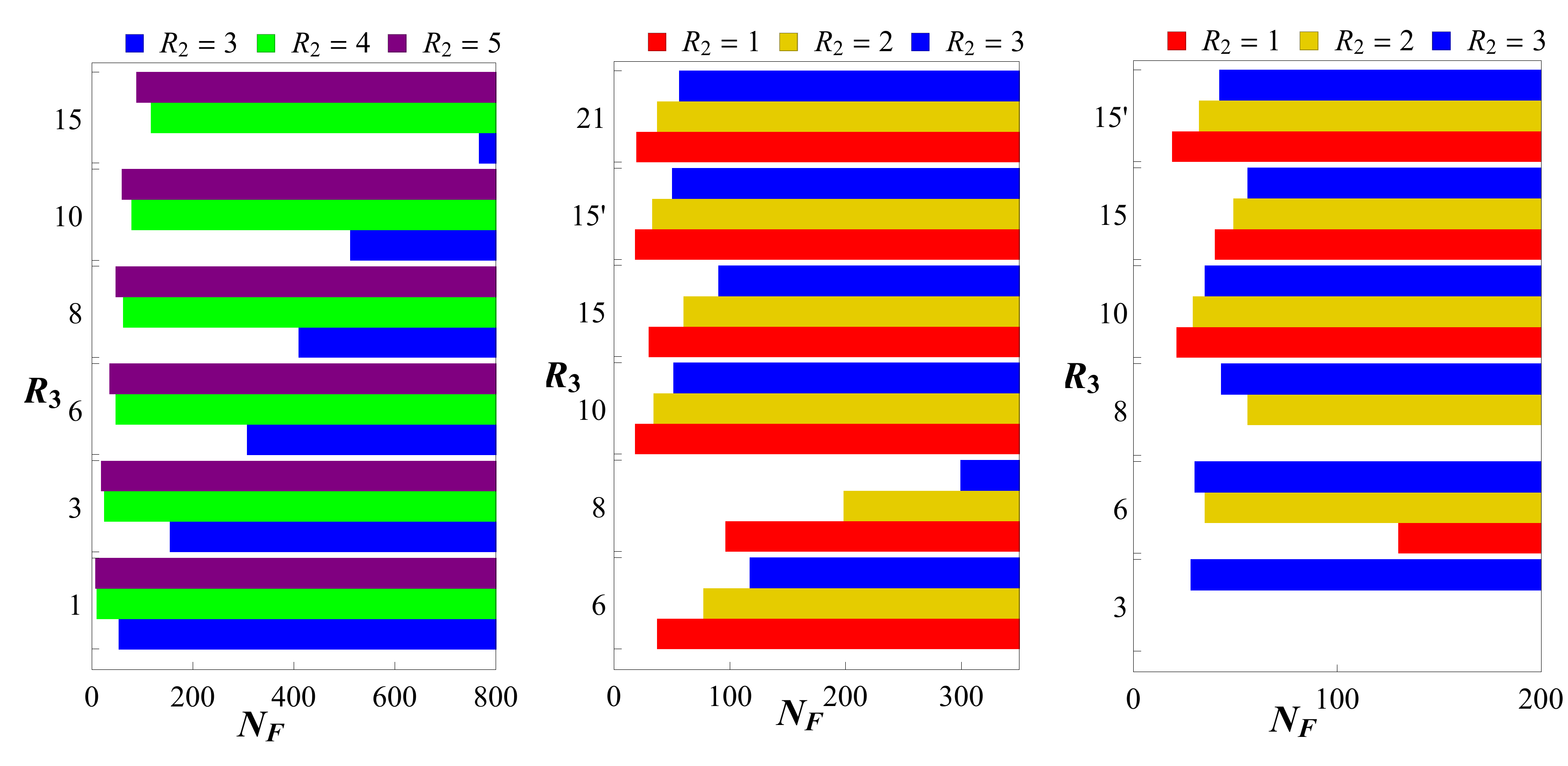}
\end{center}
\vskip-.5cm
\caption{Availability of weakly interacting UV fixed points \fp2 (left panel), \fp3 (middle panel), and \fp4 (right panel) in dependence on the representation $(R_2,R_3)$ and the flavor multiplicities $N_F$ of  BSM fermions. The pattern of results continues to higher $(R_2,R_3)$. Partially interacting fixed points \fp2 are absent for any $N_F$ as soon as $R_2=\bm1$ or $\bm2$; \fp3 is absent whenever  $R_3=\bm1$ or $\bm3$;  fully interacting UV fixed points  \fp4 are absent for $R_3=\bm1$ or  $(R_2,R_3)=(\bm1,\bm8), (\bm2,\bm3)$, and $(\bm1,\bm3)$.} 
\label{fig:fp234}
\end{figure}

{\bf  Strong strong and weak weak gauge coupling}. In Fig.~\ref{fig:asyA} we analyse the condition \eq{conditionPI} exemplarily for \fp3 where the strong coupling remains interacting in the deep UV whereas the weak coupling vanishes asymptotically. We assume that the BSM fermions carry no $SU(2)_L$ charges $(R_2=\bm{1})$, but different  $SU(3)_C$  representations $R_3=\bm{3}, \bm{6}, \bm{8}$ and $\bm{10}$. We observe the following pattern. For fermions in the fundamental, a narrow window of weakly interacting fixed points exists for a low number of flavors $N_F$.  
These low-$N_F$ solutions come out as IR fixed points in that they relate to settings with asymptotic freedom in both gauge sectors (see the discussion in Sec.~\ref{synopsis}).
With increasing $N_F$, the fixed point takes negative values and becomes unphysical. 
Conversely, for fermions in higher-dimensional representations (anything but the fundamental), we find that a  fixed point exists for sufficiently large $N_F$. No fixed points exist for intermediate values of $N_F$. Occasionally we find that fixed points can exist for exceptionally low values of $N_F$, in which case the fixed point is IR rather than UV. 
In Fig.~\ref{fig:fp234} (middle panel), we show the set of parameters $(R_2,R_3,N_F)$ for which \fp3 exists as an interacting UV fixed point. In Tab.~\ref{tFP3} we summarise the minimum number of BSM fermions $N_{\rm AS}$ which lead to a weakly coupled UV fixed point with $\alpha^*\le 1$. 

The pattern of results is easily understood once $N_F$ is sufficiently large. The necessary condition for existence \eq{conditionPI} of \fp3  turns into a quadratic polynomial in $N_F$ after inserting the explicit expressions for the loop coefficients,
\beq\label{conditionPIexp}
X\,N_F^2+Y\,N_F-\, Z<0\,,
\eeq
with coefficients
\beq\label{xyz}
\begin{array}{rl}
X&=Z\,C_2(R_3)\,[5-2C_2(R_3)]/52\,,\\
Y&=Z^2[C_2(R_3)+5]/52-C_2(R_3)\,,\\
Z&=C_2(R_3)d(R_3)d(R_2)\,,
\end{array}
\eeq
and with $C_2(R)$  and $d(R)$
defined in \eq{group}. For sufficiently large $N_F$, the sign of the coefficient $X$ dictates whether  the condition \eq{conditionPIexp} provides an upper or a lower bound on $N_F$.
If  $X>0$, the condition \eq{conditionPIexp} provides an upper bound on the number of the BSM 
fermions. However, we observe that $X>0$ if and only if  the BSM fermions transform under the fundamental representation of $SU(3)_C$ (see Tab.~\ref{tFP3} for explicit values of 
the Casimir invariant for several $R_3$ of the lowest dimension).
In this case it is readily confirmed that a solution to \eq{conditionPIexp} is incompatible with the  lower bound from \eq{af_qcd} for any choice of $R_3$, meaning that such a fixed point is necessarily an IR fixed point. We conclude that asymptotic safety via a partially interacting fixed point  cannot be achieved within the fundamental representation of $SU(3)_C$.
On the other hand, for higher-dimensional representations the coefficient $X$  becomes negative. Consequently, \eq{conditionPIexp}  provides a lower bound on the number of BSM 
fermions required to achieve asymptotic safety, $N_F\ge N_{\rm AS}$. 
The case $X=0$ has no physical solutions.  Exemplary values for the lower bound , $N_F\ge N_{\rm AS}$ for different representations $R_3$ and $R_2$ are given in Tab.~\ref{tFP3},
where we additionally require weak  coupling $\alpha_i^*<1$ at the fixed point.\\[-1.5ex]

{\bf Strong weak and weak strong gauge coupling}. Next we turn to \fp2 where the weak sector remains interacting in the deep UV whereas the strong coupling becomes asymptotically weak. Qualitatively, our findings for \fp2 are very similar to those discussed previously for  \fp3. 
The absence of asymptotic freedom in the $SU(2)_L$ gauge sector, \eq{as_qcd},  requires 
a minimal number of BSM fermion flavors $N_F\ge N_{\rm AF}$.  In Fig.~\ref{fig:fp234} (left panel), we show the set of parameters $(R_2,R_3,N_F)$ for which \fp2 exists as an interacting UV fixed point.
In Tab.~\ref{tFP2}, we provide $N_{\rm AF}$ 
for $SU(3)_C$ singlets,  triplets and sextets, and for different dimensions of the $SU(2)_L$ 
representations. Since the SM contribution to the one-loop gauge coefficient is larger for $SU(2)_L$ than for $SU(3)_C$, 
lower values for $N_F$ and lower dimensions of representations are required to lose asymptotic freedom for $SU(2)_L$. Similarly, from \eq{conditionPI} we find that asymptotic safety cannot be achieved with fermions in the fundamental representations of $SU(2)_L$.
The minimal number of BSM fermion flavors required for a weakly-coupled asymptotically safe fixed point, $N_{\rm AS}$, 
are given in Tab.~\ref{tFP2} for various choices of $R_2$ and $R_3$.

\subsection{Regaining  asymptotic freedom}

\begin{figure}[t]
\begin{center}
\vskip-.5cm
\includegraphics[width=0.875\textwidth]{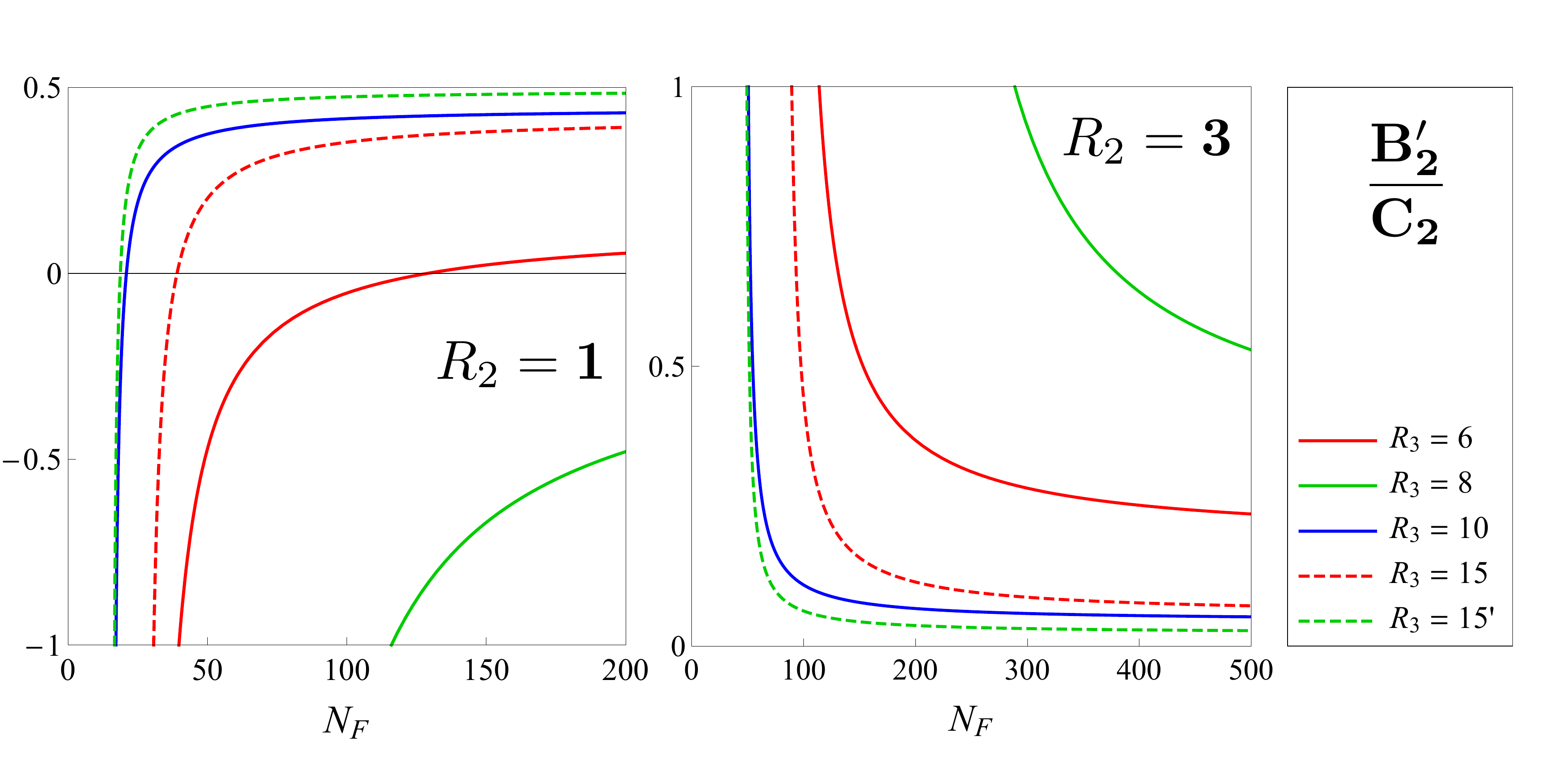}
\end{center}
\vskip-.8cm
\caption{Asymptotic freedom at partially interacting fixed points \fp3. Shown is the condition for asymptotic freedom \eq{BAF} for the effective coefficient $B_2'$ in units of the two loop coefficient $C_2>0$ at \fp3 for, exemplarily,  $R_2=\bm1$  (left panel) and $R_2=\bm3$  (right panel), and as a function of $R_3$ (color coding given in the legend). We observe that asymptotic freedom in the weak sector is regained as soon as $R_2>\bm1$ and $R_3>\bm3$, for any $N_F$. For $R_2=\bm1$, a lower bound on $N_F$ is found (left panel). Qualitatively and quantitatively similar results are obtained at \fp2 (not displayed). 
}  
\label{fig:B2'}
\end{figure}

Next we  discuss the fate of the gauge coupling which vanishes at partially interacting fixed points \fp2 or \fp3, and which we denote for notational simplicity as $\alpha_{\rm AF}$. The coupling $\alpha_{\rm AF}$ must be asymptotically free for a partially interacting fixed point to be viable, or else the UV fixed point cannot be reached by any finite RG trajectory along the $\alpha_{\rm AF}$ direction. In general, we find that $B_{\rm AF}$ becomes negative as soon as BSM fermions carry charges of both gauge groups.  However, the sign of $B_{\rm AF}$ plays no role, as  it 
no longer dictates whether this sector remains asymptotically free or not.
Rather, to leading order in the asymptotically free gauge coupling, we have
\beq\label{alphaAF}
\beta_{\rm AF}
=  \left(- B_{\rm AF}  + G_{\rm AF}\, \alpha^*_{\rm AS}-D_{\rm AF}\, \alpha^*_y\right)\,\alpha_{\rm AF}^2+{\cal O}(\alpha_{\rm AF}^3)\,,
 \eeq
showing that the one loop coefficient $B_{\rm AF}$ is replaced by  $B'_{\rm AF}=B_{\rm AF}  - G_{\rm AF}\, \alpha^*_{\rm AS}+D_{\rm AF}\, \alpha^*_y$. 
We stress that this shift is a consequence of  partially interacting fixed points. 
 It arises from residual interactions at the UV fixed point due to asymptotic safety of the gauge coupling $\alpha_{\rm AS}$ and the BSM Yukawa coupling. Their residual interactions modify the running of the asymptotically free coupling owing to fermions which carry charges under both gauge groups. Provided that the shifted one loop coefficients $B'$ take positive values, 
\beq\label{BAF}
B'_{\rm AF}>0\,,
\eeq 
the non-interacting gauge sector becomes asymptotically free in the deep UV. We also stress  that the BSM Yukawa interactions play a central role: only Yukawa couplings add negatively to the beta function  \eq{alphaAF}. 
Without them, \eq{BAF} cannot be achieved starting from $B_{\rm AF}<0$. 
Using \eq{BSM}, we have the following  expressions for the shifted one loop coefficients
\beq\label{Bprime}
\begin{array}{l}
$\fp2$\,:\quad B_3\to B_3'= B_3-G_3\,\alpha^*_2 +D_3\,\alpha^*_y\,,\\
$\fp3$\,:\quad B_2\to B_2'= B_2-G_2\,\alpha^*_3+D_2\,\alpha^*_y \,.
\end{array}
\eeq
We conclude that \eq{BAF}, \eq{Bprime} are necessary conditions for the corresponding partially interacting fixed point to qualify as UV completions of the SM.

In Fig.~\ref{fig:B2'} the condition for asymptotic freedom \eq{Bprime} at \fp3 is shown for models with $R_2=\bm1$ (left panel) and $R_2=\bm3$ (right panel) and various $R_3>\bm3$ (recall that there are no viable UV fixed points \fp3 for $R_3\le \bm3$, Fig.~\ref{fig:fp234}). If $R_2=\bm1$, we observe that $B_2'$ is positive for sufficiently large $N_F$, and negative for sufficiently low  $N_F$, thus leading to a lower bound. Conversely, if $R_2=\bm3$ (or larger), the sign of $B_2'$ is always positive. In this case asymptotic freedom is guaranteed without any further constraints as soon as $\alpha_3$ is asymptotically safe. The same pattern of results holds true for \fp2. We conclude that as soon as the BSM fermions carry a non-trivial charge under the asymptotically free coupling $R_{\rm AF}\neq \bm1$, for any $R_{\rm AS}\neq \bm1$, the condition \eq{BAF} follows from the condition for asymptotic safety for $\alpha_{\rm AS}$ \eq{conditionPI}. For BSM fermions with 
$R_{\rm AF}= \bm1$, \eq{BAF} entails an additional lower bound on $N_F$.

\subsection{Fully interacting fixed points}
Finally we consider the case \fp4. In the case where both gauge couplings and the BSM Yukawa coupling remain weakly interacting at the fixed point in the asymptotic UV the overall behaviour of the system \eq{BSM} depends on the interplay between one- and two-loop coefficients. Using the results of Tab.~\ref{tFPs}, the necessary condition for a fixed point can be stated as
\beq\label{conditionFI}
\alpha_2^*=\frac{{C'_3}B_2-B_3G'_2}{{C'_2C'_3}-G'_2G'_3}>0\,,\quad
\alpha_3^*=\frac{{C'_2}B_3-B_2G'_3}{{C'_2C'_3}-G'_2G'_3}>0\,,
\eeq
with primed two-loop coefficient given in \eq{Cs'}. Unlike the condition for partially interacting fixed points \eq{conditionPI}, those for fully interacting ones involve ratios of differences of Yukawa-shifted loop coefficients. In particular, fully interacting fixed points may exist even if only one of the conditions \eq{conditionPI} is satisfied. For the purpose of this work, we have investigated the conditions \eq{conditionFI} numerically. In Fig.~\ref{fig:fp234} (right panel), we show the set of parameters $(R_2,R_3,N_F)$ for which  \fp4 exists as an interacting UV fixed point. Our results for the lowest number of flavor multiplicities $N_F\ge N_{\rm AS}$  are summarised in Tab.~\ref{tFP4}.

\subsection{Large-$N_F$ approximation}\label{largeNF}
 Some  analytical insights about interacting UV fixed points can be obtained in the limit of many flavors of fermions $N_F\gg1$, which we discuss separately for either type of fixed point.\\[-1.5ex]

 \begin{table}[t]
         \begin{center}
         \begin{tabular}{ccccccccc}
\rowcolor{LightGreen}
\toprule
\rowcolor{LightGreen}
\multicolumn{1}{c|}{\cellcolor{LightBlue}$\bm{$\quad \fp4\quad $}$}
&\multicolumn{1}{c}{\cellcolor{LightGreen}$\   \bm {R_3= 1} \ $}
&\multicolumn{1}{c}{\cellcolor{LightGreen}$\  \bm {R_3= 3}\  $}
&\multicolumn{1}{c}{\cellcolor{LightGreen}$\  \bm {R_3= 6} \ $}
&\multicolumn{1}{c}{\cellcolor{LightGreen}$\  \bm {R_3= 8} \ $}
&\multicolumn{1}{c}{\cellcolor{LightGreen}$\  \bm {R_3= 10} \ $}
\\ 
\rowcolor{LightGreen}
\rowcolor{LightYellow}
\multicolumn{1}{c|}{ \cellcolor{LightGreen}  $\ \ \bm{R_2}\ \ $}&  $N_{\rm AS}$  & $N_{\rm AS}$  &  $N_{\rm AS}$ &  $N_{\rm AS}$  &  $N_{\rm AS}$ \\ 
   \midrule
$\bm 1$ & -- & -- & (130) 130 & -- & (21) 21\\
\rowcolor{LightGray}
$\bm 2$ & -- & -- & (29) 35 & (45) 56 & (27) 29 \\
$\bm 3$ & -- & (23) 28 & (27) 30 & (38) 43 & (33) 35\\
\rowcolor{LightGray}
$\bm 4$ &-- & (17) 18 & (26) 28 & (36) 39 & (37) 39\\
$\bm{5}$ &-- & (15) 16 & (27) 28 & (36) 38 & (40) 42\\
\bottomrule
         \end{tabular}
         \end{center}
         \caption{The minimal number of the BSM fermions flavors $N_F\ge N_{\rm AS}$ required for the fully interacting   fixed point \fp4 to exist, in dependence 
on the fermion representations $R_2$ and $R_3$ under $SU(2)_L$ and $SU(3)_C$.  The values for $N_{\rm AS}$ in brackets relate to the absolute lower bound, and those without to settings with $0<\alpha^*_3,\alpha^*_2,\alpha_y^*<1$.
}    
\label{tFP4}
\end{table}

{\bf Partially interacting fixed points.} For a partially interacting fixed point, and using the explicit solution for \fp3 as given in Tab.~\ref{tFPs}, the large-$N_F$ approximation leads to
\begin{align}\label{fp3N}
(\alpha_3^*,\alpha_2^*,\alpha_y^*)\Big|_{N_F\gg 1}=\frac{1}{X_3}\left(\frac{1}{3}\,,\,0\,,\,\frac{2C_2(R_3)}{N_F}\right) +{\rm subleading\,,} 
\end{align}
where $X_3(R_3)=2C_2(R_3)-5$. The subleading terms are at least one power in $N_F$ smaller than the leading order terms. Several observations can now be made. First of all, positivity of the fixed point couplings requires $X_3(R_3)>0$ or 
$C_2(R_3)>\s052$.
Hence, our  result confirms that asymptotic safety cannot be achieved
within the fundamental representations of $SU(3)_C$ even at large-$N_F$, owing to $X_3({\rm fund.})<0$.
Secondly, we observe that the Yukawa coupling scales like $1/N_F$ and can always be made arbitrarily small. Conversely, the size of the gauge coupling is solely determined by  the quadratic Casimir $C_2(R_3)$, and  independent of $N_F$ in the large-$N_F$ limit.
We stress that \eq{fp3N} is parametrically close to the Gaussian fixed point, provided that $X_3$ becomes parametrically large.  In addition, the necessary condition for asymptotic freedom \eq{BAF} for the weak coupling simplifies to leading order at large-$N_F$ and reads
$C_2(R_3)>\s052$. Interestingly, the condition for asymptotic freedom exactly coincides with the condition for asymptotic safety of \eq{fp3N} at large-$N_F$,
\beq\label{conditionfp3N}
C_2(R_3)>\052\,,\quad{\rm or}\quad
 R_3\ge \bm6\,.
\eeq
We conclude that higher dimensional representations under $SU(3)_C$ with \eq{fp3N}
are favoured for the theory to display a perturbative UV fixed point in the  $SU(3)_C$ coupling, and for $SU(2)_L$ sector to regain asymptotic freedom at the partially interacting UV fixed point \fp3, see Fig.~\ref{fig:asyA}. Analogous results are established for \fp2, where the large-$N_F$ expansion starts off with 
\begin{align}\label{fp2N}
(\alpha_3^*,\alpha_2^*,\alpha_y^*)\Big|_{N_F\gg 1}=\frac{1}{X_2}\left(0\,,\,\frac{1}{2}\,,\,\frac{3C_2(R_2)}{N_F}\right) +{\rm subleading\,,} 
\end{align}
and $X_2(R_2)=3C_2(R_2)-5$. Again, subleading terms are suppressed by at least one additional power in $N_F$ over the leading terms.  A necessary condition for asymptotic safety is $X_2(R_2)>0$, thus excluding the fundamental representation owing to $X_2({\rm fund.})<0$. We also conclude that \fp2 is parametrically close to the Gaussian fixed point in the limit of high-dimensional representations $X_2$. Furthermore, to leading order at large-$N_F$ the condition for asymptotic freedom \eq{BAF} for the strong coupling becomes
\beq\label{conditionfp2N}
C_2(R_2)>\053\,,\quad{\rm or}\quad R_2\ge \bm3\,.
\eeq
Once more, this secondary condition coincides with the condition for asymptotic safety of \eq{fp2N}. \\[-1.5ex]

\begin{figure}[t]
\begin{center}
\includegraphics[width=0.65\textwidth]{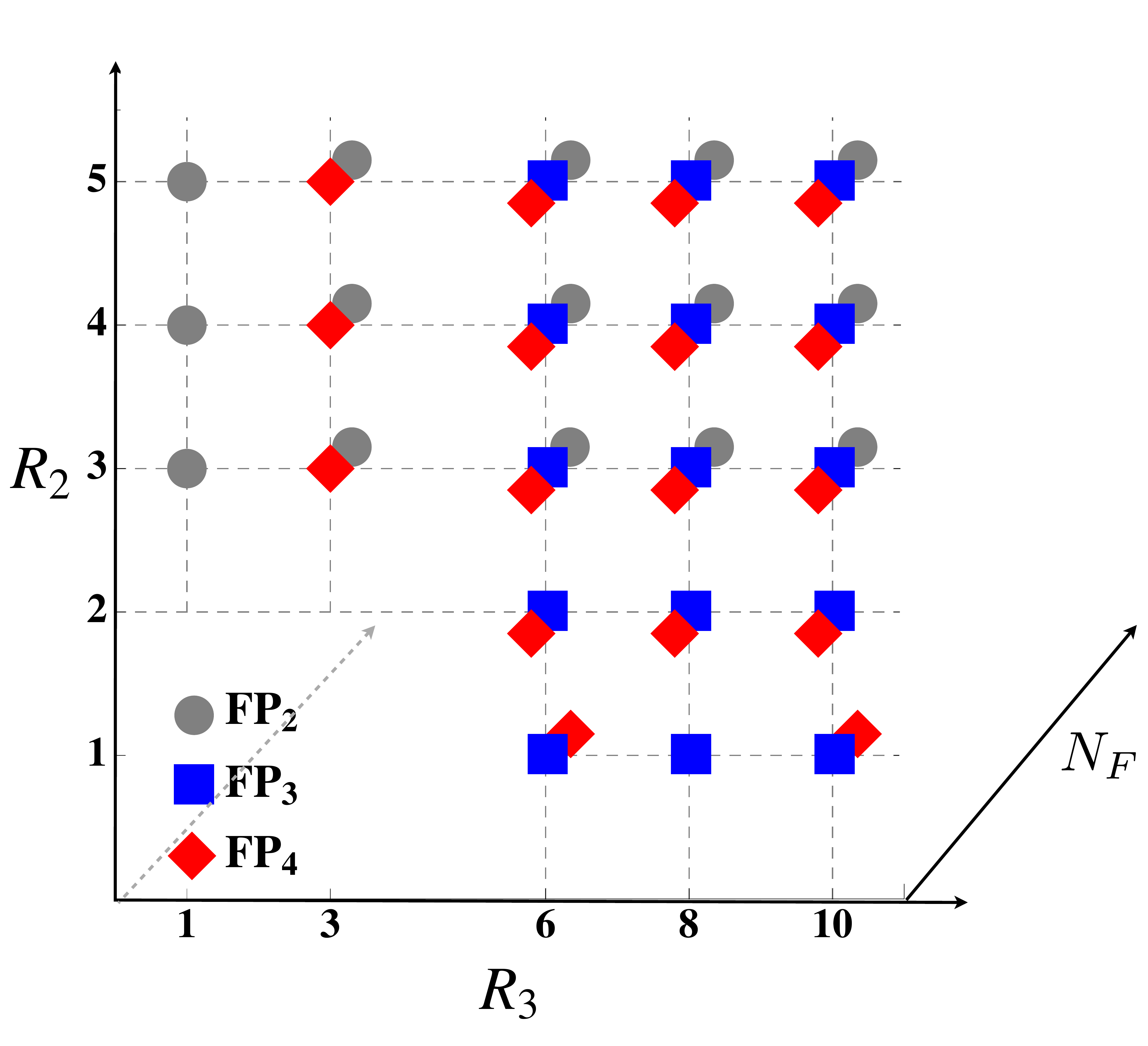}
\end{center}
\vskip-.5cm
\caption{Summary of weakly interacting UV fixed points of \eq{BSM} in dependence on the fermion representation and flavor multiplicities $(R_3,R_2,N_F)$. The different symbols relate to \fp2 (gray circle), \fp3 (blue square) and \fp4 (red diamond). 
Overlapping symbols indicate that either type of fixed point can exist, with lower-lying symbols relating to fixed points which arise at a higher number of fermion flavors $N_F$, see also Fig.~\ref{fig:fp234}.} 
\label{fSynopsis}
\end{figure}

{\bf Fully interacting fixed points.} For the fully interacting fixed point, using the explicit solution for \fp4 as given in Tab.~\ref{tFPs} and performing a large-$N_F$ limit, we find
\begin{align}\label{fp4N}
(\alpha_3^*,\alpha_2^*,\alpha_y^*)\Big|_{N_F\gg 1}=
\frac{1}{X_{32}}\left(\frac{1}{3}\,,\,\frac{1}{2}\,,\,\frac{2C_2(R_3)+3C_2(R_2)}{N_F}\right)+{\rm subleading\,,} 
\end{align}
with $X_{32}=2C_2(R_3)+3C_2(R_2)-5$. The above expression holds true provided that $R_3\neq \bm1$ and $R_2\neq \bm1$. The requirement of asymptotic safety results in an inequality $X_{32}>0$. Furthermore, the result also shows that the fully interacting fixed point is parametrically close to the Gaussian provided that $X_{32}$ is large.
The explicit result  explains why a fully interacting fixed point
with asymptotic safety can be achieved even with  BSM fermions in the fundamental representation of 
$SU(3)_C$, as long as 
they transform under $SU(2)_L$  in a representation of a dimension higher than the fundamental. Analogously, \fp4 exists for BSM fermions in the fundamental representation of $SU(2)_L$ provided that $R_3>\bm3$. Note that these large-$N_F$ estimates are in very good agreement with the numerical findings in Tab.~\ref{tFP4}.
In the special case where  $R_3>\bm1$ and $R_2=\bm1$, and instead of \eq{fp4N}, one obtains
\begin{align}\label{fp4N2}
(\alpha_3^*,\alpha_2^*,\alpha_y^*)\Big|_{N_F\gg 1}=\left(\frac{1}{3X_{32}}\,,\,\frac{19}{35}
-\frac{24}{35X_{32}}\,,\,
\frac{2C_2(R_3)}{N_F X_{32}}\right)+{\rm subleading} \,,
\end{align}
with $X_{32}=2C_2(R_3)+3C_2(R_2)-5$ as before. Notice that this fixed point is parametrically close to the  ``would-be'' Banks-Zaks fixed point in the $SU(2)_L$ sector of the SM.  The condition for existence is now given by $X_{32}>\s0{24}{19}$ which translates into $C_2(R_3)>3\s05{38}$. Solutions are given by the $R_3=\bm6$ and $R_3\geq\bm{10}$  representations under $SU(3)_C$. Curiously, the adjoint representation $R_3=\bm8$  with $R_2=\bm1$ is not a solution  of \eq{fp4N2} owing to  the  ``would-be'' Banks-Zaks IR fixed point.
Finally, for $R_3=\bm1$ and $R_2\geq\bm1$ 
one readily confirms that $\alpha_3^*$ and $\alpha_2^*$ cannot simultaneously take positive values meaning that an asymptotically safe fixed point does not arise at large $N_F$.

\subsection{Synopsis of UV fixed points}\label{synopsis}

We are now in a position to summarise the main results for weakly interacting UV fixed points in extensions of the SM of the form \eq{L}. We have observed that interacting UV fixed points can arise as partially or fully interacting ones. In either of these cases, necessary conditions for their existence have been found, providing us with constraints on the remaining BSM parameters $(R_3,R_2,N_F)$.  We have also observed that for fixed $(R_3,R_2)$, UV fixed points typically exists for all $N_F$ down to  limiting values specified in Tab.~\ref{tFP3}, \ref{tFP2} and \ref{tFP4}. 
 Fig.~\ref{fSynopsis} shows a summary of our findings, in dependence on the fermion representation $(R_3,R_2)$ under $SU(3)_C\otimes SU(2)_L$ with different symbols relating to the different fixed points \fp2, \fp3, and \fp4. Broadly speaking, results show the existence of fixed points both with increasing dimensionality of the fermion representation, and with increasing flavor multiplicities. Similarly, fixed points come out less strongly coupled the larger their dimensionality $R_2$, $R_3$ and the flavor multiplicity $N_F$.
We also observe that different  types of  fixed points might  coexist for BSM fermions with the same set of  representations $(R_2,R_3)$, starting from a lowest value for $N_F$ where the fixed point arises for the first time. In Fig.~\ref{fSynopsis} 
 the possibility of coexistence is indicated by overlapping symbols: the lower lying symbol relates to a fixed point which arises for larger $N_F$.

Hence,  four  distinct cases arise: $(i)$ For high dimensional representations, starting from $R_2=\bm 3$ and $R_3=\bm 6$ onward,  all three types of fixed points are realised starting from some lowest value for $N_F$. For fixed $(R_2,R_3)$ but with increasing $N_F$ results show  that  the fully interacting fixed point \fp4 is achieved first, followed by the partially interacting \fp3 for the strong coupling, and ultimately followed by the partially interacting fixed point \fp2 in the weak coupling.
$(ii)$ If the fermions are in the fundamental representation of one of the two gauge groups, we find that the corresponding partially interacting fixed point is absent throughout. However, the other two fixed points still exist and the order in which they appear, with increasing $N_F$, is exactly the same as the order observed for the higher dimensional representations. 
$(iii)$ If the BSM fermions are uncharged under $SU(3)_C$, only the partially interacting fixed point \fp2 can arise, starting from $R_2=\bm 3$ onwards. 
$(iv)$ If the BSM fermions are uncharged under $SU(2)_L$, we find that  \fp3 arises first, followed by \fp4, while \fp2 is absent throughout. This holds true for all $R_3=\bm 6$ or higher, except for $R_3=\bm 8$ where only \fp3 appears.

Finally, we discuss the status of interacting fixed points for low numbers of flavors $N_F$. The low-$N_F$ partially interacting fixed points  \fp3 at $R_2=\bm1$ with $R_3=\bm3,\bm6$ and $\bm8$  all have $B_2,B_3>0$, \eq{af_qcd}.
Hence the theory remains asymptotically free in both gauge group factors, and the interacting fixed point is formally an IR fixed point of the type discussed in Tab.~\ref{tUV}~{\bf b)}. 
Similarly, for \fp2 and \fp4 we find a handful of low-$N_F$ fixed points all of which occur where asymptotic freedom persists in both gauge groups \eq{af_qcd}. 
Conversely, we also have low-$N_F$ fixed points \fp3  with $R_2=\bm2,\cdots,\bm{10}$ which have asymptotic freedom only in the strong gauge coupling, while the weak sector has become infrared free. Such fixed points are not of phenomenological interest because they cannot be linked with any finite $\alpha_2\neq 0$ in the IR and shall be dropped.

This completes our investigation of weakly coupled UV fixed points of \eq{L} to the leading non-trivial order in perturbation theory, \eq{BSM}. In the next section, we explain whether and how   these fixed points are connected with the SM at low energies under the RG evolution of couplings.

\section{\bf Matching onto the Standard Model}\label{SMmatching}
In this section, we evaluate the conditions under which BSM  trajectories emanating out of interacting UV fixed points are connected with the SM at low energies.

\subsection{Matching conditions}
\begin{figure}[t]
 \begin{center}
 \includegraphics[width=0.5\textwidth]{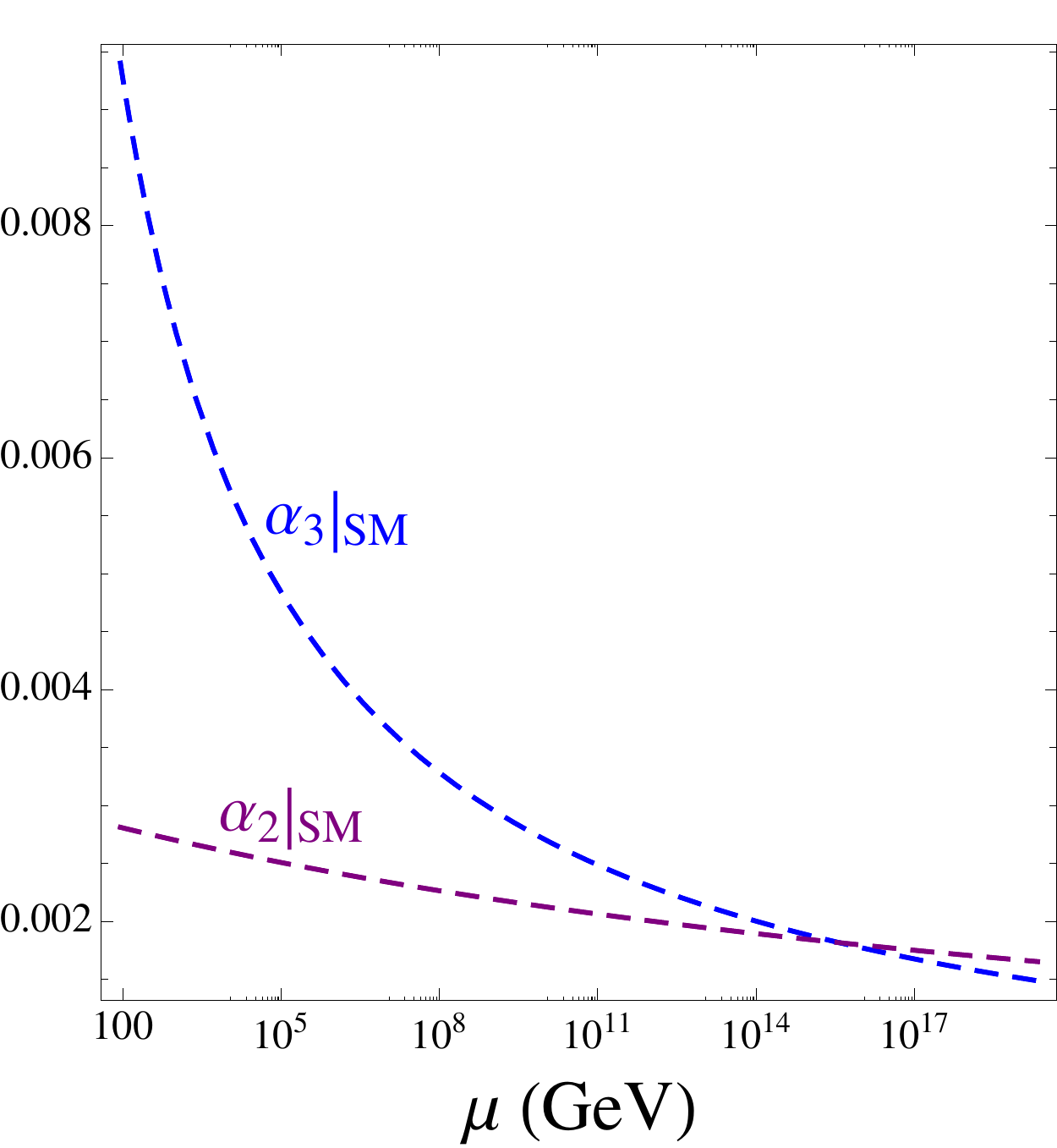}
 \end{center}
 \caption{SM running of the strong and weak gauge coupling $\alpha_3$ (blue) and  $\alpha_2$  (purple) from the $Z$ mass up to Planckian energies. The SM GUT scale reads approximately $\mu_{\rm GUT} \approx 4\times10^{15}$~GeV. UV safe trajectories have to coincide with SM values at the matching scale $\mu=M$ where BSM matter decouples.}
 \label{fig:SM}
 \end{figure}

Any RG trajectory emanating from free or interacting UV fixed points qualifies as a UV complete quantum field theory. The UV critical surface then determine the set of UV-safe trajectories. The relevant or marginally relevant couplings in the UV determines the dimensionality of the UV critical surface \eq{scaling}. Conversely, the irrelevant couplings are uniquely fixed by the relevant couplings in the UV. Consequently, the number of fundamentally free parameters which characterise the UV-safe trajectories is given by the dimensionality of the UV critical surface. At low energies,  physically viable BSM trajectories must connect with those of the SM, see Fig.~\ref{fig:SM}, as soon as the BSM matter fields have decoupled.

It remains to check whether the UV fixed points discovered in the previous section are connected through well-defined RG trajectories to the SM at low energies. 
Away from the fixed point, BSM matter fields will develop scalar and fermion masses $M_S$ and $M_\psi$ which are independent parameters of the theory. 
Phenomenological constraints for $M_S$ and $M_\psi$ are worked out in Sec.~\ref{Pheno} below. 
For RG scales much larger than the masses,
the BSM matter fields are effectively massless, and the RG flow is given by \eq{BSM}. Conversely, for RG scales much lower than the masses,
the BSM fields are taken to be infinitely heavy and decouple. The RG flow \eq{BSM} reduces to the one of the SM, also restoring confinement of QCD at low energies. 
Hence, the BSM contributions to the running gauge couplings decouple
as soon as $\mu$ is of the order of the BSM fermion mass. 
Furthermore,  threshold effects are subleading to the overall picture and will be neglected. Consequently, the RG flow of the SM is matched onto the RG flow of the BSM extension at the matching scale $M$,
\beq\label{matchingscale}
\mu=M\approx M_\psi
\eeq
below which the BSM fermions decouple. This leads to matching conditions between the RG flow of the SM at scales below \eq{matchingscale}, and BSM flows \eq{BSM} above the mass scale \eq{matchingscale},
\beq\label{matching}
\begin{array}{l}
\alpha_i(\mu=M)\Big|_{\rm SM}=\alpha_i(\mu=M)\Big|_{\rm BSM}\,,
\end{array}
\eeq
for $i=2,3$ (or $i$=AF,AS in settings with partially interacting fixed points). There will be no matching condition for the BSM Yukawa coupling since it is not part of the SM. Rather, after the decoupling of the BSM fields, the Yukawa coupling will ``freeze out'' at its value at decoupling. For the quantitative studies below, we  use PDG SM reference values at the scale of the $Z$ pole mass \cite{Olive:2016xmw},
\beq
\begin{array}{l}
\alpha_2({\mu=m_Z})= 2.7\times 10^{-3}\,,\\
\alpha_3({\mu=m_Z})=9.5\times 10^{-3}\,,
\end{array}
\eeq
together with the two loop perturbative running of gauge couplings in the SM, using \eq{BSM} with $N_F=0$. Fig.~\ref{fig:SM} illustrates the SM running between the mass of the $Z$ boson ($m_Z=91.19$~GeV) and Planckian energies. Note that equality of gauge couplings
\beq\label{GUT}
\alpha_2(\mu)=\alpha_3(\mu)
\eeq
arises in the SM at the GUT scale  $\mu_{\rm GUT} \approx 4\times10^{15}$~GeV. We emphasize that the matching of the BSM extension \eq{BSM} onto the SM \eq{matching} takes place at perturbatively small couplings.

 \begin{table}[t]
         \begin{center}
         \begin{tabular}{cccccccccc}
\rowcolor{LightGreen}
\toprule
\rowcolor{LightGreen}
\multicolumn{1}{c}{\cellcolor{LightGreen}
}
&\multicolumn{1}{c}{\cellcolor{LightGreen}\bf \ \ parameter\ \ }
&\multicolumn{3}{c}{\cellcolor{LightGreen}\bf UV fixed points}
&\bf $\quad$ type$\quad$
&
\\ 
\rowcolor{LightGreen}
\multirow{-2}{*}{\cellcolor{LightGreen}\ \ {\bf  model}\ \ }
&${(R_3,R_2,N_F)}$
&\multicolumn{1}{c}{\cellcolor{LightGreen}$\  {\alpha_3^*}\  $}
&\multicolumn{1}{c}{\cellcolor{LightGreen}$\  {\alpha_2^*} \ $}
&\multicolumn{1}{c}{\cellcolor{LightGreen}$\  {\alpha_y^*}\ $}
&
(Fig.~\ref{fSynopsis})
&
\multirow{-2}{*}{\bf $\quad$ info$\quad$}

\\
   \midrule
\rowcolor{LightGray}
\bf A
&$(\bm1,\bm4,12)$
& 0
& 0.2407
&0.3385
&\FP2\ \ {$\color{gray}\newmoon$}
&Fig.~\ref{fig:match_fp2}, low scale${}^*$\\
&
& 0.1287
& 0
&0.1158
&\FP3\ \ {$\color{blue} \blacksquare$}
&Fig.~\ref{fig:match_fp3}, low scale${}^*$\\
\multirow{-2}{*}{\bf B}
&\multirow{-2}{*}{$(\bm{10},\bm1,30)$}
&$\quad 0.1292\quad$
& $\quad0.2769\quad$
&$\quad0.1163\quad$
&$\quad$\FP4\ \ {$\color{red} \blacklozenge$}$\quad$
&Fig.~\ref{fig:sepB}, no match\\
\rowcolor{LightGray}
&
& 0.3317
& 0
&0.0995
&\FP3\ \ {$\color{blue} \blacksquare$}
&Fig.~\ref{fig:match_fp3b}, low scale${}^*$\\
\rowcolor{LightGray}
{\bf C}
&$(\bm{10},\bm4,80)$
& 0.0503
& 0.0752
&0.0292
&\FP4\ \ {$\color{red} \blacklozenge$}
&Fig.~\ref{fig:match_fp4b}, high scale\\
\rowcolor{LightGray}
& 
& 0
& 0.8002
&0.1500
&\FP2\ \ {$\color{gray}\newmoon$}
&Fig.~\ref{fig:match_fp2b}, high scale\\
&
& 0
& 0.0895
&0.0066
&\FP2\ \ {$\color{gray}\newmoon$}
&(no Fig.), low scale${}^*$\\
\multirow{-2}{*}{\bf D}
&\multirow{-2}{*}{$(\bm3,\bm4,290)$}
& 0.0416
& 0.0615
&0.0056
&\FP4\ \ {$\color{red} \blacklozenge$}
&Fig.~\ref{fig:match_fp4}, low scale\\
\rowcolor{LightGray}
\bf E
&$(\bm3,\bm3,72)$
&  0.1499
& 0.2181
&0.0471
&\FP4\ \ {$\color{red} \blacklozenge$}
&Fig.~\ref{fig:sepE}, low scale\\
\bottomrule
         \end{tabular}
         \end{center}
         \caption{UV fixed points and matching characteristics for various benchmark scenarios. An asterisk indicates that a matching is permitted at any scale including low (TeV) energy scales.} 
\label{tBenchmark}
\end{table}

 \subsection{Partially interacting fixed points}\label{piFPs}
As has been detailed in Sec.~\ref{s-partial}, at partially interacting fixed points \fp2 and \fp3, one of the two gauge couplings becomes asymptotically free, 
while the other one becomes asymptotically safe.
Moreover, the asymptotically (free) safe 
coupling is (marginally) relevant and, hence, the UV critical surface is invariably  two-dimensional.
On the other hand, the BSM Yukawa coupling $\alpha_y$ is irrelevant and fully specified by the asymptotically safe coupling
in the UV. 

In this light, a convenient  choice for the two fundamentally free dimensionless parameters which characterise UV-safe trajectories running out of the fixed point are the deviations of the gauge couplings from their UV fixed point values  at some high scale $\mu=\Lambda$,
\beq\label{deltaP}
\begin{array}{l}
\delta\alpha_i(\Lambda)=\alpha_i^*-\alpha_i(\Lambda)\,,
\end{array}
\eeq
with $i=2,3$ (or $i$=AF,AS). We take the practical view that the high scale is essentially given by the Planck scale. Quantum gravity effects should  be retained at scales close to and above $\Lambda$. The BSM Yukawa coupling is an irrelevant coupling and entirely dictated by the UV hypercritical surface relating it with $\alpha_{\rm AS}$ and $\alpha_{\rm AF}$,
 \beq
 \alpha_y=F_y(\alpha_{\rm AS},\alpha_{\rm AF})\,.
\eeq
The parameters \eq{deltaP} will be used to match trajectories onto the SM. Specifically, the parameter $\delta\alpha_{\rm AS}$ controls at which energy scale the asymptotically safe coupling is crossing over from the UV fixed point towards the Gaussian IR fixed point of \eq{BSM}. For $0<\delta \alpha_{\rm AS}\ll 1$, $\alpha_{\rm AS}$ will start out of the UV fixed point along the separatrix which connects the UV fixed point with the Gaussian in the IR. In the immediate vicinity of the UV fixed point the RG flow is of the power-law type and thus fast, controlled by the relevant scaling exponent. Further away from the fixed point, as soon as $\alpha_{\rm AS}\approx \s023 \alpha^*_{\rm AS}$ and below \cite{Litim:2015iea}, we observe a cross-over whereby the running becomes logarithmically slow instead, dominated by the ``would-be'' Gaussian  IR fixed point of \eq{BSM}. Hence, the parameter $\delta \alpha_{\rm AS}$ allows us to chose at which scale $\alpha_{\rm AS}(M)$ has reached the desired SM value. Notice that this discussion is largely  independent of $\alpha_{\rm AF}$ provided the latter remains small.

The running of $\alpha_{\rm AF}$ out of the UV fixed point is controlled by the RG flow \eq{alphaAF}, which in turn  is largely determined by the parameter $\delta\alpha_{\rm AF}$, together with the coefficients $B_{\rm AF}$ and $B_{\rm AF}'$, \eq{Bprime}. Integrating \eq{alphaAF} close to the UV fixed point gives
\beq\label{AFone}
\frac{1}{\alpha_{\rm AF}(\mu)}=\frac{1}{\delta\alpha_{\rm AF}(\Lambda)}+B_{\rm AF}'\,\ln (\mu/\Lambda)\,.
\eeq
One might expect that the two free parameters \eq{deltaP} are sufficient to match the RG flow in the UV to two preset values at low energies. We stress, however, that a matching may fail  if the ``would-be'' asymptotically free coupling $\alpha_{\rm AF}$ runs into Landau poles at intermediate energies. Thus, we must explain how Landau poles are avoided.
 In the deep UV, we have that  $B_{\rm AF}'>0$, \eq{BAF}. However, the coefficient  $B_{\rm AF}'$ depends on the asymptotically safe gauge coupling $\alpha_{\rm AS}$ and on the Yukawa coupling. Both of these run  out of the UV fixed point and induce an effective running of  $B_{\rm AF}'\to B_{\rm AF}'(\mu)$ owing to \eq{alphaAF}. For sufficiently small $\alpha_{\rm AS}$ such as close to the matching scale $\mu=M$, the coefficient $B_{\rm AF}'$ falls back onto the BSM one loop coefficient  $B_{\rm AF}'\to B_{\rm AF}$. Close to the matching scale the one loop approximation is viable and we have  
\beq\label{AFoneM}
\frac{1}{\alpha_{i}(\mu)}=\frac{1}{\alpha_{i}(M)}+B_{i}\,\ln (\mu/M)
\eeq
for both $i={\rm AF, AS}$. 
If $B_{\rm AF}>0$, meaning that the gauge sector remains asymptotically free, we have that $B_{\rm AF}'(\mu)>0$ at all intermediate scales and matching will always be possible.
If $B_i<0$, the one loop running \eq{AFoneM} for the gauge couplings reach a ``would-be'' Landau pole at 
\beq
\frac{\mu_i}{M}=\exp\left(-\frac1{B_i\,\alpha_i(M)}\right)\,.
\eeq
For $\alpha_{\rm AS}$ the Landau pole is avoided automatically owing to  the two loop Yukawa terms: with growing energy, once the scale $\mu_{\rm AS}$ is reached, the two loop terms kick in and $\alpha_{\rm AS}$ settles into its UV fixed point, see Sec.~\ref{sec:as}.
For $\alpha_{\rm AF}$ it is not guaranteed  that $B'_{\rm AF}(\mu)$ changes sign in time for  $\alpha_{\rm AF}$ to avoid the Landau pole. We find that $\alpha_{\rm AF}$  avoids a Landau pole provided that
\begin{align}
\label{gas_upper}
B_{\rm AF}\cdot\alpha_{\rm AF}(M)<{B_{\rm AS}}\cdot\alpha_{\rm AS}(M).
\end{align}
The condition \eq{gas_upper} ensures that the ``would-be'' one loop Landau pole for $\alpha_{\rm AF}$ arises at higher scales $\mu_{\rm AF}>\mu_{\rm AS}>M$ than the one for $\alpha_{\rm AS}$. 
The crucial point about scales $\mu\approx \mu_{\rm AS}$ is that the two loop terms have become active. Two loop terms also contribute to the running of $\alpha_{\rm AF}$ and thereby ensure that the sign of $B'_{\rm AF}$  has become positive. We conclude that  \eq{gas_upper} is sufficient to
provide an upper bound for viable matchings to the SM, ensuring that neither of the couplings escapes a successful matching  through a Landau pole at intermediate energies.
Within the confines of \eq{gas_upper}, this enables us to match SM and BSM running onto each other essentially at any scale between TeV and Planckian energies.

\subsection{Fully interacting fixed points}\label{fully}

All fully interacting UV fixed points are characterized by a stability matrix \eq{linear} with a single relevant eigenvalue. This important result states that a linear combination of the gauge groups' kinetic terms together with the BSM Yukawa interaction term in the fundamental Lagrangean \eq{L} correspond to the sole UV relevant operator in the theory. This result has important implications. Unlike in asymptotically free theories (or in asymptotically safe theories at partially interacting fixed points \fp2 or \fp3) where  every gauge coupling corresponds to a UV relevant direction, here, instead, the UV critical surface is 
of a lower dimensionality. This new effect is a  consequence of competing gauge interactions in the UV. Most notably, it entails that the number of fundamentally independent parameters is reduced, leading to an enhanced level of predictivity.

In our models, the UV critical surface at fully interacting UV fixed points becomes
one-dimensional, parametrised by a single free parameter. Consequently, only one out of the three couplings $(\alpha_3,\alpha_2,\alpha_y)$ may be considered as an independent variable. For \fp4, and without loss of generality we chose this to be $\alpha_3$. The UV critical surface then uniquely determines
the weak and the Yukawa coupling as functions of the strong coupling,
\beq\label{delta}
\begin{array}{l}
\alpha_i=F_i(\alpha_3)\quad\quad{\rm for}\quad i=2,y\,.
\end{array}
\eeq
Most importantly, the UV critical surface imposes a relation between the two gauge couplings which arises as a strict consequence of asymptotic safety at a fully interacting fixed point \fp4. 
We may then use the dimensionless parameter 
\beq\label{deltaF}
\delta\alpha_3(\Lambda)=\alpha_3^*-\alpha_3(\Lambda)
\eeq
at the high scale $\Lambda$ to parametrise all UV safe trajectories running out of the fully interacting UV fixed point \fp4. The  UV-IR connecting separatrix 
\beq\label{separatrix}
(\alpha_3,\alpha_2,\alpha_y)(\mu)\equiv  (\alpha_3,F_2 (\alpha_3),F_y (\alpha_3))(\mu)
\eeq 
uniquely determines the relation between the strong and the weak gauge coupling for all scales above the matching scale. The role of the free parameter $\delta\alpha_3(\Lambda)$ is to determine at which scale the curves \eq{separatrix}
display a cross-over from UV dominated running towards IR dominated running. The task to identify trajectories which can be matched onto the SM at some matching scale $\mu=M$  reduces to analysing the separatrix \eq{separatrix}. Given that the set of determining equations is over-constrained, a successful matching cannot be guaranteed from the outset, meaning that the viability needs to be checked for each \fp4. 
On the other hand, if trajectories emanating out of \fp4 can be matched, the BSM extension implies a fundamental relation between both gauge couplings, which would not exist otherwise.
In settings where \fp4 exists alongside \fp2, or \fp3, or both, either of the partially interacting fixed points is the more relevant UV fixed point. Their UV critical dimensions are larger,  UV-IR connecting trajectories can be found which link \fp2 or \fp3 with \fp4. An example for this is discussed below in Fig.~\ref{fig:match_fp2b}.

To understand more explicitly how a matching of \fp4 onto the SM depends on the fermion representations and flavor multiplicities,  we evaluate the link between the gauge couplings as dictated by the UV critical surface, \eq{scaling}.  Since in the general case the separatrix cannot be resolved analytically, 
we  use the critical surface approximation of \eq{scaling}, see App.~\ref{AppA} for the technicalities, keeping in mind that far from the UV fixed point the critical surface may deviate from the one given by the separatrix. The relation \eq{delta} then takes the simple linear form
\beq
\label{cs}
 \begin{array}{rl}
 \alpha_2(M)&=-X+Y\,\alpha_3(M)\,,\\
Y&=-V_2/V_3\,,\\
X&=-\alpha_2^*+Y \,\alpha_3^*\,.
\end{array}
\eeq
The parameters $V_i$
are related to the UV relevant eigendirection   \eq{scaling} which characterises the UV critical surface.
Using \eq{BSM}, we find the explicit expressions
\beq\label{XY}
 \begin{array}{l}
\di X=\frac{B_3D_2-B_2D_3}{C_{2}D_3-G_{3}D_2}\,,
\quad\quad
\di  Y=\frac{C_{3}D_2-G_{2}D_3}{C_{2}D_3-G_{3}D_2}\,,
\end{array} \eeq
in terms of the perturbative loop parameters. Whether a matching of trajectories $(\alpha_2,\alpha_3)(\mu)$ with \eq{cs} onto the SM is possible or not depends on the signs and magnitude of $X$ and $Y$. In the large-$N_F$ limit, we find
\beq
\begin{array}{l}
\di
X\Big|_{N_F\gg 1}=
\frac{21}{5}\frac{C_2(R_2)-\s0{19}{112}C_2(R_3)}{C_2(R_3)C_2(R_2)d(R_3)d(R_2)}\,\frac{1}{N_F}+{\rm subleading}\,,\\[2.5ex]
\di
Y\Big|_{N_F\gg 1}= \frac{3}{2}+{\rm subleading}\,.
\end{array}
\eeq
With increasing $N_F$ we  observe that $Y>0$ rapidly approaches the large $N_F$ limit $Y=\0{3}{2}$.\footnote{Note that this ratio is a direct consequence of the $SU(3)_C$ and $SU(2)_L$ gauge groups of the SM.} Conversely, $X$ may have either sign depending on the representation $(R_3,R_2)$, though not on $N_F$. Also, $X$  becomes parametrically small for high dimensional representations and for large flavor multiplicities $N_F$. If $X<0$ we have $\alpha_2(M)>\alpha_3(M)$  indicating that a matching below GUT-type scales is impossible. Furthermore, \eq{cs} also implies that $\alpha_2(M)>-X$, stating that a matching becomes impossible at any scale if $-X$ becomes too large. On the other hand, if $X>0$ matchings can be found to SM values, in particular at low scales where $\alpha_2(M)<\alpha_3(M)$. For intermediate and large values of $N_F$ the condition $X>0$ becomes
\beq\label{azero}
C_2(R_2)-\s0{19}{112}C_2(R_3)>0\,.
\eeq
  A few comments are in order:

\begin{figure}[t]
 \begin{center}
 \includegraphics[width=0.49\textwidth]{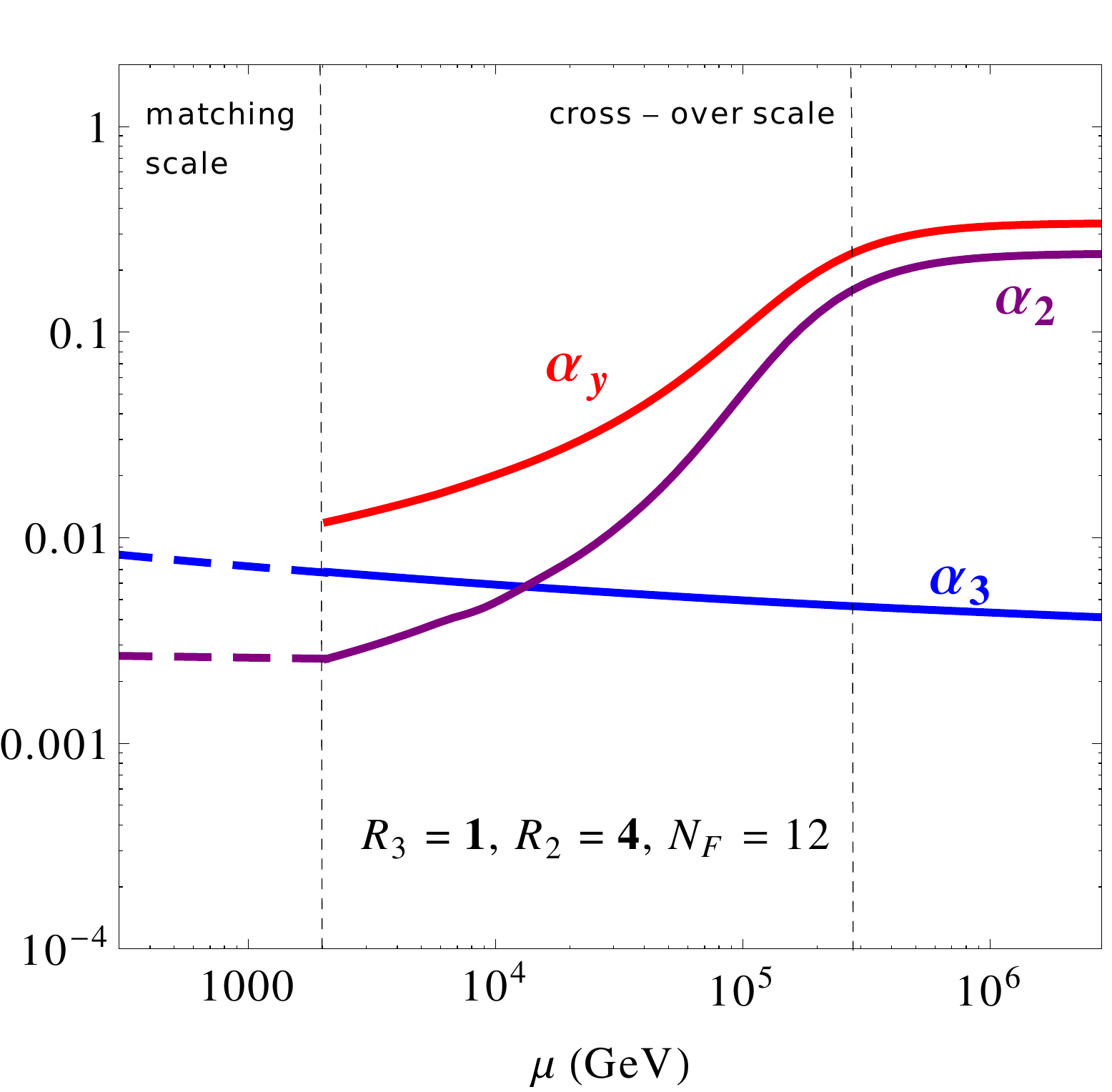}
 \end{center}
 \vskip-.5cm
 \caption{Low scale matching of the partially interacting fixed point \fp2 onto the SM  for the benchmark scenario $A$ with  $(R_3,R_2,N_F)=(\bm{1},\bm{4},12)$ and $M=2$ TeV. BSM (SM) running is shown by full (dashed) lines. Around the cross-over scale the BSM running of $\alpha_2$ and $\alpha_y$ slows down from power-law to  logarithmic.
Notice that the running of the strong coupling is not modified by BSM fermions.}
 \label{fig:match_fp2}
 \end{figure}

 $(i)$ If $R_3<\bm{10}$, the condition \eq{azero} is satisfied for any $R_2$.
Non-trivial constraints arise from \eq{azero} once $R_3=\bm{10}$ or higher (for example, $R_3=\bm{10}$ necessitates $R_2\geq\bm3$, and  similarly for  higher dimensional representations $R_3>\bm{10}$).

$(ii)$ An increasing flavor multiplicity $N_F$ is required  for models with a low dimensional $SU(3)_C$ representation $R_3$ or  high dimensional $SU(2)_L$ representation $R_2$ to ensure that the magnitude of $X$ stays within the limits compatible for a matching onto SM values. Also, increasing 
 the dimension of the matter representations lowers $\alpha_3(M)$. For these cases it then follows that the matching can only take place at a high scale. 

$(iii)$ The linear approximation for the separatrix \eq{cs} becomes exact once $R_2=\bm{1}$. This can be seen as follows.  
Using \eq{XY}  we find $X=-\s0{19}{35}$ at \fp4  for any viable  $(R_3,N_F)$. The exact same result is found if \eq{cs} is evaluated at  the Banks-Zaks IR fixed point of the weak gauge coupling in the SM $(\alpha_2^*,\alpha_3^*)=(\s0{19}{35},0)$, see \eq{wouldbe}. 
This result indicates that the  UV critical surface \eq{cs} at \fp4  coincides with the critical surface at the Banks-Zaks IR fixed point. The latter therefore controls the running of the weak coupling away from the fully interacting UV fixed point  by directing all UV safe trajectories straight into the Banks-Zaks fixed point. As is shown more explicitly below, it is for this reason that a matching of \fp4 with  $R_2=\bm1$  onto the SM is impossible.  

$(iv)$ For all scenarios considered in this work, we find that the condition \eq{azero} is a  good estimator for the availability of a matching with the SM.  

This completes the general discussion of matching conditions for UV safe trajectories onto the SM.

    \begin{figure}
 \begin{center}
 \includegraphics[width=0.5\textwidth]{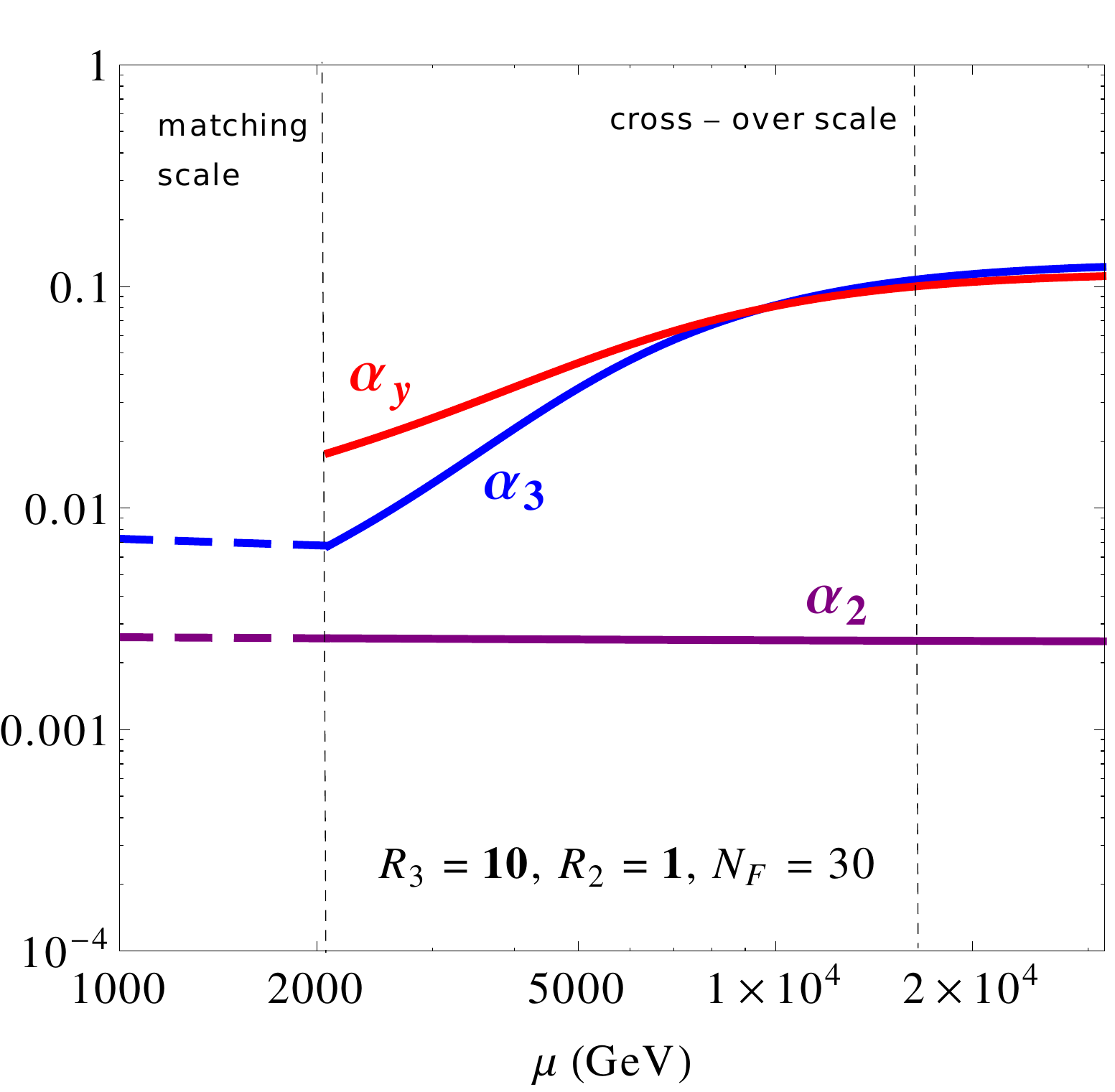}
 \end{center}
 \vskip-.5cm
 \caption{Low scale matching from the partially interacting fixed point \fp3 onto the SM  for the benchmark scenario $B$ with  $(R_3,R_2,N_F)=(\bm{10},\bm{1},30)$ and $M=2$ TeV. BSM (SM) running is shown by full (dashed) lines. Around the cross-over scale the BSM running of $\alpha_3$ and $\alpha_y$ slows down from power-law to  logarithmic.
Notice that the running of the weak coupling is not modified by the BSM fermions.}
 \label{fig:match_fp3}
 \vskip1cm
 \begin{center}
  \includegraphics[width=0.49\textwidth]{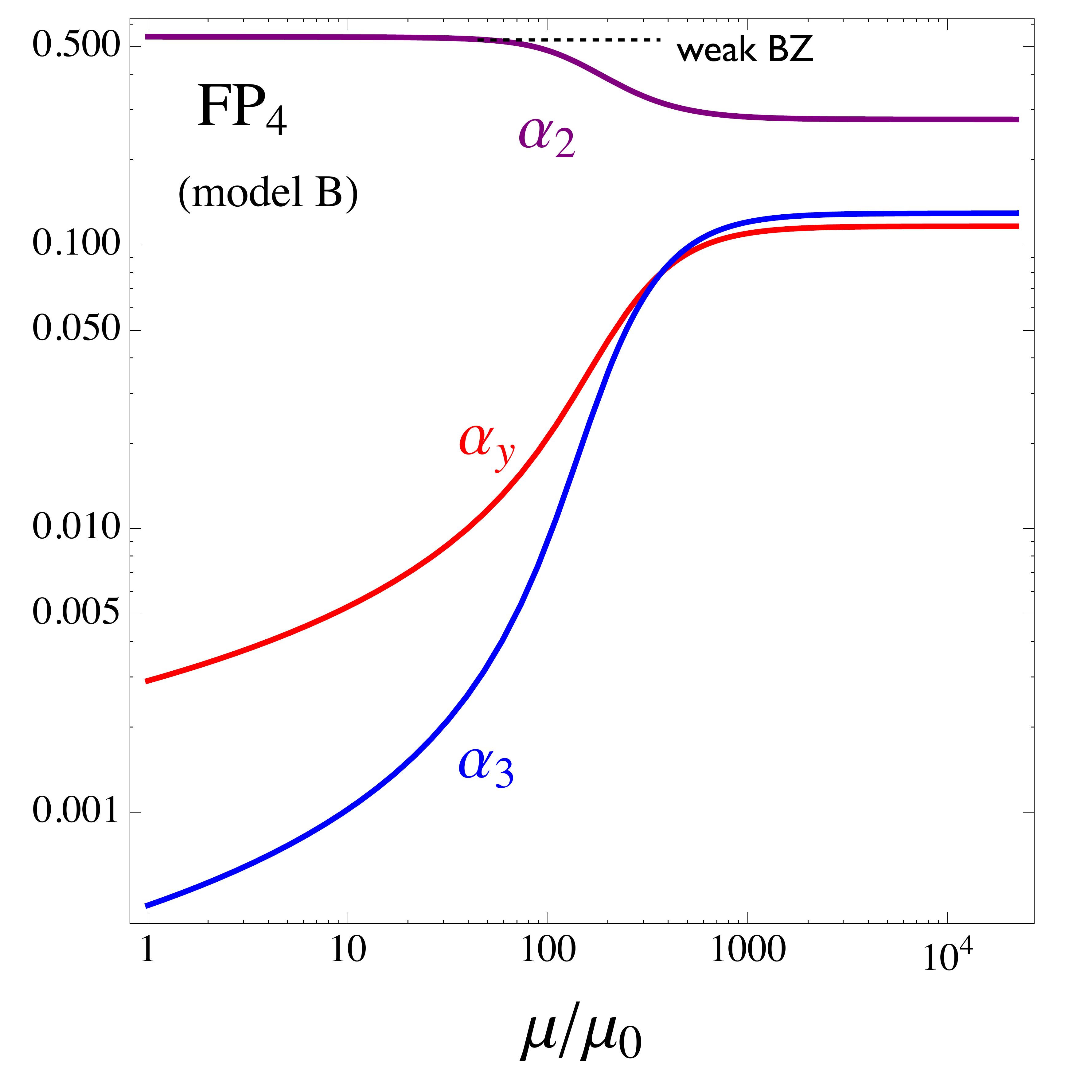}
 \end{center}
 \vskip-.5cm
 \caption{Shown is the running of the gauge and BSM Yukawa couplings along the UV-IR connecting separatrix emanating from \fp4 for the benchmark model $B$, \eq{B}. The scale $\mu_0$ may take any value determined by the free parameter $\delta\alpha_3(\Lambda)$. The weak gauge coupling is attracted towards its ``would-be'' Banks-Zaks fixed point (weak BZ) , see \eq{wouldbe}, indicated by the dashed line. Consequently, the interacting UV fixed point cannot be matched onto the SM.}
 \label{fig:sepB}
 \end{figure}
 
\subsection{Benchmark scenarios}

 Let us now illustrate how the matching works in practice for a selection of benchmark scenarios, summarised in Tab.~\ref{tBenchmark}, covering low scale and high scale matchings. \\[-1.5ex]
 
 {\bf Benchmark scenario A}. 
 For this setting we assume that the BSM fermions do not carry  $SU(3)_C$ charges. Following on from our earlier discussion, \fp2 is the sole UV fixed point which may arise and neither  \fp3 nor \fp4 are available,  see
 Fig.~\ref{fSynopsis}. We consider the parameters
 \beq \label{A}
 (R_3,R_2,N_F)=(\bm{1},\bm{4},12)
 \eeq
with RG trajectories displayed in Fig.~\ref{fig:match_fp2}. The matching scale $M$ may take any value between TeV and Planckian energies. In Fig.~\ref{fig:match_fp2}, for illustration, we have set it to the low value $M=2$~TeV (vertical dashed line).
Evidently, the running of the strong coupling is not modified by BSM matter and remains SM-like throughout. Once the matching scale is fixed, the model predicts the 
value of the (otherwise unconstrained) Yukawa coupling. For $M=2$ TeV one obtains $\alpha_y(M)=0.022$. We have also indicated the cross-over scale where the running of the asymptotically safe couplings changes from power-law behavior in the deep UV to logarithmic behavior towards the IR (dashed vertical line). 
The cross-over 
scale is found to be around $\mu_{\rm cr}\sim 3\times10^5$ GeV and much larger than the matching scale.\\[-1.5ex]

       \begin{figure}[t]
 \begin{center}
 \includegraphics[width=0.49\textwidth]{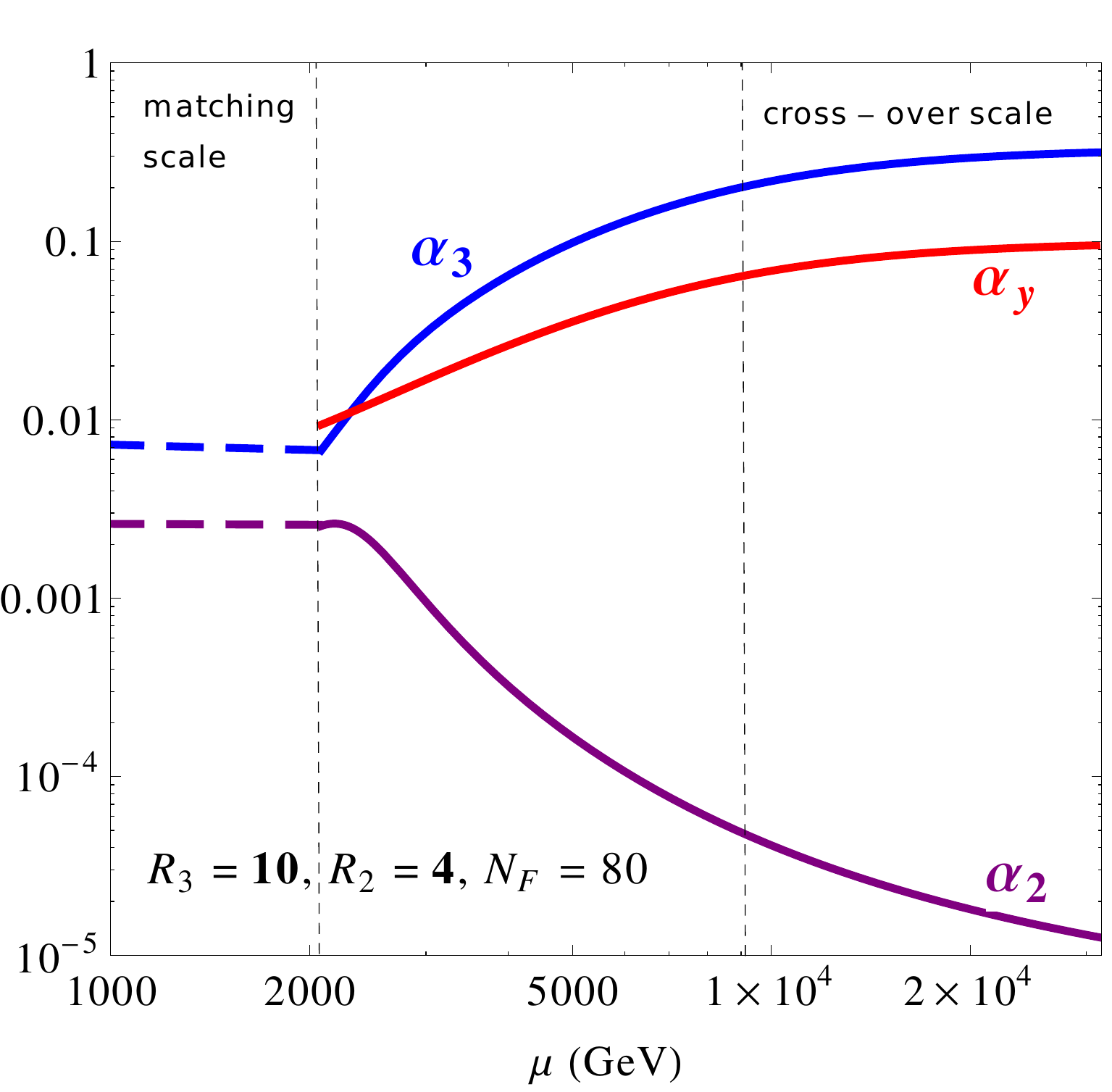}
 \end{center}
 \caption{Low scale matching from the partially interacting fixed point \fp3 onto the SM  for the benchmark scenario $C$ with $(R_3,R_2,N_F)=(\bm{10},\bm{4},80)$ and  $M=2$ TeV. BSM (SM) running is shown by full (dashed) lines. Around the cross-over scale the BSM running of $\alpha_3$ and $\alpha_y$ slows down from power-law to  logarithmic.
Notice that the approach towards asymptotic freedom of the weak coupling is enhanced by BSM fermions.}
 \label{fig:match_fp3b}
 \end{figure}
 \begin{figure}[t]
 \begin{center}
 \includegraphics[width=0.5\textwidth]{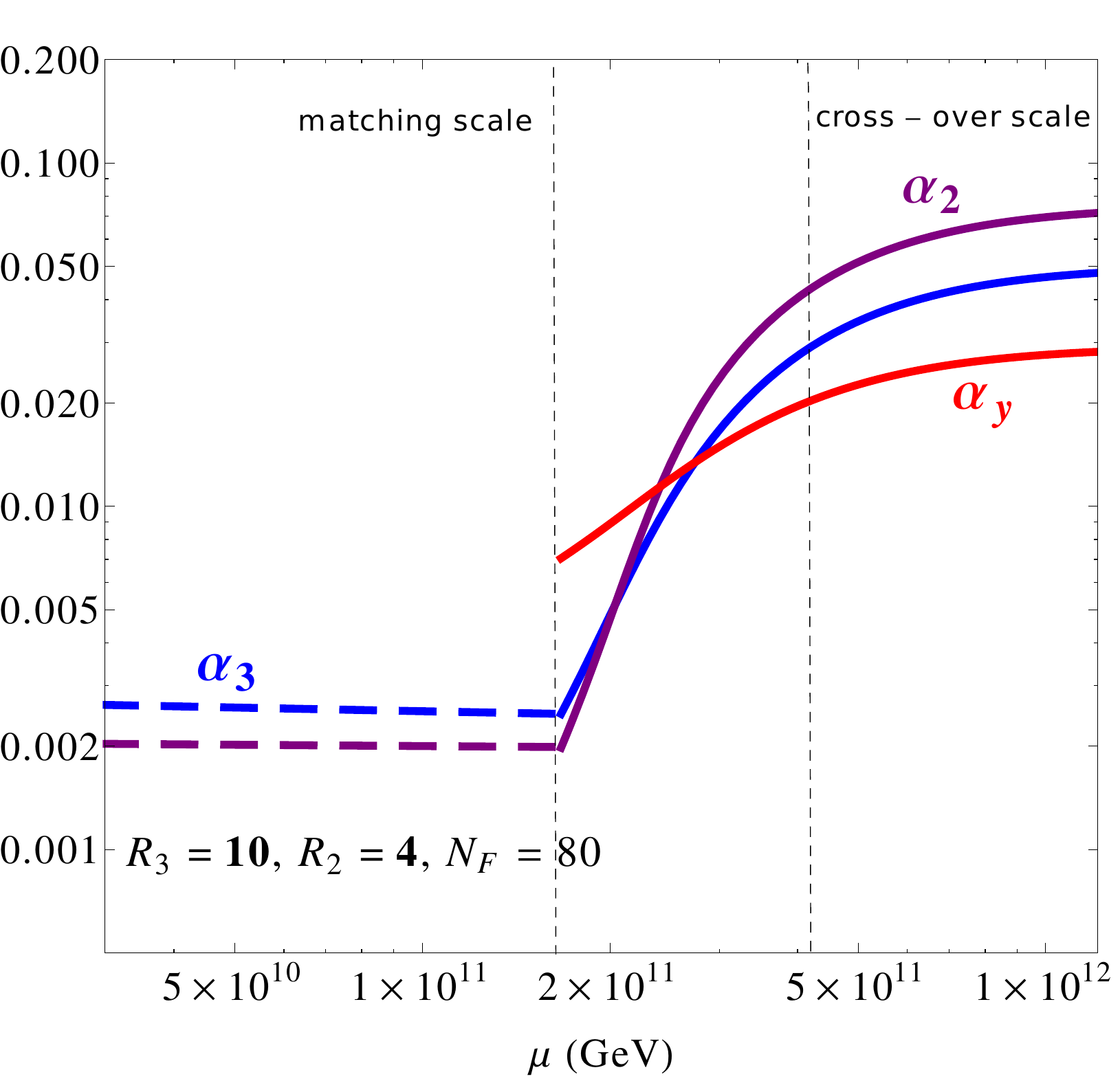}
 \end{center}
 \vskip-.5cm
 \caption{High scale matching of a fully interacting fixed point \fp4 onto the SM  for the benchmark scenario $C$ with 
 $M=2 \times 10^{11}$ GeV. Once BSM matter fields are active, the weak coupling approaches asymptotic safety more rapidly than the strong coupling. See also Fig.~\ref{fig:match_fp3b} and Fig.~\ref{fig:match_fp2b}.}
 \label{fig:match_fp4b}
 \end{figure}
 
 \begin{figure}[t]
 \begin{center}
\includegraphics[width=0.5\textwidth]{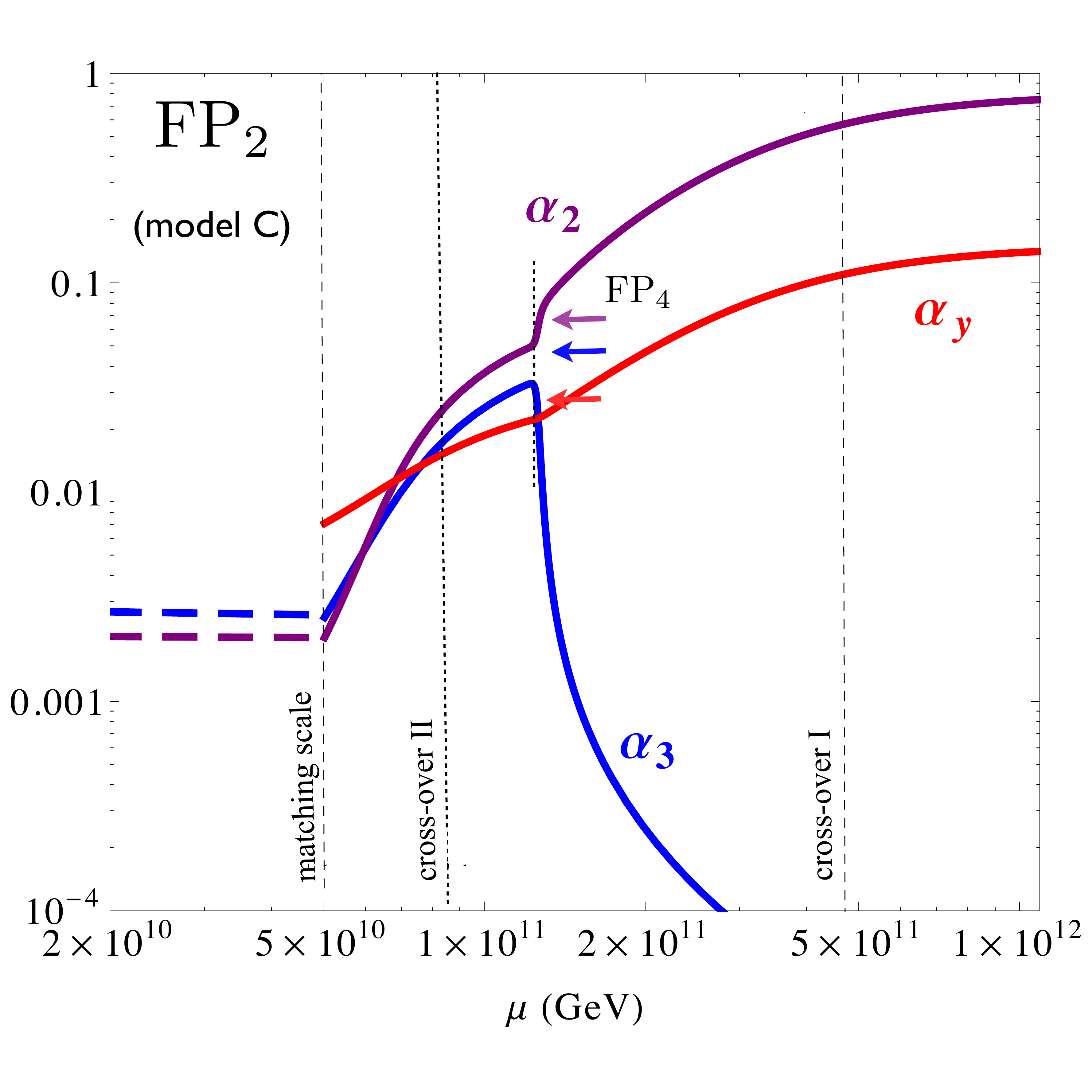}
 \end{center}
 \vskip-.5cm
 \caption{High-scale matching of a partially interacting fixed point \fp2 onto the SM  for the benchmark scenario $C$ with $(R_3,R_2,N_F)=(\bm{10},\bm{4},80)$. Trajectories emanate out of the UV fixed point \fp2 and initially cross over into the vicinity of \fp4 (indicated by arrows) at around $1.5\times 10^{11}$~GeV.  Subsequently, trajectories display a second cross over to match with the SM at about  $M=5 \times 10^{10}$ GeV.} 
 \label{fig:match_fp2b}
 \end{figure}

  {\bf Benchmark scenario B}. 
For this case we assume that the BSM fermions do not carry $SU(2)_L$ charges. From Fig.~\ref{fSynopsis} it follows that solely \fp3, possibly in conjunction with \fp4 can arise. Informed by the results of Tabs.~\ref{tFP3} and~\ref{tFP4} we chose the parameters   
\beq\label{B}
(R_3,R_2,N_F)=(\bm{10},\bm{1},30)
\eeq 
to ensure that the model has both types of UV fixed points, \fp3 and \fp4. 
The partially interacting fixed point \fp3 can always be matched onto the SM at any scale.  In Fig.~\ref{fig:match_fp3}, this is illustrated for a  low matching scale $M=2$ TeV. In this model, the running of the weak gauge coupling is not modified by BSM matter to the leading orders in perturbation theory.  The crossover scale $\mu_{\rm cr}\sim 1.6 \times 10^{4}$ GeV is an order of magnitude larger than the matching scale. Elsewise the same reasoning as in  Fig.~\ref{fig:match_fp2} applies. 
In contrast, the impossibility for a matching at \fp4 is illustrated in Fig.~\ref{fig:sepB}. 
The scale $\mu_0$ is arbitrary and can take any value upon tuning the UV parameter $\delta\alpha_3(\Lambda)$. 
However, the UV safe trajectory emanating out of \fp4 
is attracted towards the strongly coupled domain, owing to the Banks-Zaks IR fixed point at $(\alpha_2^*,\alpha_3^*)=(\s0{19}{35},0)$ in the weak sector, \eq{wouldbe}. 
Consequently, the weak coupling cannot become weak in the IR and a matching to SM values is impossible at any intermediate scale due to the dominance of the Banks-Zaks fixed point. We stress that this pattern is a direct consequence of the BSM fermions being uncharged under $SU(2)_L$. The non-availability of \fp4 persists for all models with $R_2=\bm1$, in line with our discussion in Sec.~\ref{fully}.
\\[-1.5ex]

 We now turn to benchmark scenarios where  the BSM fermions carry  both $SU(2)_L$ and $SU(3)_C$ charges. In these cases we find realisations for either of the partially interacting fixed points \fp2 and \fp3, as well as for the fully interacting fixed point \fp4. \\[-1.5ex]

 {\bf Benchmark scenario C}.   As soon as $R_2\ge \bm{2}$ and $R_3\ge \bm3$, and for sufficiently large $N_F$,  all three types of UV fixed points arise, see Fig.~\ref{fSynopsis}. 
To illustrate such settings, we consider the benchmark scenario $C$ with parameters 
 \beq\label{C}
 (R_3,R_2,N_F)=(\bm{10},\bm{4},80)
 \eeq 
which displays \fp2, \fp3 and \fp4 within the perturbative domain, see Tab.~\ref{tBenchmark}. In Fig.~\ref{fig:match_fp3b}, we begin with \fp3 where $\alpha_2^*=0$ in the deep UV. Once more we observe that the matching condition \eq{gas_upper} on the one loop BSM parameters for the strong and the weak coupling 
can be satisfied at any scale between a few TeV and Planckian energies including the
 low matching scale $M=2$~TeV chosen in Fig.~\ref{fig:match_fp3b}. 
Furthermore,  the weak coupling $\alpha_2$ continues to decrease even directly below the matching scale, for any matching scale. For the weak gauge coupling the approach towards asymptotic freedom is accelerated over the SM rate owing to the two loop BSM Yukawa contributions which are winning over the contributions by the strong gauge coupling along the entire UV-IR connecting separatrix into the fixed point. This pattern is consistent with the matching condition \eq{gas_upper}, which is fullfilled for any intermediate scale. For the example shown in  Fig.~\ref{fig:match_fp3b}, the cross-over scale and the matching scale are separated by an order of magnitude.

 In Fig.~\ref{fig:match_fp4b} we turn to the matching of the fully interacting fixed point \fp4 corresponding to the same parameter set 
\eq{C}.  Couplings run out of the UV fixed point \fp4 along a unique separatrix \eq{separatrix} connecting \fp4 with the Gaussian fixed point. The separatrix thereby imposes a link between $\alpha_2$ and $\alpha_3$.  On the separatrix, and close to the fully interacting fixed point, the weak coupling is genuinely stronger than the strong coupling. At crossover, it becomes rapidly weaker than the strong coupling. The gauge couplings also weaken more rapidly than the BSM Yukawa coupling. For the parameters \eq{C}, the separatrix dictates that the unique matching scale onto SM values comes out comparatively high, with $M\approx 2 \times 10^{11}$ GeV.

In Fig.~\ref{fig:match_fp2b} we  consider the matching with \fp2 where $\alpha_3^*=0$ in the  UV. In this model, we find that a matching at \fp2 is more strongly constrained compared to a generic partially interacting fixed point. The reason for this is the influence of \fp4 on UV-IR connecting trajectories and the necessity to avoid an early Landau pole in the strong sector. Specifically, starting at some high scale $\Lambda$ we observe that the weak and the Yukawa couplings decrease with energy, while the strong coupling increases towards the IR.  For too small $\delta\alpha_3(\Lambda)$ \eq{deltaP} the strong coupling does not grow fast enough.
For too large $\delta\alpha_3$ the strong coupling runs into a Landau pole at intermediate scales. Within a narrow window for $\delta\alpha_3$, however, the growth of $\alpha_3$ is tamed due to \fp4. Then, trajectories are close to the separatrix connecting \fp2 with \fp4, with a  cross-over scale $\mu\approx 5 \times 10^{11}$~GeV, see
Fig.~\ref{fig:match_fp2b}. 
Below the cross-over scale, couplings are attracted towards \fp4 (see Tab.~\ref{tBenchmark} for the fixed point values) which, however, is not reached exactly. Instead, at scales about $\mu\approx 1.2 \times 10^{11}$~GeV the couplings are driven away from \fp4, now following the separatrix which connects \fp4 with the Gaussian. 
In consequence, we find that couplings can be matched onto the SM  at a high matching scale of about $M=5\times 10^{10}$~GeV, close to the matching scale found for \fp4. 

This result is consistent with the ``Landau pole avoidance condition'' \eq{gas_upper} derived in Sec.~\ref{piFPs},
which for the parameters \eq{C} has no solutions for low matching scales.
For example, at $M=2$~TeV the SM predicts $\alpha_3^{\rm SM}(M)\approx 0.0067$, yet the upper boundary \eq{gas_upper} reads $\alpha_3(M)\approx 0.004$.  
Only for sufficiently high  matching scales such as $M\approx 5\times 10^{10}$ GeV, the condition \eq{gas_upper} eases up and allows a consistent matching without the strong coupling prematurely running into a perturbative Landau pole at intermediate scales. 

We conclude that all three fixed points qualify as UV completions for the SM, although the specifics of the UV completion differ due to finer details of the fixed point structure. \\[-1.5ex]

\begin{figure}[th!]
 \begin{center}
 \includegraphics[width=0.49\textwidth]{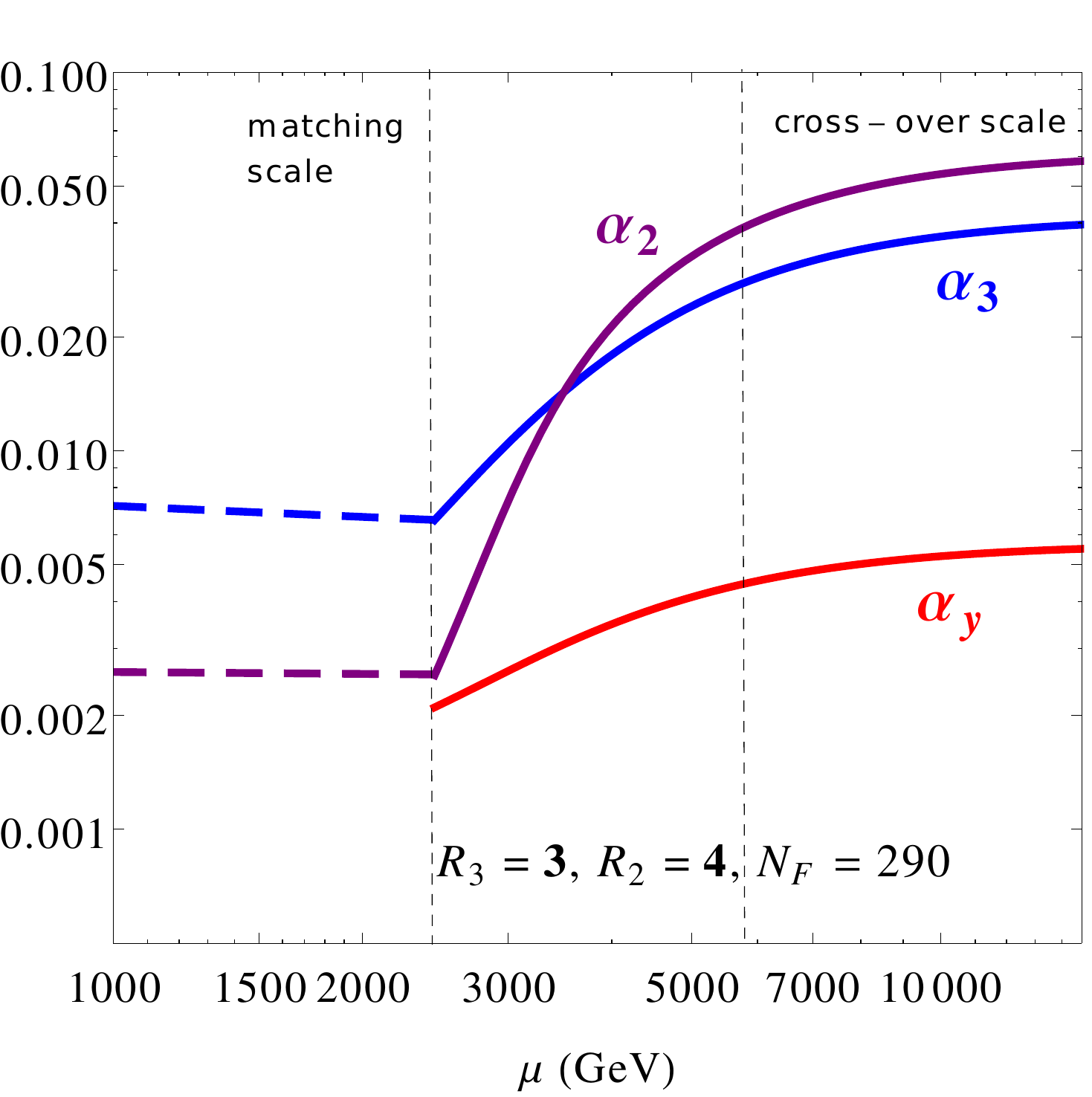}
 \caption{Low scale matching from the fully interacting fixed point \fp4 onto the SM  for the benchmark scenario $D$ with $(R_3,R_2,N_F)=(\bm{3},\bm{4},290)$ and matching scale $M=2.4$ TeV.
Once BSM matter fields are active, the weak coupling approaches asymptotic safety more rapidly than the strong coupling.
}
 \label{fig:match_fp4}
\vskip1.3cm
  \includegraphics[width=0.51\textwidth]{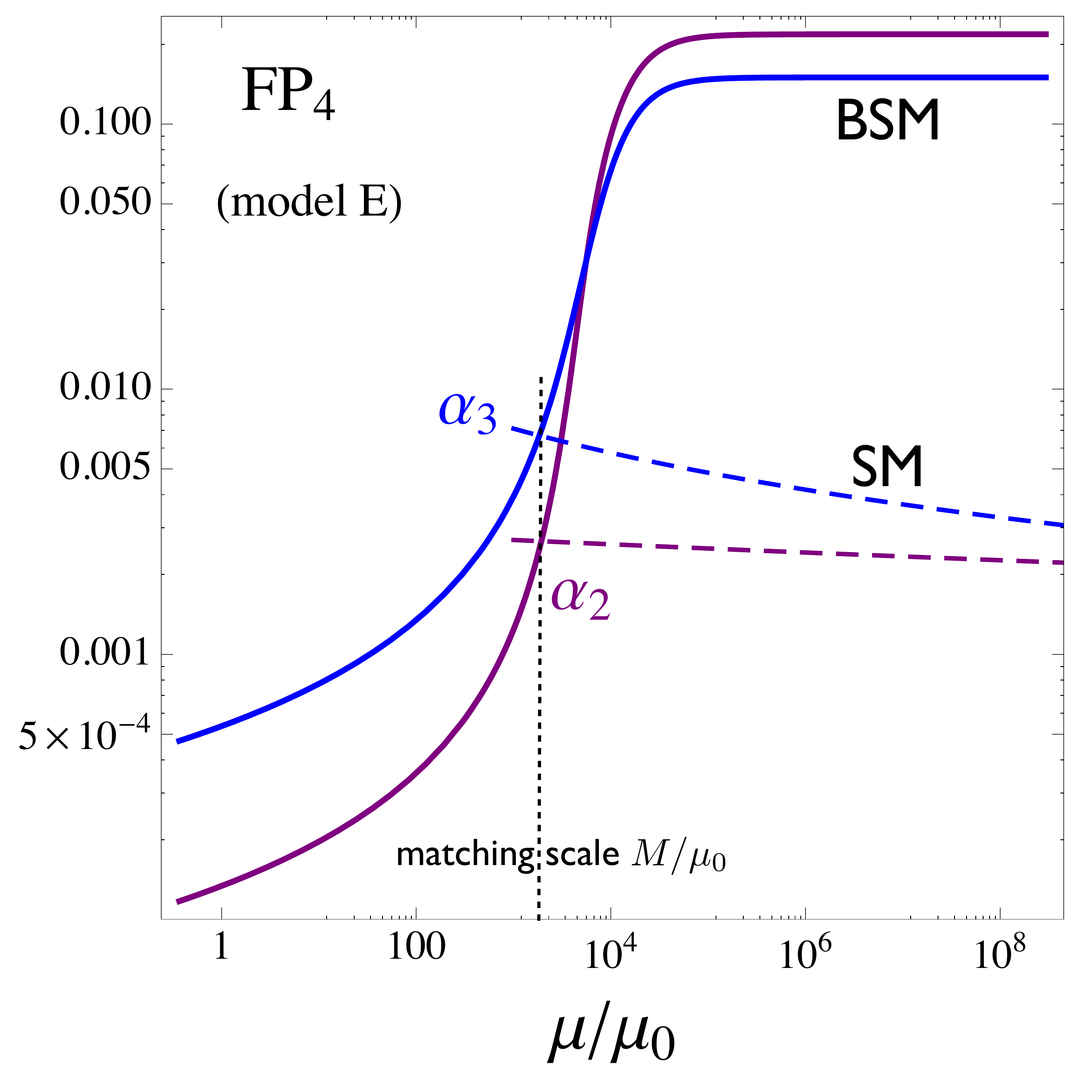}
 \end{center}
 \vskip-.5cm
 \caption{The matching procedure at \fp4 for the benchmark model $E$ with parameters \eq{E}. The thick lines show the BSM running along the UV-IR connecting separatrix  as a function of the scale $\mu_0$, which may take any value determined by the free parameter $\delta\alpha_3(\Lambda)$.  The dashed lines show the SM running of couplings from Fig.~\ref{fig:SM} provided that $\mu_0=1$~GeV. It is observed that SM and BSM values both coincide at the matching scale $M\approx 1.6$~TeV, see \eq{matching}, indicated by the short-dashed vertical line.} \label{fig:sepE}
 \vskip-.5cm
 \end{figure}

 {\bf Benchmark scenario D}.   For models with $R_3=\bm{3}$ and $R_2\ge \bm3$, and for sufficiently large $N_F$,  we have observed that the UV fixed points \fp2 and \fp4 coexist, see Fig.~\ref{fSynopsis}. 
To illustrate the matching procedure for these settings, we set the
parameters as 
 \beq\label{D}
 (R_3,R_2,N_F)=(\bm{3},\bm{4},290)\,.
 \eeq 
The partially interacting  fixed point \fp2 can be matched onto the SM, particularly at low matching scale $M$ (not displayed). Results for \fp4 are displayed in Fig.~\ref{fig:match_fp4}. Couplings run out of the UV fixed point along the separatrix \eq{separatrix}. Unlike the previous example of benchmark $C$ \eq{C}, in this case couplings display a cross over at much lower energies. In particular, a matching to SM values is possible at a low scale of about $M\approx 2.4$~TeV. As explained after \eq{XY},
the price to pay is the necessity for a large multiplicity of flavors,
significantly larger than $N_{\rm AS}=18$ as minimally required for weakly coupled asymptotic safety to arise (see Tab.~\ref{tFP4}). It is worth contrasting the successful matching at \fp4 with the failure for the benchmark scenario $B$ \eq{B}: unlike the weak sector of the SM, the strong sector does not display a ``would-be'' Banks Zaks IR fixed point, see \eq{SM}. Consequently, trajectories emanating out of \fp4 are attracted towards the Gaussian fixed point rather than being diverted by an interacting fixed point as in \eq{B}. We conclude that the (non)-availability of a matching with \fp4  in the benchmark scenario $D$ ($B$) is dictated by features of the SM rather than the specifics of the BSM extension. \\[-1.5ex]
 
\begin{figure}[t]
\begin{center}
\includegraphics[width=0.50\textwidth]{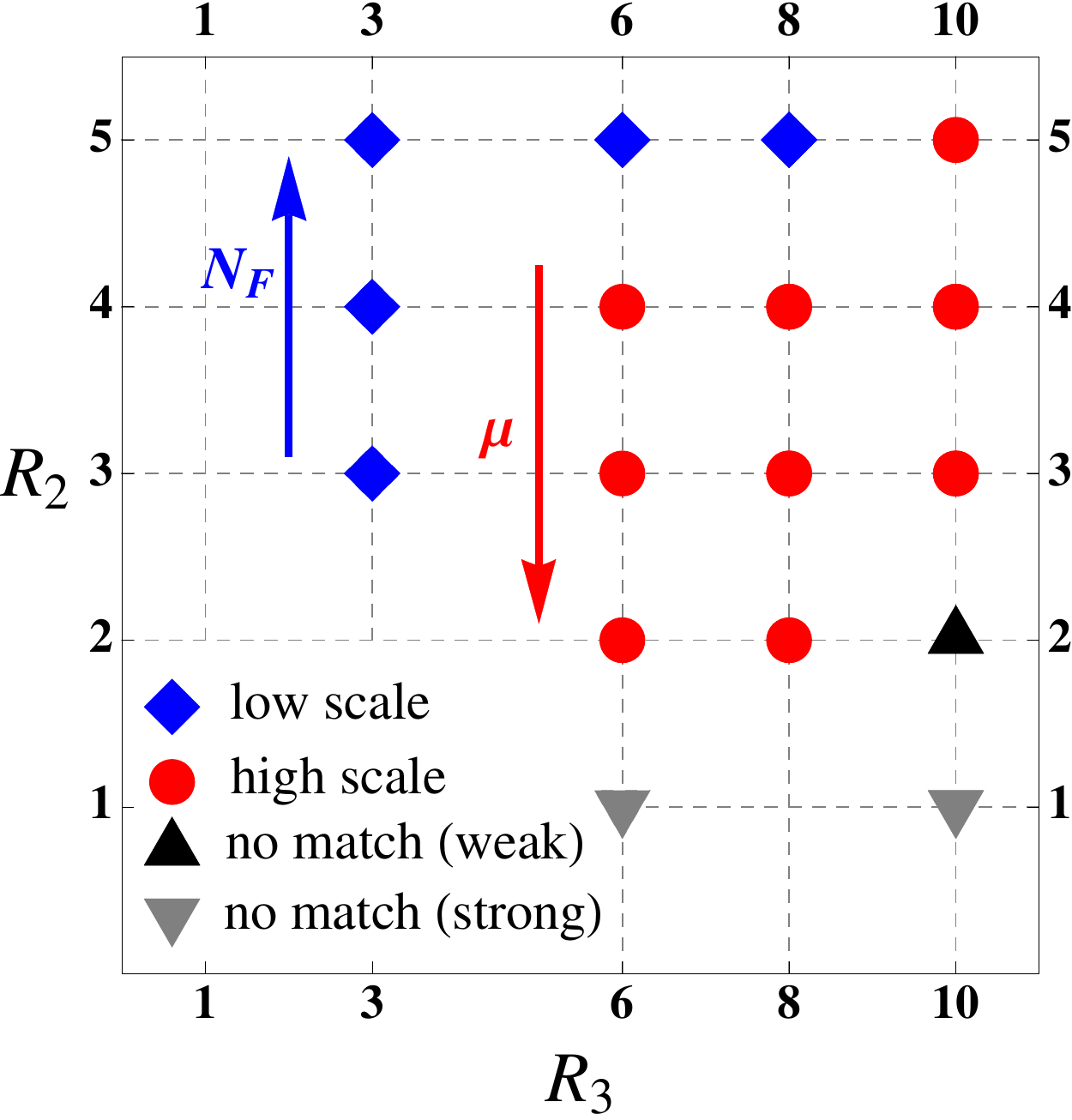}
\end{center}
\vskip-.5cm
\caption{Summary of matching conditions at  fully interacting UV fixed points \fp4 of the RG system \eq{BSM} in dependence on the BSM parameters $(R_3,R_2,N_F)$. 
Blue diamonds indicate low-scale matchings in the multi-TeV regime.
Red dots stand for a high matching scale beyond the reach of present day colliders. Black triangles indicate scenarios where a matching onto SM values is not available despite of both gauge couplings approaching the Gaussian. Gray triangles indicate the unavailability of a matching due to strong coupling phenomena in the weak gauge sector  ($R_2=\bm1$). 
Arrows additionally illustrate how the number of BSM fermion flavors $N_F$ (blue arrow) and the matching scale $\mu$ (red arrow) vary with the representation to ensure a successful matching.}
\label{fig:asyGYF_SM}
\end{figure}

  {\bf Benchmark scenario E}.  
For all settings with $R_2\ge \bm{2}$ and $R_3\ge \bm3$, the theory can display all three types of interacting UV fixed points. In any of these cases, \fp4 arises at the lowest possible value for $N_F$, see Fig.~\ref{fSynopsis}. It is then interesting to evaluate scenarios where \fp4 is the sole UV fixed point. Using our results from Tab.~\ref{tFP4}  we consider exemplarily the case
\beq
\label{E}
(R_3,R_2,N_F)=(\bm3,\bm3,72)\,. 
\eeq
The UV-IR connecting separatrix is displayed in Fig.\ref{fig:sepE}. We observe that both gauge couplings decrease towards the IR. The scale $\mu_0$ is a free parameter and solely fixed by the free parameter $\delta\alpha_3(\Lambda)$ in the deep UV. We confirm once more that the hierarchy $\alpha_2>\alpha_3$ in the deep UV invariably transforms into $\alpha_2<\alpha_3$ once the RG flow falls below the cross-over scale.   
Also shown is the SM running of gauge couplings (dashed lines) taken from Fig.~\ref{fig:SM}, with data points starting from $\mu=1$ TeV (corresponding to the choice $\mu_0=1$~GeV on the lower axis). Tuning the value of $\delta\alpha_3(\Lambda)$ (or $\mu_0$) along the separatrix amounts to shifting the separatrix in its entirety parallel to the lower axis. In Fig.\ref{fig:sepE}, values have been chosen to exemplify that the separatrix can match  SM values, \eq{matching}, at the matching scale $M\approx 1.6$~TeV. We stress once more that the shape of the separatrix, and hence the fixed point behavior of the theory in the deep UV, uniquely dictates the scale at which a matching to the SM can be made.

\begin{figure}[t]
 \begin{center}
 \includegraphics[width=0.7\textwidth]{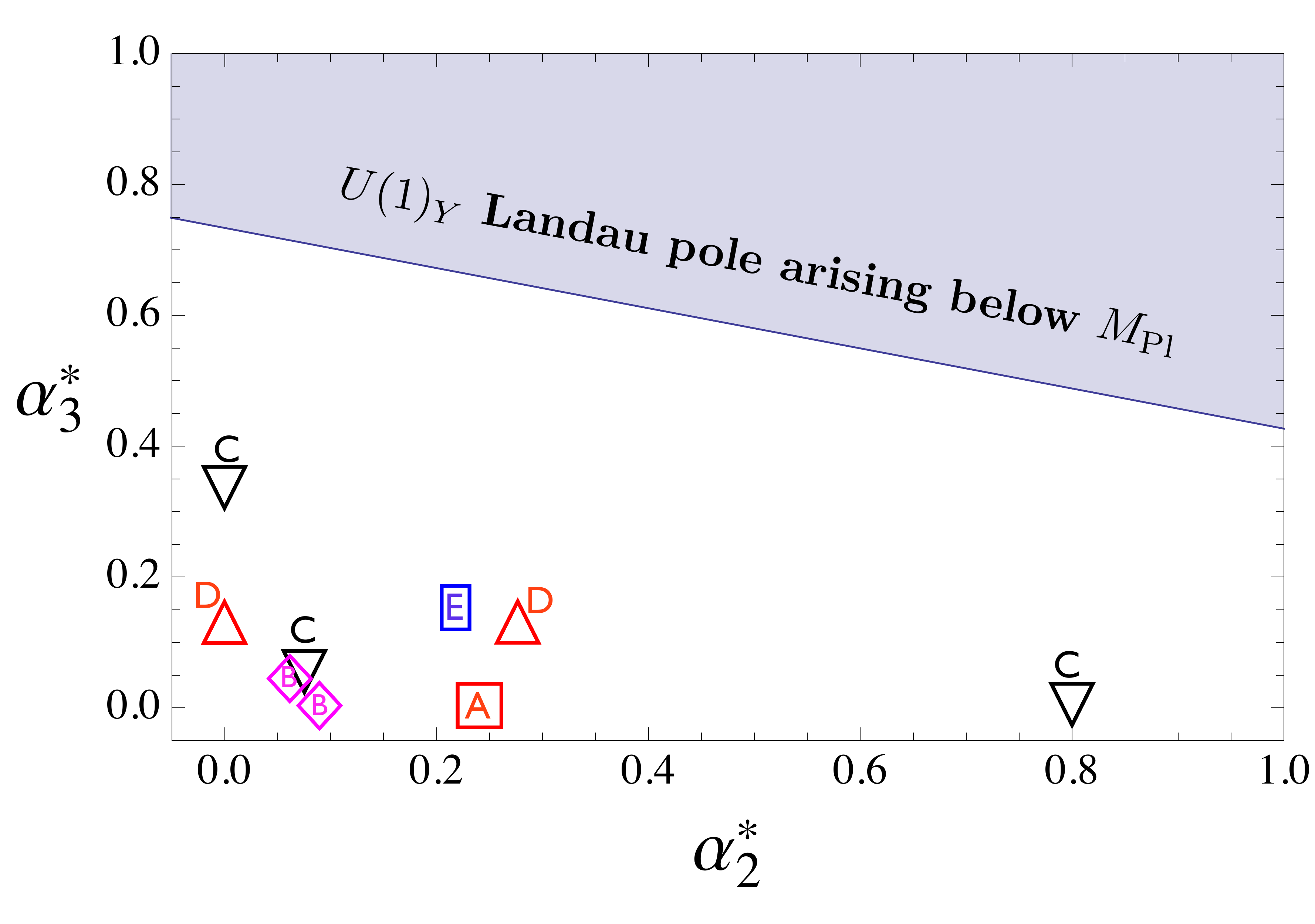}
 \end{center}
 \vskip-.5cm
 \caption{Shown is a conservative estimate for the exclusion area (gray) of fixed point values for the gauge couplings $(\alpha_2,\alpha_3)$ to ensure the absence of a Landau pole for the $U(1)_Y$ hypercharge below Planckian energies. Equally shown are the partially and fully interacting fixed point values for the benchmark models $A,B,C,D$ and $E$ given in Tab.~\ref{tBenchmark}, for comparison. All benchmark models are UV-safe.}
 \label{pExclusion}
 \end{figure}

\subsection{Synopsis of matching conditions}\label{synopsis}

To summarise, we have established that partially interacting fixed points \fp2 and \fp3 can comfortably be connected with the SM at low energies provided there are no nearby competing fixed points in the phase diagram of the theory. Moreover, in these cases the matching scale and thus the masses of BSM matter fields remain  freely adjustable parameters. The underlying reason for this is that  both gauge couplings remain relevant couplings in the deep UV. Typical examples for this are shown in Fig.~\ref{fig:match_fp2}, \ref{fig:match_fp3}, and~\ref{fig:match_fp3b} for \fp2 of benchmark $A$, and \fp3 of benchmark $B$ and  $C$, respectively. On the other hand, a matching to the SM becomes more contrived, or even  impossible, if nearby  competing fixed points 
influence the running of couplings. An example for this is shown in Fig.~\ref{fig:match_fp2b} for \fp2 of benchmark $C$, where the nearby  fully interacting fixed point \fp4 impacts on the UV-safe trajectories emanating out of the partially interacting fixed point \fp2, thereby enforcing a high matching scale. 

The matching of fully interacting fixed points \fp4 to the SM is qualitatively different. The reason for this is that only one of the gauge couplings remains a relevant coupling, which reduces the number of freely adjustable parameters in the  UV by one. Unlike partially interacting ones, fully interacting fixed points predict a relation between the gauge couplings.  The availability of a matching to the SM is then encoded in the UV-IR connecting separatrix and must be checked on a case by case basis, see~Fig.~\ref{fig:match_fp4}. 
In Fig.~\ref{fig:asyGYF_SM} we summarise  our results for the matching conditions at \fp4 in dependence on the BSM paramaters $(R_3,R_2,N_F)$. Low-scale matchings in the multi-TeV regime are indicated by blue diamonds. Examples for this relate to benchmark $D$ and $E$ displayed in   Fig.~\ref{fig:match_fp4} and \ref{fig:sepE}, respectively. 
  This is  contrasted with matchings  at a high scale, beyond the reach of present day colliders (red dots). An example for the latter is furnished by benchmark $C$ as shown  in Fig.~\ref{fig:match_fp4b}. On the other hand, matchings fail if gauge couplings along the separatrix never hit SM values, Fig.~\ref{fig:SM}, despite both gauge couplings approaching the Gaussian.  In Fig.~\ref{fig:asyGYF_SM} such scenarios are indicated by black triangles. 
Finally, nearby competing fixed points may distort the UV-safe separatrix and disallow a matching to the SM.  In Fig.~\ref{fig:asyGYF_SM}, the unavailability of matching due to strong coupling in the weak gauge sector  ($R_2=\bm1$) is indicated by gray triangles.
  An example for this is benchmark $B$ in Fig.~\ref{fig:sepB} where the competing fixed point  is the ``would-be'' Banks-Zaks fixed point of the SM weak sector. 
Arrows have been added in Fig.~\ref{fig:asyGYF_SM} to illustrate how the number of BSM fermion flavors $N_F$ (blue arrow) and the matching scale $\mu$ (red arrow) vary with the BSM matter representation to ensure a successful matching.

We briefly come back to the perturbativity of interactions  in the fixed point regime. We have found that the gauge couplings for all benchmarks take small values $\alpha\approx 0.04 - 0.8$, see Tab.~\ref{tBenchmark}.   Moreover, in the large-$N_F$ and large representation limit, fixed point couplings are parametrically small, \eq{fp3N} --
\eq{fp4N2}. In models which permit an asymptotic large-$N$ Veneziano limit, it has also been shown that perturbativity in $N_F\cdot\alpha\ll 1$ is guaranteed \cite{Litim:2014uca}.
Here, at finite $N_F$, the products $N_F\cdot\alpha$ come out  of order ${\cal O}(1-10)$  for all benchmarks, hinting towards the onset of strong coupling.  Ultimately, this pattern of result reflects the unavailability of a  Veneziano limit because fixed points necessitate representations higher than the fundamental, Fig.~\ref{fSynopsis}. Future studies should therefore include loop corrections beyond the leading orders, and non-perturbative effects.

Finally, we  comment on the role of the $U(1)_Y$ hypercharge. The SM predicts a Landau pole for the hypercharge many orders of magnitude beyond the Planck scale $M_{\rm Pl}\sim 10^{19}$ GeV. In our setup the BSM fields do not carry hypercharge. The interesting case where BSM fields carry hypercharge will be detailed elsewhere. Nevertheless, the running of the hypercharge  nevertheless differs from SM running above the matching scale because the strong or the weak or both gauge coupling(s) will grow and eventually settle at interacting fixed points. Interacting fixed points accelerate  the running of the hypercharge due to contributions at two loop. To exclude that a Landau pole may arise below Planckian energies, we assume a ``worst case'' scenario in which $(\alpha_2,\alpha_3)$ take fixed point values already at a very low scale of $1.5$~TeV (bounds are softened if fixed point values are reached at higher scales). This leads to a conservative exclusion plot shown in Fig.~\ref{pExclusion} where the shaded area indicates the forbidden region of values for the UV fixed point. We observe that gauge sectors must become strongly coupled already at low energies to inflict a Landau pole below Planckian energies for the hypercharge. For comparison, we also indicate the location of UV fixed points for the benchmark models $A,B,C,D$ and $E$ as given in Tab.~\ref{tBenchmark}. 
Quantitatively, none of the benchmark models reach a Landau pole for the hypercharge  below $10^{26}$ GeV. We conclude that all benchmark models are deeply in the UV-safe region of parameter space.

\section{\bf Phenomenology}\label{Pheno}

In this section we discuss experimental  signatures of   asymptotically safe SM extensions.
We assume  that the BSM sector can at least  partially be accessed at the  LHC, which implies a low matching scale and
 masses of the BSM matter fields  in the multi-TeV range. An order of magnitude heavier states can be considered at future colliders \cite{Shiltsev:2015tta}.
Because of the flavor symmetry the BSM fermions are stable in the model \eq{L}. Allowing the flavor symmetry to be broken, the lightest BSM fermion is still stable
as long as Yukawa interactions with SM fermions are absent. The latter holds except for a few low-dimensional representations with tuned hypercharge of the BSM fermions.
As we assume that the BSM fields do not carry hypercharge, these exceptional cases cannot be realized.
Without mixing with SM fermions, flavor physics constraints are not relevant to our models.
We further assume $R_3 \neq \bm1$. If the new fermions would be colorless, their production at hadron colliders  would be of higher order and suppressed. Scenarios with $R_3 = \bm1$ 
are  certainly suitable for study at  an $e^+ e^-$-machine operating  at  high energies \cite{Gomez-Ceballos:2013zzn,Shiltsev:2015tta,dEnterria:2016sca}.

An obvious search strategy is to look for asymptotically safe BSM physics by probing the strong running coupling evolution and the weak interaction. We discuss various constraints and opportunities from the running gauge couplings, from the weak sector,
from direct searches for long-lived QCD-bound states composed out of BSM fermions and SM partons,
and from LHC diboson searches, offering further constraints on BSM matter
including the $(M_\psi,M_S)$ parameter space.

\subsection{Strong coupling constant evolution \label{sec:run}}

The presence of a large number of fermions charged under $SU(3)_C \times SU(2)_L$  changes  the running of the corresponding gauge couplings drastically, as illustrated 
in Figs.~\ref{fig:match_fp2}-\ref{fig:sepE}. The deviation from the SM, shown in Fig.~\ref{fig:SM}, kicks in rather quickly with an order one increase in slope of the asymptotically safe coupling 
and provides  a smoking gun signature of BSM physics  considered in this work. Threshold corrections are not expected to change this picture qualitatively, although the onset of BSM effects may be  somewhat smoother. 

The CMS collaboration has extracted the value of the QCD running coupling up to the scale 2 TeV \cite{Khachatryan:2014waa}
using the measurement of the inclusive jet cross section for proton-proton collisions at a centre-of-mass energy 
of 7 TeV with data corresponding to an integrated luminosity of 5.0 $\textrm{fb}^{-1}$ \cite{Chatrchyan:2012bja}. This determination is consistent with the SM.
Also other measurements, including the one of the inclusive 3-jet production differential cross section \cite{CMS:2014mna}, have not
observed deviations from the two-loop running predicted in the SM up to $1.5$ TeV.
Therefore, modulo threshold corrections, the lower limit on the mass of new  colored matter reads
\begin{eqnarray}
\label{eq:aslimit}
M_\psi \gtrsim 1.5 \,  \mbox{TeV} \, .
\end{eqnarray}
In the following subsections  we work out further experimental constraints on $M_\psi$, which in turn
 imply limits on the matching scale. We recall that scenarios with partially interacting fixed points \fp2 and \fp3 can generically be matched at any scale (except in specific circumstances, see Fig.~\ref{fig:match_fp2b}), whereas the matching scale is uniquely fixed  for fully interacting fixed points \fp4. The latter scenarios are therefore subject to stronger experimental constraints.

 \begin{figure}[t]
 \hskip-1.3cm\includegraphics[width=0.77\textwidth]{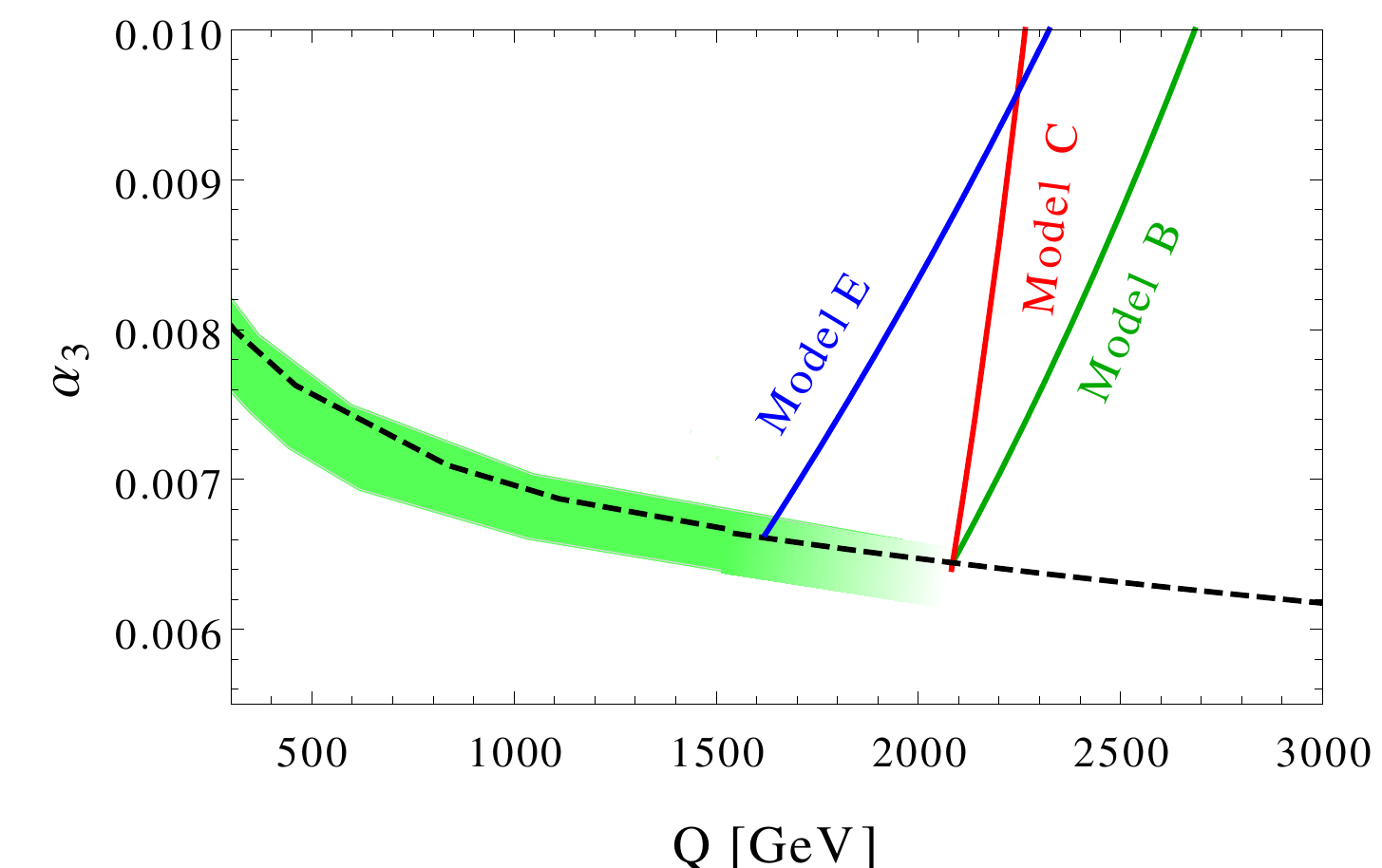}
 \caption{\small SM running of the QCD  coupling constant (black dashed line) and its uncertainty (green band) as determined by CMS  \cite{Khachatryan:2014waa}.
 Colored solid lines indicate the running of $\alpha_3$  in asymptotically safe benchmark scenarios $B$, $C$ and $E$ summarized in Tab.~\ref{tBenchmark} 
 with a low matching scale  around $1.5$ to 2 TeV. }
 \label{fig:qcd_running}
 \end{figure}

In Fig.~\ref{fig:qcd_running} we show the running of the strong coupling (black dashed line) and its uncertainty (green band) as determined by CMS \cite{Khachatryan:2014waa}.
In addition, we show  the running of $\alpha_3$ in the  asymptotically safe benchmarks $B$, $C$ and $E$ introduced in Tab.~\ref{tBenchmark}
for low matching scales around $1.5$ to 2 TeV. Note that benchmark $E$ (blue curve)  relates to a  fully interacting UV fixed point whose  matching scale  is fixed at $1.6$ TeV. As can be seen, benchmark $E$ is already being  probed experimentally. Threshold corrections may allow to evade the CMS limit, as the data near 2 TeV are also losing statistics.

 \begin{figure}[t]
  \hskip-1.5cm
  \includegraphics[width=0.77\textwidth]{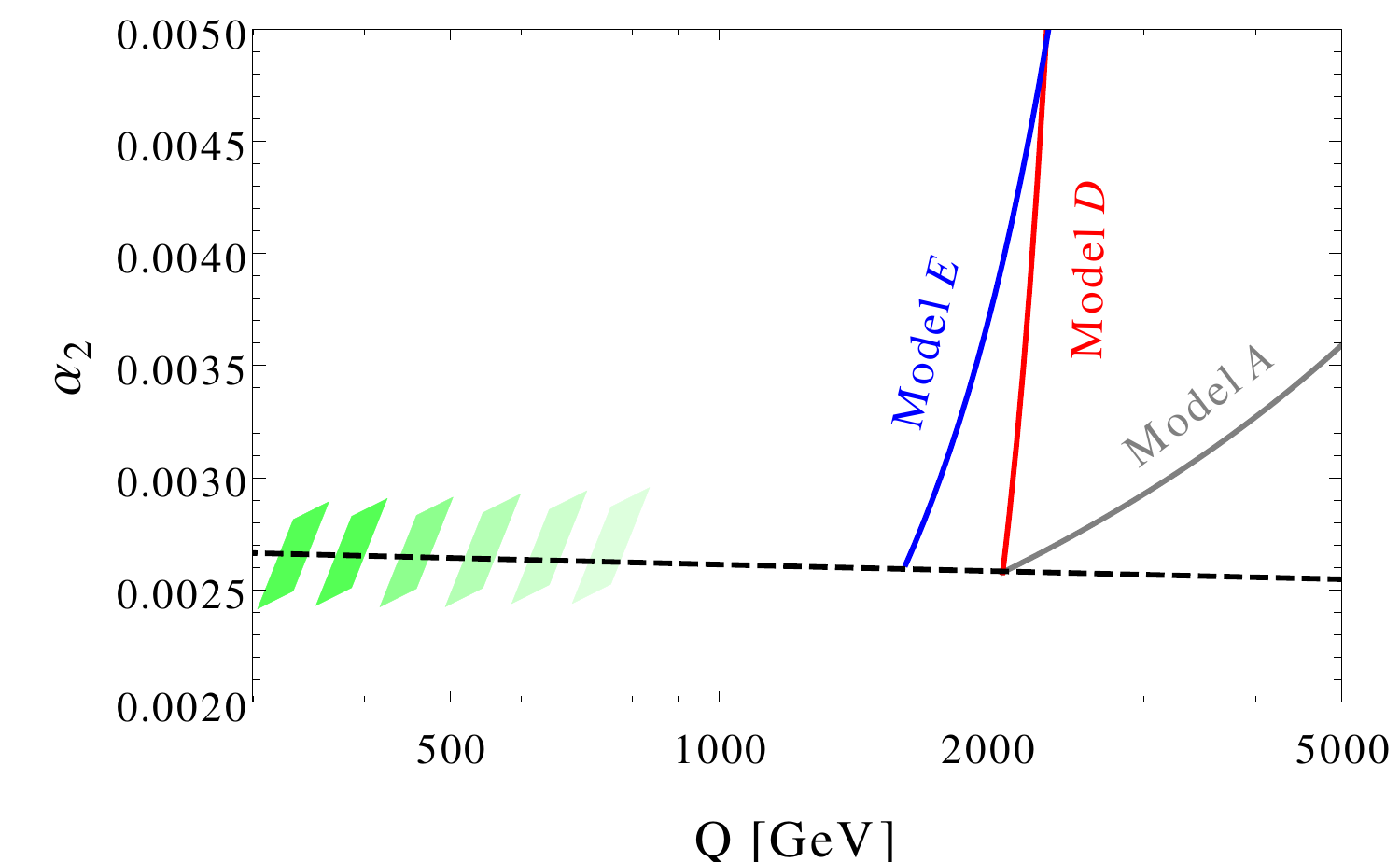}
 \caption{\small SM running of the weak coupling constant (black dashed line) and schematically indicated by the green hatched band the region where the weak sector of the SM has
 passed experimental tests, see text for details.
 Colored solid lines indicate the running of $\alpha_2$  in asymptotically safe benchmark scenarios introduced in Tab.~\ref{tBenchmark} 
 that allow for a low matching scale  around $1.5$ to 2 TeV.}
 \label{fig:weak_running}
 \end{figure}

\subsection{The weak sector \label{sec:weak}}

The experiments at the LEP collider have  probed  the SM's electroweak sector with scrutiny and found no significant deviation up to  $\sim 209$ GeV \cite{Schael:2013ita}. 
The LHC has extended related  SM tests into  the several ${\cal{O}}(100)$ GeV regime \cite{Azzurri:2016cej}, still allowing for  weakly-interacting uncolored vector-like fermions below the TeV-scale.
Within asymptotically safe models, this can happen, for instance, in benchmark $A$ by noting that since  the fixed point is partial only, the matching scale can be different than the one shown in Fig.~\ref{fig:match_fp2}.
Electroweak vector boson scattering at the LHC and at future lepton colliders \cite{dEnterria:2016sca} is  sensitive to such  BSM effects.

For  $R_2 \neq \bm1$  contributions to the $\rho$-parameter arise if the BSM fermions encounter $SU(2)_L$ breaking due to mass splitting $\delta M \ll M_\psi$  in the fermion multiplet.
This implies  \cite{Olive:2016xmw}
\begin{eqnarray}
N_F \,d(R_3)\, S(R_2) \,  \delta M^2 \lesssim ( 40 \, \mbox{GeV})^2 \, ,
\end{eqnarray}
a splitting below percent level for TeV-ish fermion masses and higher.

In Fig.~\ref{fig:weak_running} we show the running of the weak  coupling (black dashed line) and, schematically, the region with agreement with the SM's weak theory
denoted by the hatched green band. The solid colored lines correspond to the   asymptotically safe benchmarks $A$, $D$ and $E$ summarized in Tab.~\ref{tBenchmark} for a low  matching scale around $1.5$ to 2 TeV.

Constraints from rare decays  can be evaded as long as the BSM fermions do not couple directly to SM Higgs, quarks, or leptons -- as is the case in our setup.
For BSM fermions with $R_2>\bm1$ a contribution to the anomalous magnetic moment of the muon arises at 2-loop in the electroweak interactions. We estimate this as
$\Delta a_\mu \sim d(R_3)\,S(R_2)\, N_F\, [ \alpha/(4 \pi) (m_\mu/M_\psi) ]^2$. Comparison with data 
$\Delta a_\mu^{\rm exp} \sim (2-3) \cdot 10^{-9}$  \cite{Olive:2016xmw} yields the constraint
\begin{eqnarray}
d(R_3) \,S(R_2)\,N_F\,  \left( \frac{ \mbox{TeV}}{M_\psi} \right)^2 \lesssim 10^{4} \,  ,
\end{eqnarray}
which is satisfied for all our benchmarks  in Tab.~\ref{tBenchmark} and  for $M_\psi$ above a TeV.

Below the BSM mass threshold the effects of the BSM fermions can be studied indirectly through electroweak precision tests. 
Charged and neutral current Drell-Yan (DY) processes offer a promising way to test  such corrections~\cite{Alves:2014cda}, as they are both experimentally clean and very well understood theoretically. The oblique parameter $W$~\cite{Barbieri:2004qk,Cacciapaglia:2006pk} is of particular interests since its impact increases with energy allowing high precision studies at present and future colliders \cite{Farina:2016rws}.
$W$ is directly related to the BSM contribution that modifies the electroweak beta function,
\begin{equation}
\label{parW}
 W=-\frac{\alpha_2}{10}\, \frac{M_W^2}{M_\psi^2}\,B_2^{\scriptscriptstyle\rm BSM},
\end{equation}
where $B_2^{\scriptscriptstyle\rm BSM}$ denotes the BSM contribution to the 1-loop coefficient $B_2$ of $\beta_2$, see \ref{Bs}.

\begin{figure}[t]
\begin{center}
\includegraphics[height=7.5cm]{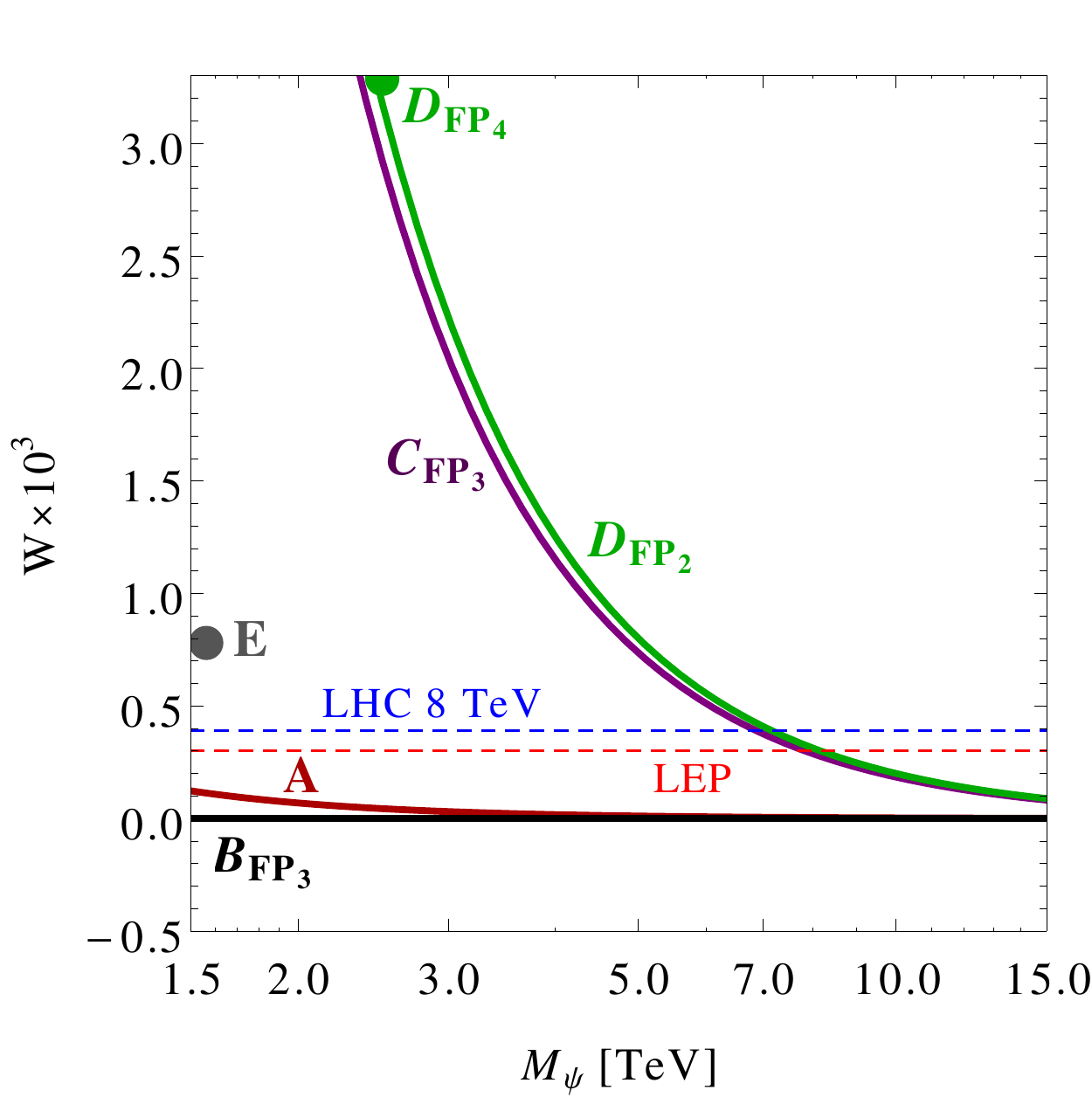}
\includegraphics[height=7.5cm]{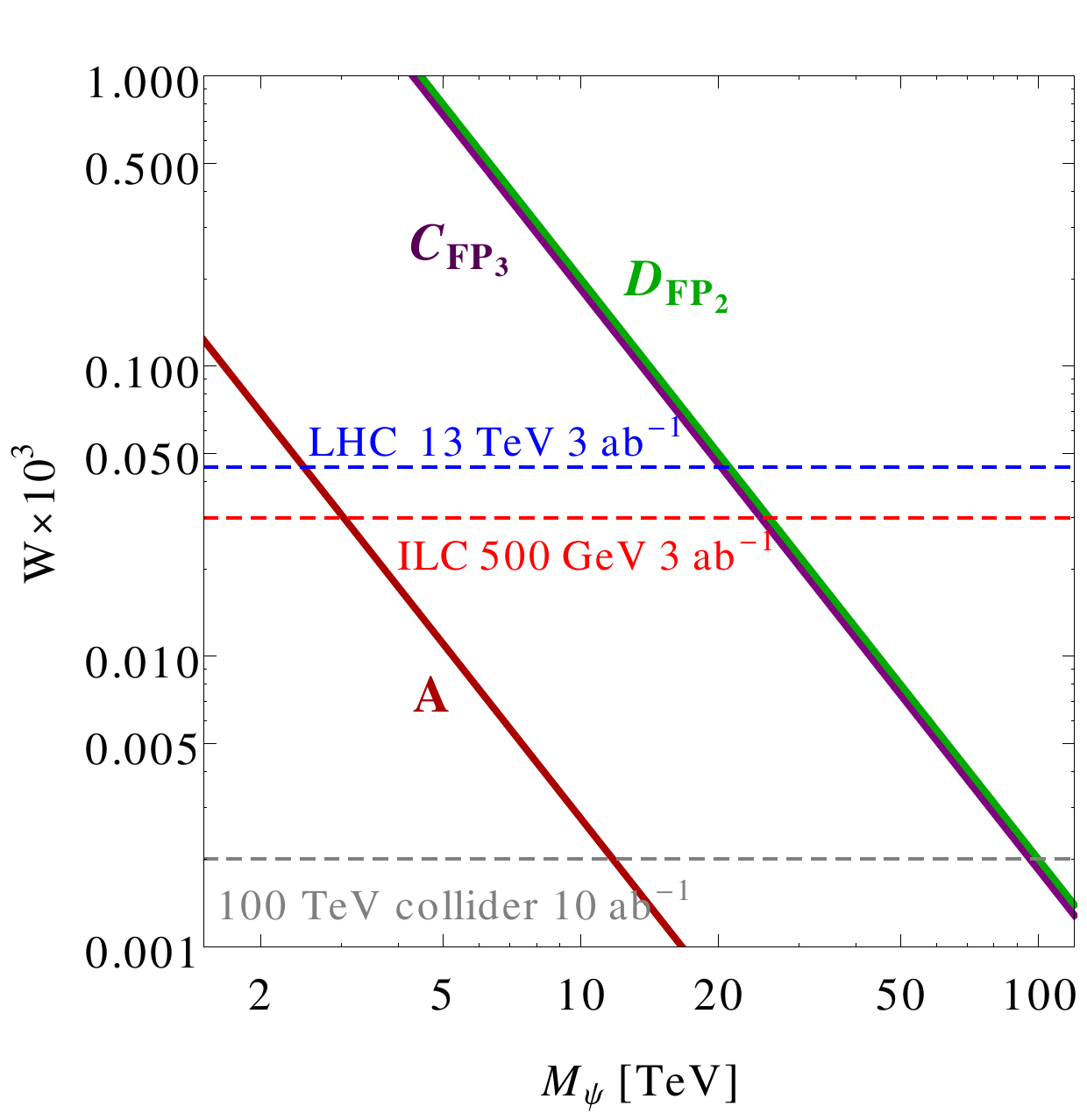}
\end{center}
\vskip-.5cm 
\caption{Shown is the electroweak precision parameter $W$ (\ref{parW}) as a function of the BSM fermion mass for the low-scale benchmark models given in Tab.~\ref{tBenchmark}. In the left panel dashed lines  show 95\% C.L. upper  limits obtained from  
LEP (red)  and  the LHC at 8 TeV (blue).
In the right panel dashed lines indicate the projected reach of the  LHC at 13 TeV (blue),  the   ILC 500 GeV (red), and a  100 TeV collider (gray). Experimental limits are taken from  \cite{Farina:2016rws}.}
\label{fig:dyewk}
\end{figure}

Fig.~\ref{fig:dyewk} displays $W$ versus the matching scale $M_\psi$ for the  benchmark scenarios defined in Tab.~\ref{tBenchmark}. In the left panel the experimental 95\% C.L. upper limits  \cite{Farina:2016rws} from LEP (red dashed line) and   LHC 8\,TeV (blue dashed line) are shown.  Constraints on negative $W$ exist but are not relevant here.
For the fully-interacting fixed points \fp4 of benchmark $D$ and $E$ (full dot),  a low matching scale  at around 2\,TeV dictates a high multiplicity of BSM fermions, and a large contribution to \eq{parW}. These scenarios turn out to be excluded. For scenarios with partially interacting fixed points  (\fp2 or \fp3) the matching scale is a free parameter. For benchmark models $C$ (magenta line) and  $D$ (green line) the matching scale must be larger than about 8\,TeV to satisfy  LEP constraints. Benchmark $A$ (red line) is not yet constrained by the data owing  to the low number of BSM fermion species $N_F$, while benchmark $B$ (black line)  is not probed at this order because its BSM fermions are $SU(2)_L$ singlets. 

The right panel of Fig.~\ref{fig:dyewk}  shows the projected sensitivities of the  LHC at 13\,TeV with $3\,\mbox{ab}^{-1}$ integrated luminosity (dashed blue line), the ILC 500 with $3\,\mbox{ab}^{-1}$ (dashed red line), and a 100\,TeV collider with $10\,\mbox{ab}^{-1}$ (dashed gray line). The precision of the $W$ determination is expected to increase by two orders of magnitude, requiring the matching scale in all allowed benchmark scenarios to be above around 10\,TeV.

While this analysis demonstrates the importance of  DY measurements for  the type of BSM scenarios outlined here,  constraints based on  (\ref{parW})  must be taken with a grain of salt. The reason for this  is that the corrections to $W$ are only known to one loop order. 
In our framework, competing two loop corrections in the gauge beta functions play an important role as they are responsible for the fixed point. To estimate two loop effects, we replace $B_2^{\scriptscriptstyle\rm BSM}$ in \eq{parW} by the  effective coefficient $B_2^\prime$ evaluated on the RG trajectory near the matching scale. For benchmark models $C$ where $\alpha_3$ becomes asymptotically safe, 
we find that the bound on the matching scale softens, from about 8\,TeV to about 5\,TeV. Also, benchmark  $B$ now contributes negatively to $W$ and can be probed in the future. Similar two loop effects are expected for the other benchmark models.
A complete two loop analysis  of the oblique parameter $W$, although desirable,  is beyond the scope of this work.

\subsection{$R$-hadrons \label{sec:R}}

We assume that at least some of the BSM fermions can be pair-produced,
\beq
2 M_\psi < \sqrt{s} \, ,
\eeq
where $\sqrt{s}$ denotes the accessible center of mass energy at the collider.
 At least the lightest of the fermions has a long life, longer than
 a typical hadronization time scale, and  forms colorless QCD bound states with ordinary partons (quarks and gluons), the so-called 
$R$-hadrons.

Both the ATLAS and CMS collaborations searched for  heavy long-lived charged $R$-hadrons using a data sample corresponding to 3.2 fb$^{-1}$ of proton-proton collisions at $\sqrt{s}=13$ TeV.
No significant deviations from the expected background have been  observed which allowed to put a model-independent 95\% confidence level (C.L.)  upper limits
on the production cross section of long-lived $R$-hadrons. In the framework of supersymmetry, those results have been translated into a lower bound on the mass of
the fermionic partner of the gluon (gluino), which read 1.5 TeV for CMS \cite{Khachatryan:2016sfv} and 1.6 TeV for ATLAS \cite{Aaboud:2016uth}. 
Recently CMS has updated their analysis to 12.9 fb$^{-1}$ of data \cite{CMS-PAS-EXO-16-036}, with the corresponding limit on the gluino mass increased to 
1.7 TeV.

At the LHC any colored and hypercharge-neutral BSM fermion would be produced in the same way. Therefore one can easily recast the experimental limits for gluino searches in the framework of asymptotically safe scenarios considered in this study.
In the leading order the $\psi\bar{\psi}$ pairs can be produced by gluon fusion or by quark-anti-quark annihilation, where the former mechanism is a dominant
one. We can additionally assume 
that the main contribution to $pp\to \psi\bar{\psi}$ comes from a $t$-channel exchange of a BSM fermion.
In this case the production cross section $ \sigma_{\psi\bar{\psi}}$ depends on $(R_3,R_2,N_F)$  and scales proportionally to the 
 factor $\mathcal{C}_3$,
\begin{align}\label{sigmaC3}
 \sigma_{\psi\bar{\psi}}\sim N_F\, \mathcal{C}_3  \quad{\rm with}\quad \mathcal{C}_3= [C_2(R_3)]^2\,d(R_3)\,d(R_2)\,.
\end{align}
We can then put lower limits on the BSM fermion mass using the experimental limits for gluinos provided in \cite{Khachatryan:2016sfv} and \cite{Aaboud:2016uth},
rescaling the gluino production cross section by $\mathcal{C}_3$. Notice also that it is possible for real representations, such as those with  $(p,p)$ for $R_3$,  that $\psi$ is a Majorana fermion; in all other cases we ignore the differences with respect to the
gluinos in our estimates. 

In Tab.~\ref{tab:Rhad} we show the lower bounds on $M_{\psi}$ in dependence of $R_2$ and $R_3$ together with  $\mathcal{C}_3$ for $N_F=1$.
The lower bound increases with increasing $d(R_3)$. For $R_3=\bm{15}^\prime$ and $d(R_2) >1$ it reads  {2.3} TeV.
We conclude that colored BSM fermions must be heavier than at least {1.5} TeV, consistent with (\ref{eq:aslimit}).
For larger $N_F$, the bounds get stronger. For example, for benchmarks $B$, $D$ and $E$ defined in Tab.~\ref{tBenchmark} we find $M_\psi^{\min}= 2.6, 2.4$ and $2.1$ TeV, respectively.
The limit for benchmark $C$ is beyond {2.8}~TeV.
For benchmarks $C$ and $D$ the constraints  from DY processes obtained in Sec.~\ref{sec:weak} are  stronger than the $R$-hadron ones. 

\begin{table}[t]
         \begin{center}
         \begin{tabular}{ccccccc}
\toprule
\rowcolor{LightGreen}
\rowcolor{LightGreen}
\multicolumn{1}{c}{\cellcolor{LightBlue}$\bm{\psi(R_3,R_2)}$}
&\multicolumn{2}{c}{$\ \ \ \ \bm{R_2= 1}\ \ \ \ $}
&\multicolumn{2}{c}{$\ \ \ \ \bm{R_2= 2}\ \ \ \ $}
&\multicolumn{2}{c}{$\ \ \ \ \bm{R_2= 3}\ \ \ \ $}\\ 
\midrule
\rowcolor{LightYellow}
\rowcolor{LightYellow}
\cellcolor{LightGreen}
   $\ \ \bm{R_3}\ \ $      
   & $\ \ \mathcal{C}_3$ & $M_{\psi}^{\rm min}$ (TeV)   & $\ \ \mathcal{C}_3$ & $M_{\psi}^{\rm min}$ (TeV)  & $\ \ 
\mathcal{C}_3$ & $M_{\psi}^{\rm min}$ (TeV) \\ 
\midrule
\rowcolor{LightGray}
$\bm 3$ & $5\s013
$ & (1.3) & $10\s023
$ & (1.4) & 16 & 1.5\\
$\bm 6$ & $66\s023
$ & 1.7 & $133\s013
$ & 1.8 & 200 & 1.9  \\
\rowcolor{LightGray}
$\bm 8$ &  72 & 1.7 & 144 & 1.8 & 216 &  1.9 \\
$\bm{10}$ & 360 & 2.0 & 720 & 2.1 & 1080 & 2.2\\
\rowcolor{LightGray}
$\bm{15}$ & $\ 426\s023
\ $ & 2.0 & $\ 853\s013
\ $ & 2.1   & $\ 1280\ $& 2.2 \\
$\bm{15}^\prime$  & $1306\s023
$ & 2.2 & $2313\s013
$ & 2.3 & 3920 & 2.4 \\
\bottomrule
         \end{tabular}
         \end{center}
         \caption{Lower limits on the mass of the lightest BSM fermion, $M_\psi^{\rm min}$, as derived from the searches for long-lived charged particle 
         by CMS~\cite{CMS-PAS-EXO-16-036,Khachatryan:2016sfv} and ATLAS~\cite{Aaboud:2016uth} for $N_F=1$. We make explicit a dependence 
on the fermion representations $R_2$ and $R_3$ under $SU(2)_L$ and $SU(3)_C$, respectively.
The dominant contribution to the production cross section  is proportional to   $\mathcal{C}_3$, \eq{sigmaC3}, which is also given.
Values in parentheses correspond to scenarios $(R_3,R_2)$ with no weakly interacting UV fixed points, see Fig.~\ref{fSynopsis}.  }
\label{tab:Rhad}
\end{table}

\subsection{Diboson spectra and resonances \label{sec:di}}

\begin{figure}[t]
\begin{center}
\includegraphics[width=0.7\textwidth]{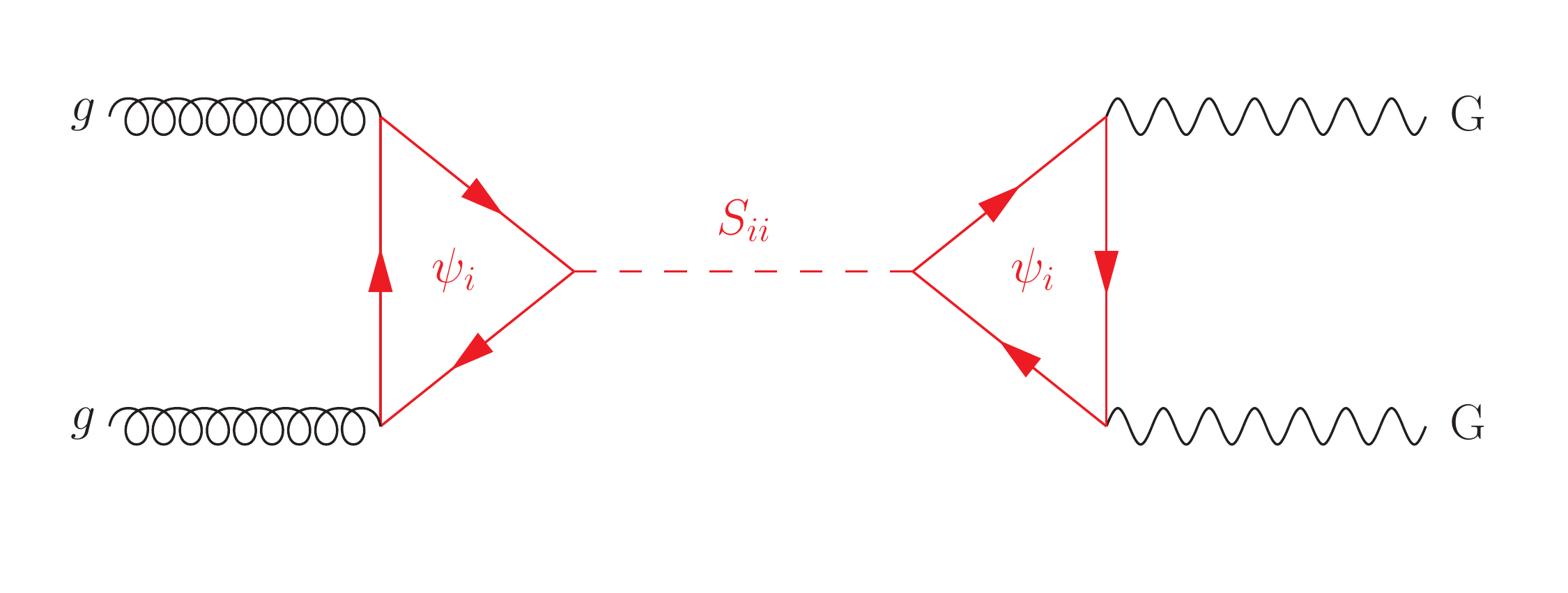}
\end{center}
\vskip-1cm 
\caption{\small Production via gluon fusion and decay of the scalar resonance $S$ in the asymptotically safe SM extension \eq{BSM}. Here, $i$ denotes the BSM flavor index and $GG$ stands for any combination of SM gauge bosons \eq{GG}.}
\label{fig:feynman}
\end{figure}

Consider the situation where  the BSM scalars are lighter than twice the mass of the fermions and can be resonantly produced,
\beq \label{eq:diboson}
M_S < 2 M_{\psi} \, , \quad {\rm and} \quad M_S < \sqrt{s} \, .
\eeq 
In such a case  the scalars cannot decay on-shell to any of the BSM fermions and  because its mixing with the SM Higgs boson is negligible, 
the only possible decay channels are loop-mediated decays into pairs of gauge bosons 
\beq\label{GG}
GG=gg,\,\gamma \gamma,\, ZZ,\, Z \gamma,\ {\rm or}\ WW\,.
\eeq
The cross sections for these, as well as their relative strengths, depend directly on transformation properties of the BSM fermions under $SU(3)_C$ and $SU(2)_L$.
Since $S$ does not couple directly to the SM fermions 
its dominant production mechanism is  gluon fusion which proceeds through the loops containing $\psi_i$. This process is schematically depicted in Fig.~\ref{fig:feynman}.
Due to the particular flavor structure of the asymptotically safe BSM sector, 
one needs to consider a simultaneous production of $N_F^2$ scalars $S_{ij}$, each of them coupled to exactly one fermion pair $\bar{\psi}_i\psi_j$. 
However, since  flavor is conserved in the fermion-gauge boson interactions, only diagonal couplings are allowed in this process and the number of 
simultaneously produced scalars is reduced to $N_F$. 
Due to interference effects between the $N_F$ diagrams, it is useful to investigate separately the limiting cases of maximal and no interference.\\[-1.5ex]

{\bf Maximum interference.}
Provided that all scalars $S_{ii}$ have the same mass $M_S$ and total decay width $\Gamma_S$, the interference between them is  maximal and 
 the  cross section for a diboson signal $GG$   \eq{GG} is given by \cite{Gunion:1989we,Djouadi:2005gj}
 \begin{align}\label{Max}
 \sigma(pp\to S_{1,...,N_F}\to GG)=N_F^2\,\sigma(pp\to S_{1}) \textrm{BR}(S_{1}\to GG)=
 \frac{N_F^2\,\pi^2}{8M_{S}^3\,\Gamma_{S}} I_{\textrm{pdf}}\,\Gamma^1_{GG}\,\Gamma^1_{gg}\,.
 \end{align}
Notice that \eq{Max} scales as $N_F^2$ times the  cross section for one individual flavor. In the above, $\Gamma_{GG}^1$ denotes the partial decay width into two gauge bosons with only one generation of BSM fermions in the loop. Similarly, $\Gamma_{gg}^1$ stands for 
the corresponding partial width into two gluons,
and $I_{\textrm{pdf}}$ is the 
integral of parton (gluon) distribution function in proton, evaluated at the energy scale $\mu=M_S$ with the center of mass energy 
$\sqrt{s}$.
\\[-1.5ex]

{\bf No interference}. Interference effects are absent provided that the masses of the scalars $S_{ii}$ are narrowly spaced with mass differences $\Delta M$ below the detector mass resolution $\Gamma_{\rm det}$, 
and provided that individual widths do not overlap. Consequently, the production cross section becomes
 \begin{align}\label{Min}
 \sigma(pp\to S_{1,...,N_F}\to GG)=N_F\,\sigma(pp\to S_{1}) \,\textrm{BR}(S_{1}\to GG)=
 \frac{N_F\pi^2}{8M_{S}^3\Gamma_{\rm det}}\, I_{\textrm{pdf}}\,\Gamma^1_{GG}\,\Gamma^1_{gg}\,.
 \end{align}
Notice that the total cross section scales as $N_F$ times the cross section for one individual flavor. As such, the absence of interference effects reduces the total cross section parametrically by a factor of $N_F$ over \eq{Max}. We expect that settings with partial interference effects are well covered within the limits \eq{Max} and \eq{Min}.\\[-1.5ex]

The relevant energy scale in the process is $\mu \sim M_S$, that is, the diboson invariant mass. The matching scale $M$ is of the order $M_\psi$. 
For $M_S$ below $M$,  roughly $M_S \lesssim M_\psi$ (``low $M_S$"), 
the gauge couplings assume SM evolution.
For $M_S$ above $M$ (``high $M_S$"), the gauge couplings follow  the BSM fixed-point trajectory. The kinematical range for the high $M_S$ scenario
is 
\beq\label{highMs}
M_\psi \lesssim M_S \leq 2 M_\psi\,.
\eeq
Since  characteristic diboson signatures can arise in many BSM scenarios, they have been intensively searched for at the LHC.

\begin{figure}[t]
 \begin{center}
\hskip-1cm
 \includegraphics[width=0.65\textwidth]{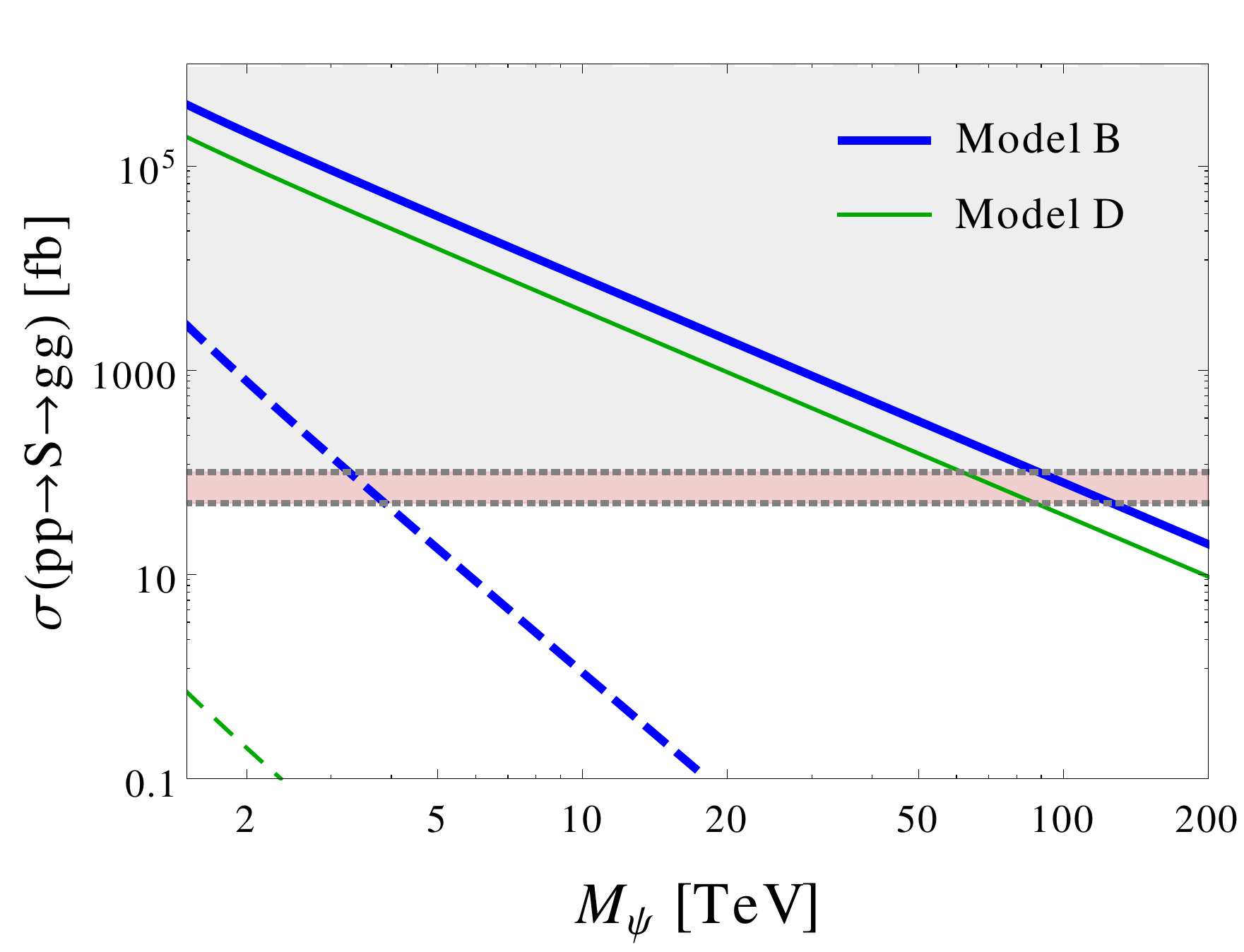}
 \end{center}
\vskip-.5cm 
\caption{
Dijet cross section as a function of the BSM fermion mass $M_\psi$ for  benchmark $B$ (thick blue curves), 
 and benchmark $D$  (thin green curves) for $M_S=1.5$ TeV. Solid curves correspond to the maximal interference between $N_F$ scalars, while dashed ones to no
 interference. The upper and lower horizontal dashed line denotes the ATLAS 95\% C.L. limit \cite{Aaboud:2017yvp}  on the dijet cross section assuming 50\% and 100\% acceptance, respectively.}  \label{fig:dijet_limit}
 \end{figure}

Recently
 both ATLAS and CMS updated their 95\% C.L.~limits on the fiducial cross section times branching ratio ($\sigma\times BR\times$ acceptance $A$ for a dijet analysis) for a general scalar resonance decaying into
$gg$  \cite{Aaboud:2017yvp,CMS-PAS-EXO-16-056}  
(updating
\cite{ATLAS-CONF-2016-030,ATLAS-CONF-2016-069,Sirunyan:2016iap}), 
$Z\gamma$  \cite{ATLAS-CONF-2016-010,CMS-PAS-EXO-16-034},
$ZZ$ \cite{ATLAS-CONF-2016-082,CMS:2016noo},
$WW$  \cite{ATLAS-CONF-2015-075,ATLAS-CONF-2016-021}
and
$\gamma\gamma$ \cite{ATLAS-CONF-2016-059,Khachatryan:2016yec}.
The exact limit in each case depends on the mass of the resonance, as well as on its total width. 
In the following we  choose $M_S=1.5$ TeV unless otherwise  stated, and
$\Gamma_S \leq  \Gamma_{\rm det}$.
First we consider limits on $\sigma(pp\to GG)$ provided by dijet searches. The partial width of $S$ into gluons \cite{Cvetic:2015vit} reads
\begin{align}\label{width:gg}
\Gamma_{gg}=\frac{\alpha_{s}^2M_{S}^3}{32\pi^3}\,
\left|\frac{yS_2(R_3)d(R_2)}{M_{\psi}} A_{1/2}(x)\right|^2,
\end{align}
where $\alpha_s=4\pi \alpha_3$ and the loop function is defined as $A_{1/2}(x)=\frac{2}{x^2}[x+ (x-1)\arcsin (\sqrt{x})^2]$ and $x=M_{S}^2/(4M_\psi^2)$. 
In Fig.~\ref{fig:dijet_limit} we show $\sigma(pp\to S\to gg)$ versus the mass of the BSM fermions $M_{\psi}$ for $M_S=1.5$ TeV. Solid curves correspond to the limit of maximal interference, while dashed ones refer to  no interference. In the latter case the total width 
is assumed  to be $\Gamma_{\rm det}=0.02M_S$, while in the former we calculate $\Gamma_S =\sum_{GG} \Gamma_{GG} \simeq \Gamma_{gg}$, which is below $\Gamma_{\rm det}$. Moreover, benchmark $B$ 
 $(R_3=\bm{10},R_2=\bm{1}, N_F=30$, thick blue lines) is  contrasted with benchmark $D$ 
$(R_3=\bm{3},R_2=\bm{4},N_F=290$, thin green lines) in Fig.~\ref{fig:dijet_limit}.  The upper and lower horizontal 
dashed line indicates the ATLAS \@ 95\% C.L.~dijet limit for an acceptance $A=50\%$ and $100\%$, respectively.
We use the NNLO parton distribution functions MSTW2008NNLO \cite{Martin:2009iq} with $I_{\textrm{pdf}}=0.5$ at 13 TeV. 

For maximal interference the non-observation of an excess in the dijet mass distribution
puts very strong  bounds on the mass of the BSM fermions, $M_\psi \gtrsim 87\, (62)$ TeV for benchmark $D$
and  
$M_\psi \gtrsim 125 \, (89)$ TeV for 
benchmark $B$
using $A=100\%  \,  (50 \%)$.
The bounds gradually become weaker when the interference decreases.
In the limit of no  interference, the respective  lower limits drop to 
$M_\psi \gtrsim 3.9 \, (3.2)$ TeV for 
benchmark $B$.
For 
benchmark $D$ the bounds drop
below the  ones from $R$-hadron  searches given in Tab.~\ref{tab:Rhad}.
Depending on $N_F$, the two limiting cases may differ by several orders of magnitude.
Moreover, since  $\Gamma_{gg}\ll\Gamma_{\rm det}$, the cross section  $pp\to S\to gg$  without  interference  is additionally reduced 
relative to the maximal one.
We conclude that, whenever applicable \eq{eq:diboson}, the dijet searches can provide significantly stronger limits on the BSM fermion mass $M_\psi$ 
than the 
DY and $R$-hadron limits worked out in Sec.~\ref{sec:weak} and Sec.~\ref{sec:R}, respectively.
   \begin{figure}[t]
 \begin{center}
\hskip-1cm
 \includegraphics[width=0.7\textwidth]{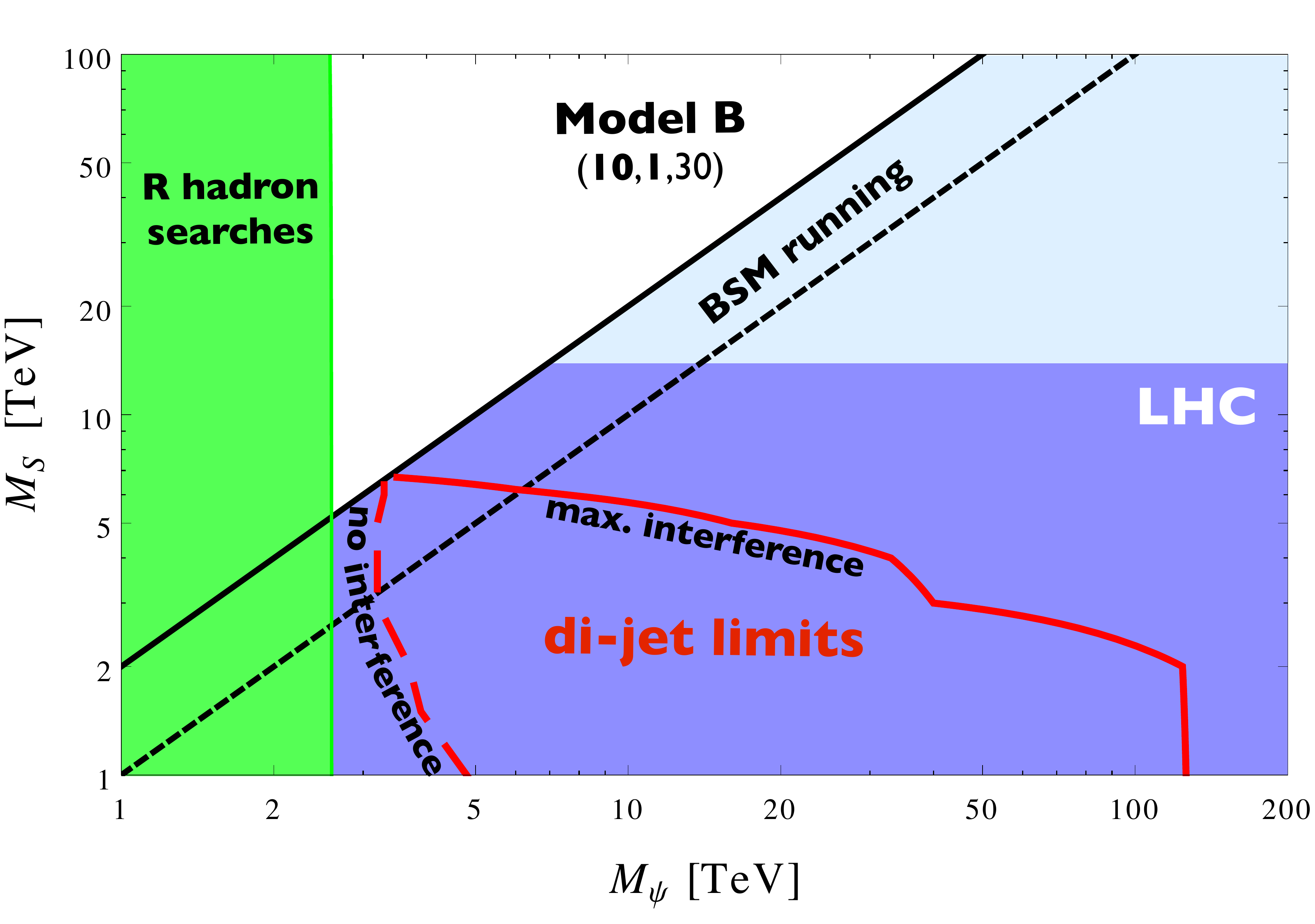} 
 \end{center}
\vskip-.5cm 
 \caption{
 Excluded regions in the $M_\psi-M_S$ plane combining $R$-hadron searches (green horizontal stripe) with dijet cross section limits from ATLAS 95\% C.L.~\cite{Aaboud:2017yvp} for the benchmark $B$ $(R_3=\bm{10}$, $R_2=\bm{1}$, $N_F=30)$.
 Excluded regions are also given for the limiting cases of maximal interference  (below red solid curve) and no interference (left of red dashed curve), with $A=100 \%$. The 
 dark (light) blue area indicates the searchable parameter space   \eq{eq:diboson} at the LHC (future colliders).
The strip  \eq{highMs} 
with $M_\psi   \lesssim M_S < 2 M_\psi$ 
where an enhancement of $\alpha_s$ due to  asymptotically safe  running 
 takes place corresponds to the region between the full and dashed black lines.}
 \label{fig:pheno}
 \end{figure}

 In Fig.~\ref{fig:pheno} we present exclusion limits in the $M_S-M_\psi$ plane from $R$-hadron searches (green horizontal stripe) and the dijet cross section limit from ATLAS  95\% C.L. --- exemplarily for the  benchmark $B$ $(R_3=\bm{10}$, $R_2=\bm{1}$, $N_F=30)$ ---, also comparing settings with maximal interference  (red solid line) and no interference (red dashed line) and $A=100 \%$. Moreover, solid and dashed black lines represent the borders of \eq{highMs} where $M_S =2 M_\psi$ and $M_S = M_\psi$, respectively.
 The excluded areas are below  the solid black lines and to the left and below  the red lines.
 The 
 dark and light blue areas indicate, respectively, the searchable parameter space   \eq{eq:diboson} at the LHC (with $\sqrt{s}=14$ TeV) and at future colliders \cite{Shiltsev:2015tta}.
For masses within the range \eq{highMs} 
--- corresponding to the band between the full and dashed black lines ---,
the strong coupling $\alpha_s$ is already  of non-SM type and enhanced,
$\alpha_s (M_S) \gtrsim  \alpha_s( M_\psi)$. This regime allows to probe higher values of $M_S$. 

If the BSM fermions transform non-trivially under $SU(2)_L$, two additional effects arise. Firstly, the lower dijet bound on the fermion mass increases 
since the cross section $\sigma(pp\to S\to gg)$
scales with $d(R_2)$. Secondly, decays into electroweak gauge bosons $VV=\gamma \gamma, WW,ZZ,Z \gamma$ become possible. 
In order to discuss  decays into weak gauge bosons in more detail, it is convenient to introduce the reduced decay widths 
\beq\label{reduced}
\bar \Gamma_{VV}=\frac{1}{F}\frac{\Gamma_{VV}}{\Gamma_{gg}}\,,\quad{\rm with}\quad
{F}=
\left(\frac{4}{3}\frac{C_2(R_2)}{C_2(R_3)}\right)^2 \,,
\eeq 
which expresses  the widths $\Gamma_{VV}$  in units of $\Gamma_{gg}$ together with a group theoretical factor $F$ which takes into account the quadratic Casimirs of the BSM fermions. 
In terms of \eq{reduced}, we find
\beq\label{GammaReduced}
\bar \Gamma_{WW}=
\frac{\alpha_2^2}{\alpha_3^2}\,,\ \ 
\bar \Gamma_{ZZ}
=
\frac{\alpha^4_{2}}{ 2 (\alpha_1+\alpha_2)^2\,\alpha_3^2}\,,\ \
\bar \Gamma_{Z\gamma}
=
\frac{\alpha_{1}\,\alpha^3_{2}}{(\alpha_1+\alpha_2)^2\,\alpha_3^2}\,,\ \
\bar \Gamma_{\gamma\gamma}
=
\frac{\alpha^2_1\,\alpha^2_{2}}{2(\alpha_1+\alpha_2)^2\,\alpha_3^2}
\,,
\eeq
where $\alpha_1=g^2_Y/(4\pi)^2$ is the hypercharge coupling. 
The  reduced decay widths depend, in general, on the three gauge couplings. We note that $\bar \Gamma_{WW}$ stands out in that it is independent of $\alpha_1$, and only sensitive to the ratio of the other two gauge couplings.

\begin{figure}[t]
 \begin{center}
 \includegraphics[width=0.475\textwidth]{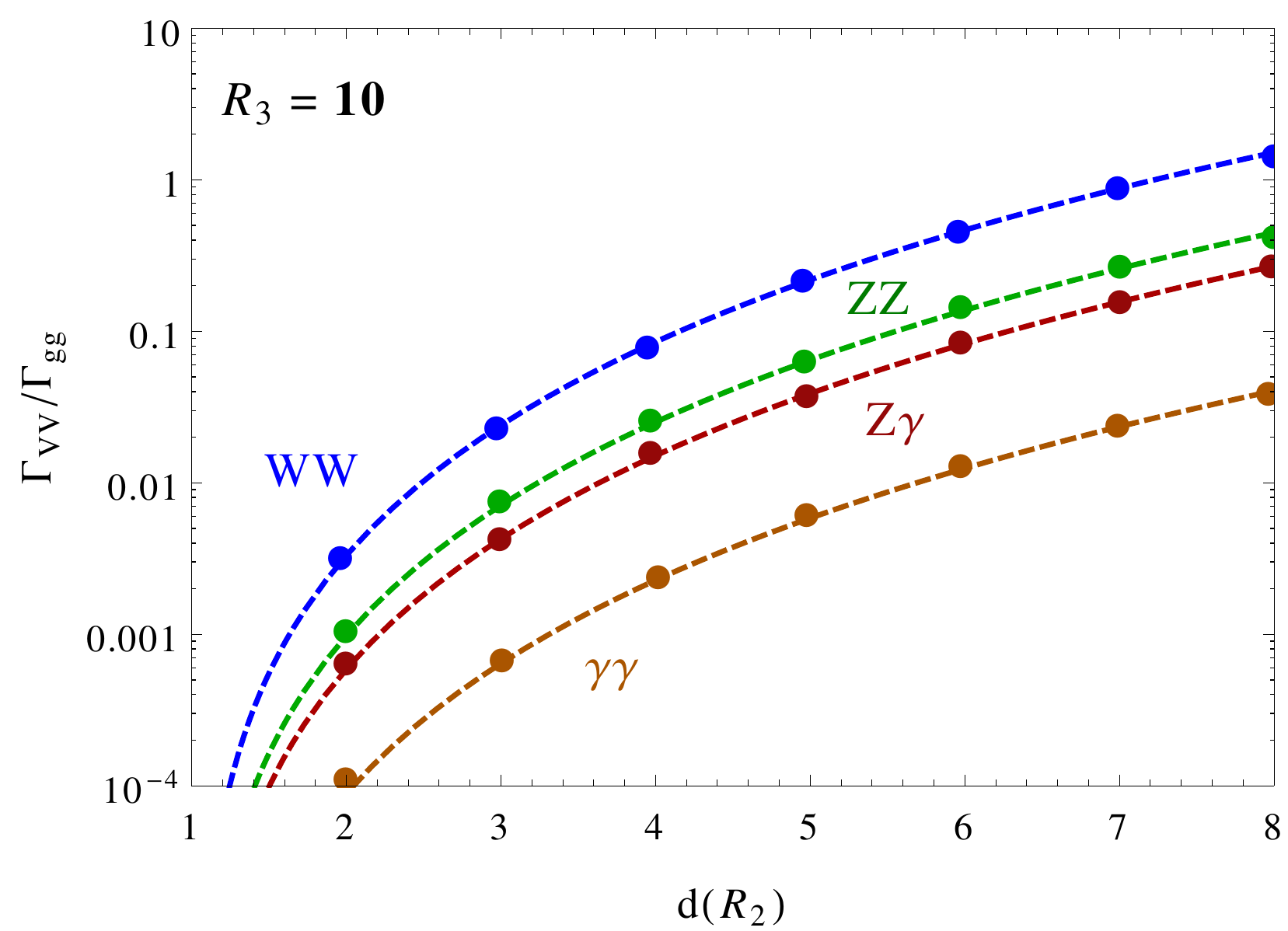}
 \includegraphics[width=0.475\textwidth]{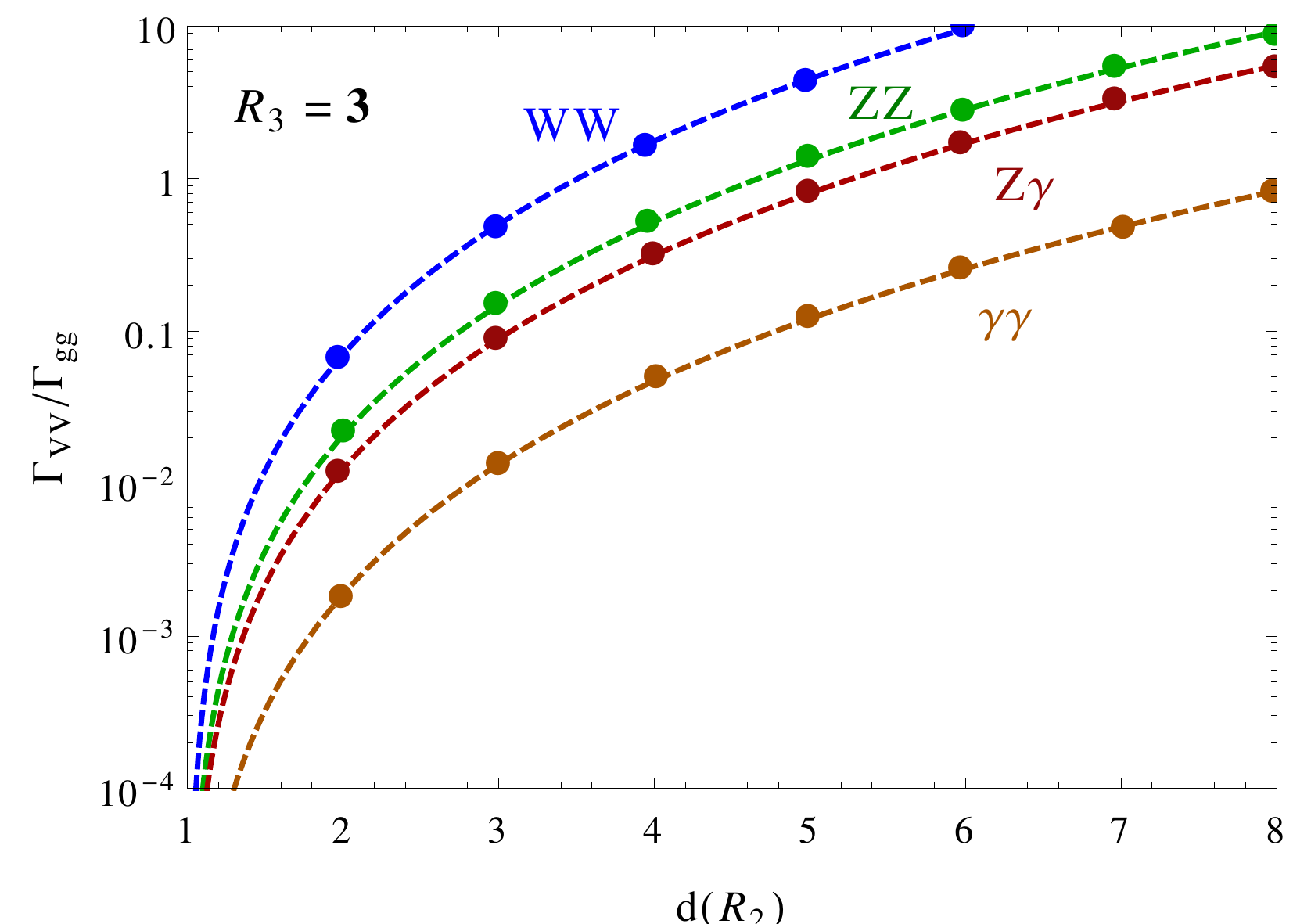}
 \end{center}
 \caption{\small The ratios $\Gamma_{WW}/\Gamma_{gg}$ (blue), $\Gamma_{ZZ}/\Gamma_{gg}$ (green),
$\Gamma_{Z\gamma}/\Gamma_{gg}$ (red), and $\Gamma_{\gamma\gamma}/\Gamma_{gg}$ (orange) versus  $d(R_2)$ for $R_3=\bm{10}$ (left) and $R_3=\bm{3}$ (right)
for low $M_S$ (see main text).}
 \label{fig:dibos_limit}
 \end{figure}

For low $M_S$ below the matching scale the ratios $\bar\Gamma_{VV}$ are solely determined by 
 the SM gauge couplings. In this case  $F$ can be determined from any of the $VV$ modes, providing information about $R_2$ and $R_3$.
 Measuring more than one mode serves as a consistency check.
  In Fig.~\ref{fig:dibos_limit} we show  the ratios $\Gamma_{WW}/\Gamma_{gg}$ (blue), $\Gamma_{ZZ}/\Gamma_{gg}$ (green),
$\Gamma_{Z\gamma}/\Gamma_{gg}$ (red), and $\Gamma_{\gamma\gamma}/\Gamma_{gg}$ (orange),  
depending on $d(R_2)$ for  $R_3=\bm{10}$ (left panel) and $R_3=\bm{3}$ (right panel) for low $M_S$. The hierarchies among the different modes
are fixed  in this regime. Deviations   can arise from switching on hypercharges of the fermions, or from running above the matching scale,  at high $M_S$, where
the gauge couplings experience  BSM running.
This modifies the running of $\alpha_2$ and $\alpha_3$ whereas the running of $\alpha_1$ remains SM-like. We may neglect its slow logarithmic running and can take $\alpha_1$ as constant for the following considerations. 

To discuss further BSM effects, we simplify \eq{GammaReduced} by exploiting that  $(\alpha_1/\alpha_2)^2\lesssim 0.08$ at TeV energies below the matching scale. 
We find
\beq\label{GammaApprox}
\bar \Gamma_{WW}=
\frac{\alpha_2^2}{\alpha_3^2}\,,\quad
\bar \Gamma_{ZZ}
\approx
\012\frac{\alpha^2_{2}}{\alpha_3^2}\,,\quad
\bar \Gamma_{Z\gamma}
\approx
\frac{\alpha_{1}}{\alpha_3}
\frac{\alpha_{2}}{\alpha_3}\,,\quad
\bar \Gamma_{\gamma\gamma}
\approx
\012\frac{\alpha^2_1}{\alpha_3^2}
\,,
\eeq
where ``$\approx$'' means equality up to relative corrections of order ${\cal O}(\alpha^2_1/\alpha^2_2)$.  We observe  that $\bar \Gamma_{\gamma\gamma}$ is no longer sensitive to $\alpha_2$ as it has reduced to a ratio of the other two gauge couplings, similarly to 
$\bar \Gamma_{WW}$. 
It follows that
$\bar \Gamma_{ZZ}\propto \bar \Gamma_{WW}$ and that $\bar \Gamma_{Z\gamma}\propto 
(\bar \Gamma_{ZZ}\cdot\bar \Gamma_{\gamma\gamma})^{1/2}$.

For models with fully interacting fixed points \fp4 we recall  that the weak and strong gauge couplings  start growing with  RG  scale above the matching scale,  dictated by the underlying separatrix into the UV fixed point, see~Fig.~\ref{fig:match_fp4}. Their ratio increases from $\alpha_2/\alpha_3<1$ below the matching scale to $\alpha_2/\alpha_3\to3/2$ sufficiently above the matching scale, invariably inverting the SM hierarchy. 
In particular, we have that $\alpha_2(\mu)/\alpha_3(\mu)>\alpha_2(M)/\alpha_3(M)$ for $\mu>M$ which implies that both $\bar \Gamma_{WW}$  and $\bar \Gamma_{ZZ}$ increase accordingly with increasing  $\mu > M$. On the other hand $\bar \Gamma_{\gamma\gamma}$ becomes suppressed. For $\bar \Gamma_{Z\gamma}$, the situation is ambiguous: the growth of $\bar \Gamma_{ZZ}$ competes with the suppression of $\bar \Gamma_{\gamma\gamma}$ and the outcome in the cross-over region will be model-dependent. 
 Quantitatively, for  the fully interacting fixed points of benchmark $D$ (benchmark $E$) we find that both $\bar \Gamma_{WW}$ and $\bar \Gamma_{ZZ}$ 
grow  from $\mu=M$ to $\mu=2  M$ by factors of about $12$ ($3$), and that $\bar \Gamma_{\gamma\gamma}$ is suppressed by factors of about 13 (2).  $\bar \Gamma_{Z\gamma}$  is very mildly suppressed only.

For models with partially interacting fixed points \fp2 we generically observe $\alpha_2(\mu > M) > \alpha_2(M)$ and $\alpha_3(\mu >M)< \alpha_3(M)$, e.g.~Fig.~\ref{fig:match_fp2}. This implies that all  reduced decay widths  increase with increasing  $\mu > M$, albeit with different factors, see \eq{GammaApprox}. Conversely,  for models with \fp3 we have  $\alpha_3(\mu >M) > \alpha_3(M)$ and $\alpha_2(\mu > M) < \alpha_2(M)$. As can be deduced from the explicit expressions in \eq{GammaReduced} and \eq{GammaApprox}, all four reduced decay widths decrease  relative to Fig.~\ref{fig:dibos_limit}.

 We conclude that  diboson searches involving pairs of  electroweak gauge
bosons  can provide stronger limits than the dijet ones if $d(R_2)$ is sufficiently large.
Due to 
the a priori unknown hierarchy between $M_S$ and $M_\psi$,  correlations of $VV$ with dijet limits cannot be interpreted unambigously.
On the other hand, an observation of a $GG$-resonance determines $M_S$, while a breakdown of SM-running of $\alpha_3$, perhaps together with a similar effect
in the weak coupling,
determines $M_\psi$. In these cases, extracting $F$ is feasible at low $M_S$.

Resonance-induced diboson signatures can arise as well from decays of $(\psi \bar \psi)$-bound states, which are expected to form somewhat below center-of mass energies of $2 M_\psi$  for $R_3 \neq \bm1$ \cite{Kats:2012ym}. In our model such $\psi$-onia can start at about $2 M_\psi \gtrsim 3$TeV, which is within LHC limits \cite{Kats:2016kuz}.  Relative decay widths are as in the case of the decays of the scalar $S$ resonance (\ref{GammaReduced}).
Further analysis is beyond the scope of this work.

\section{\bf Summary}\label{conclusion}

The concept of an interacting UV fixed point in quantum field theory is of  high interest per se; for  particle physics it opens up ``theory space''  for model building. Here, we have investigated asymptotically safe extensions of the Standard Model by adding new fermions  and scalar singlet fields. The new matter fields also interact, minimally, via a single Yukawa coupling to help generate  interacting UV fixed points. A  
large variety of stable high energy fixed points emerges where 
either the strong, or the weak, or both couplings assume finite values,
see Figs.~\ref{fig:fp234},~\ref{fSynopsis}. 
Those where one of the gauge couplings remains asymptotically free can flow into the Standard Model at any scale above ${\cal O}(1-2)$ TeV,
modulo nearby competing fixed points.  Many of the fully interacting fixed points can also be matched onto the Standard Model including at TeV scales, Fig.~\ref{fig:asyGYF_SM}. 
Specifically, with fermions charged under $SU(3)_C \times SU(2)_L$, we found that they must carry representations higher than the fundamental in at least one of the gauge sectors, Fig.~\ref{fSynopsis}. Also, fully interacting fixed points 
cannot arise if the fermions are charged under $SU(2)_L$ only. 
An intriguing feature of models 
with fully interacting UV fixed points is a relation between gauge couplings, dictated by asymptotic safety. The number of fundamentally free parameters is thereby reduced offering an enhanced degree of predictivity compared to the Standard Model, quite similar to the idea of unification.
Our results have been obtained at two loop accuracy where couplings remain small for all scales, though not parametrically small such as in the Veneziano limit \cite{Litim:2014uca}.
Of course, further study is needed to explore the full potential of this new direction.

There are several opportunities to look for asymptotically safe BSM physics at colliders. The presence of a  large number of new fermionic degrees of freedom from higher representations
of $SU(3)_C \times SU(2)_L$
with large multiplicities implies striking new physics at the corresponding mass despite being weakly coupled, e.g.~Figs.~\ref{fig:match_fp2}-\ref{fig:sepE}. Irrespective of the choice of benchmark models, the
qualitative features from the model ansatz laid out in Sec.~\ref{sec:bsm} are rather generic. For low scale matching  BSM physics can be just around the corner, as close as
${\cal O}(1-2)$ TeV:
$R$-hadron signals  arise and  the strong coupling evolution itself is altered and further collider tests should be pursued, see Fig.~\ref{fig:qcd_running}. For $SU(2)_L$-charged fermions
the weak interaction is modified,   schematically  shown in Fig.~\ref{fig:weak_running}. Corresponding shifts in electroweak observables, including  $WW$-production appear
above threshold. Loop-induced diboson spectra  involving the scalar resonance Fig.~\ref{fig:feynman} are sensitive to about an order of magnitude  higher scales 
Fig.~\ref{fig:pheno}. While the actual limits are rather model-dependent, this demonstrates that the phenomenology of 
asymptotically safe BSM can be probed at the LHC at Run 2 and beyond. Tests of the weak interaction are also
 encouraged at high energy $e^+ e^-$ colliders  \cite{Gomez-Ceballos:2013zzn,Shiltsev:2015tta,dEnterria:2016sca}.\\[4ex]

\centerline{\bf Acknowledgements}
${}$\\ We thank Joachim Brod, John Donoghue, Veronica Sanz, Martin Schmaltz, and Enrico Sessolo for useful discussions. AB is grateful to the Physics Department at Boston University  for
its hospitality and stimulating environment while this project was pursued. 
AB is supported by an STFC studentship, GH and KK in part by the DFG Research Unit FOR 1873 ``Quark Flavor Physics and Effective Field Theories'', and DL by the Science and Technology Facilities Council (STFC) under grant number ST/L000504/1.

\appendix

\renewcommand{\thesection}{{\bf \Alph{section}}}

\section{\bf Technicalities}\label{AppA}
The appendix summarises group theoretical formul\ae\ and loop coefficients, together with a discussion of UV-IR connecting separatrices.

\subsection*{Loop coefficients and group theoretical factors}\label{AppA}

We summarise  formul\ae\ for perturbative loop coefficients. 
We have exploited general expressions as given  in \cite{Machacek:1983tz,Machacek:1983fi,Machacek:1984zw,Luo:2002ti}.
We consider the SM matter fields, together with $N_F$ vector BSM fermions in the $R_3$ and $R_2$ representation under $SU(3)_C$ and $SU(2)_L$, respectively. The beta functions are stated in \eq{BSM}.  We reproduce them here for completeness,
\bea
\beta_3&&
= 
 (- B_3 + C_{3}\, \alpha_3  + G_{3}\, \alpha_2-D_3\, \alpha_y)\,\alpha_3^2\,,
\nonumber\\
\label{beta23y}
\beta_{2}&&
=  (- B_2 + C_{2}\, \alpha_2  + G_{2}\, \alpha_3-D_2\, \alpha_y)\,\alpha_2^2\,,
 \\
 \nonumber
\beta_{y}&&
= (E\, \alpha_y  -F_2\, \alpha_2-F_3\, \alpha_3)\,\alpha_y\,.
\eea
The gauge one loop coefficients read
\beq\label{Bs}
\begin{array}{l}
\di
B_3 = 14-\-\frac{8}{3}N_F\,S_2(R_3)\,d(R_2)\,, \\[1.5ex]
\di
B_2 =\frac{19}{3} - \frac{8}{3} N_F\, S_2(R_2)\,d(R_3) \,.
\end{array}
\eeq
At two loop level for the gauge couplings we have the ``diagonal'' gauge contributions
\beq\label{Cs}
\begin{array}{l}
\di
C_3=-52+ 4 N_F \,S_2(R_3)\,d(R_2)\left(2 C_2(R_3)+10\right)\,,\\[1.5ex]
\di
C_2=\frac{35}{3}+ 4 N_F\, S_2(R_2) \,d(R_3)  \left( 2 C_2(R_2)+  \frac{20}{3}\right)\,,
\end{array}
\eeq
together with the ``mixing'' gauge contributions
\beq\label{Gs}
\begin{array}{l}
\di
G_3=\;\, 9+ 8 N_F \,S_2(R_3) \,C_2(R_2) \,d(R_2)\,,\\
\di
G_2=24+ 8 N_F\, S_2(R_2) \, C_2(R_3)\, d(R_3)\,.
\end{array}
\eeq
Furthermore, the BSM Yukawa couplings also contribute at two loop level to the running of the gauge couplings with coefficients
\beq\label{Ds}
\begin{array}{l}
\di
D_3 ={4}{N_F^2}\, S_2(R_3)\,d(R_2)\,, \\
\di
D_2 ={4}{N_F^2}\, S_2(R_2)\,d(R_3) \,.
\end{array}
\eeq
The running of the BSM Yukawa coupling receives one loop contributions from itself as well as from the gauge couplings with coefficients
\beq\label{EFs}
\begin{array}{l}
\di
E =2[N_F+ d(R_2)\, d(R_3)]\,, \\
\di
F_3=12 C_2(R_3)\,,\\
\di
F_2= 12 C_2(R_2) \,.
\end{array}
\eeq
In the above expressions, $C_2(R)$, $S_2(R)$ and $d(R)$ denote the quadratic Casimir invariant, the Dynkin index and  dimension of the representation $R$, respectively. They are related by
\beq
S_2({R})=d({R}) \,C_2({R})/d(\textrm{Adj})\,,
\eeq
to the dimension of the adjoint representation $d(\textrm{Adj})$. It is also convenient to parametrize the loop coefficients through the weights $(p,q)$ for irreducible $SU(3)$ representations $R_3$, and, similarly, through the highest weight $\ell$ for  $SU(2)$ representations $R_2$, 
\beq\label{group}
\begin{array}{l}
d(R_3)=\frac12(p+1)(q+1)(p+q+2)\,,\\[1ex]
C_2(R_3)=p+q+\frac13 (p^2+q^2+p q)\,,\quad {\rm with}\quad p,q=0,1\cdots\,,\\[1ex]
d(R_2)=2\ell+1\,,\\[1ex]
C_2(R_2)=\ell(\ell+1) \,, \quad{\rm with}\quad  \ell=0,\frac{1}{2},1\cdots\,.
\end{array}
\eeq
Evidently, in the absence of BSM matter fields $(N_F=0)$, the perturbative loop coefficients reduce to their SM values
\beq\label{SM}
\begin{array}{ll}
B_3^{\rm SM}=14\,,\quad &\di B_2^{\rm SM}={19}/{3}\,,\quad \\ [.5ex]
C_3^{\rm SM}=-52\,,\quad\quad &\di C_2^{\rm SM}={35}/3\,,\quad \\ [.5ex]
G_3^{\rm SM}=9\,,\quad &G_2^{\rm SM}=24\,,
\end{array}
\eeq
together with $E^{\rm SM}=F_2^{\rm SM}=F_3^{\rm SM}=0$.  In this limit and at two loop accuracy, we observe that the $SU(2)_L$ sector displays a ``would-be'' Banks-Zaks type IR fixed point at 
\beq\label{wouldbe}
(\alpha_2^*,\alpha^*_3)=\left(\0{19}{35},0\right)\,.
\eeq 
The $SU(3)_C$ sector does not display signs of a Banks-Zaks type fixed point owing to $B_3/C_3 <0$. Fingerprints of \eq{wouldbe} become visible for scenarios with BSM matter uncharged under $SU(2)_L$. 

At places, primed two loop coefficients arise. They relate to their unprimed counterparts as
\beq\label{Cs'}
\begin{array}{l}
\di
C'_3=C_3-D_3\, F_3/E\,,\\
\di
C'_2=C_2-D_2\,F_2/E\,,\\
\di
G'_3=G_3-D_3\,F_2/E\,,\\
\di
G'_2= G_2-D_2\,F_3/E\,.
\end{array}
\eeq
In the gauge beta functions, the primed terms  arise as shifts of the unprimed coefficients induced by the BSM Yukawa coupling which takes a fixed point proportional to the gauge couplings.

\subsection*{UV-IR connecting separatrices}

Next, we summarise formul\ae\ and results related to the running of couplings along UV safe trajectories emanating out of an interacting UV fixed point. We are particularly interested in the running of the relevant gauge coupling along the UV-IR connecting hypercritical surface down to energy scales $\mu$ close to the mass $M$ of the new matter fields, $\mu\approx M$, where the model connects with the SM. We approximate the beta functions \eq{beta23y} as 
\beq\label{da}
\begin{array}{rcl}
\partial_t\alpha&=&\alpha^2(-B+C\,\alpha-D\alpha_y)\,,\\
\partial_t\alpha_y&=&\alpha_y(E\alpha_y-F\alpha)\,.
\end{array}
\eeq
These equations are applicable for partially interacting fixed points where one of the gauge couplings becomes asymptotically safe and the other one asymptotically free. The contribution of the latter is neglected. Quantitatively,  corrections from the asymptotically free coupling are subleading in the UV and numerically small for the models discussed here.  The coefficients $B, C,D$ and $F$ then take the values corresponding to the asymptotically safe gauge coupling. 
To find the exact UV-IR connecting separatrix, the system \eq{da} must be solved numerically. However, for most cases of interest, approximate estimates can be obtained as well. We  discuss two strategies. \\[-1.5ex]

{\bf UV critical surface approximation}. Firstly, we may approximate $\alpha_y$ along the separatrix through its values along the UV critical surface. The virtue of this approximation is that it becomes sufficiently exact close to the UV fixed point. 
Quantitatively, on  the UV hypercritical surface, the BSM Yukawa coupling is determined via the gauge coupling as
\beq\label{hyper}
\alpha_y=
C_y\, (\alpha-\alpha^*)+\alpha^*_y\,.
\eeq
In this expression, $C_y$ is defined via the relevant eigendirection at the UV fixed point, with $(1,C_y)^T$ denoting the eigenvector with negative eigenvalue of the stability matrix $M$ at the fixed point in the basis $(\alpha,\alpha_y)^T$. In terms of the perturbative loop coefficients, it reads
\beq
\label{Cy}
C_y=
2\0{F}E
\left(1+\sqrt{1-\0{2BF(C^2E^2-3CDEF+2D^2F^2)-B^2C^2E^2}{F^2(DF-CE)^2}}
+\0{BCE}{F(DF-CE)}\right)^{-1}
\,.
\eeq
 In order to map out  the UV-IR connecting separatrix we insert \eq{hyper} into \eq{da} to find
\beq\label{dhyper}
\partial_t\alpha=\alpha^2(-\tilde B+\tilde C\,\alpha)\,,
\eeq
with
\beq\label{tilde}
\begin{array}{rcl}
\tilde B&=&B+D(\alpha^*_y-C_y\alpha^*)\,,\\
\tilde C&=&C -D\,C_y\,.
\end{array}
\eeq
Solving the RG flow \eq{dhyper} analytically, we find $\alpha(\mu)$ at any RG scale $\mu$ in terms of $\alpha(M)$ determined at some reference mass scale $M$, 
\beq\label{mualpha}
\left(\frac{\mu}{M}\right)^{-\vartheta}=
\frac{\alpha^*-\alpha(M)}{\alpha^*-\alpha(\mu)}\frac{\alpha(\mu)}{\alpha(M)}
\exp\left(\0{\alpha^*}{\alpha(M)}-\0{\alpha^*}{\alpha(\mu)}\right)\,.
\eeq
Furthermore, the UV relevant scaling exponent $\vartheta$ is given by
\beq\label{relevant}
\vartheta={\tilde B^2/\tilde C}<0
\eeq
in terms of \eq{tilde}. The result \eq{mualpha} can be resolved for $\alpha(\mu)$ with the help of the Lambert function,
\beq\label{almu}
\alpha(\mu)=\frac{\alpha_*}{1+W(\mu,M,\vartheta)}\,,
\eeq
where $W(\mu,M,\vartheta)\equiv W_L[z(\mu,M,\vartheta)]$ with  $W_L[z]$ denoting the Lambert function, defined implicitly through $z=W_L\exp W_L$.   The variable $z(\mu,M,\vartheta)$ is given explicitly by 
\beq\label{z}
z(\mu,M,\vartheta)=\left(\frac{\mu}{M}\right)^{\vartheta}
\left(\frac{\alpha^*}{\alpha_M}-1\right)
\exp\left(\frac{\alpha^*}{\alpha_M}-1\right)\,,
\eeq
with the relevant scaling exponent $\vartheta$ given in \eq{relevant}. For any value of $\alpha_M\equiv\alpha(\mu=M)>0$ there is a unique branch of the Lambert function connecting $\alpha(\mu)$ inbetween $\alpha_M$ and $\alpha^*$. The expression \eq{almu}
can now be used to approximately determine the mass scale $M$ by matching it to values of the SM.\\[-1.5ex]

 \begin{figure}[t]
 \begin{center}
 \hskip-2cm
 \includegraphics[width=0.55\textwidth]{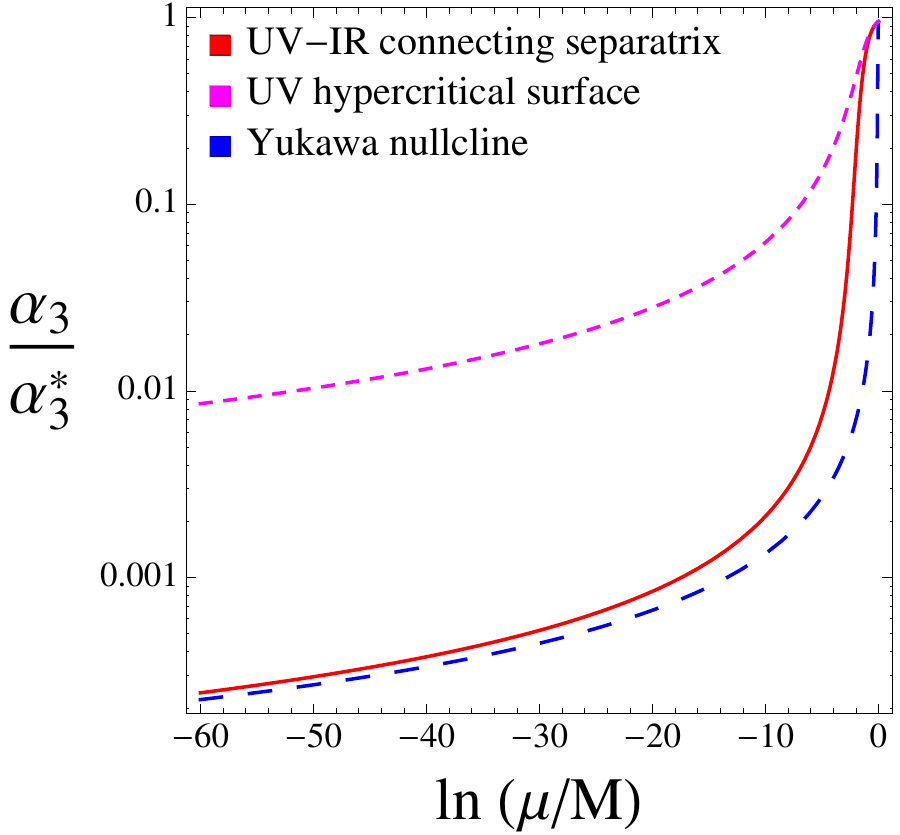}
  \end{center}
 \caption{\small UV-IR connecting trajectories at the example of a partially interacting fixed point \fp3 and parameters
 $R_2= \bm 1, R_3=\bm{10}, N_F=30$, showing the exact  separatrix (red line) in comparison with the UV hypercritical surface approximation (magenta), \eq{hyper}, and the Yukawa nullcline approximation (blue), \eq{nully2}.  The Yukawa nullcline offers a good global approximation for the exact UV-IR connecting separatrix.}
 \label{fig:separatrix}
 \end{figure}

{\bf Yukawa nullcline approximation}.  Alternatively, we may use the Yukawa nullcline to estimate $\alpha_y$ along the UV-IR connecting separatrix. In the system \eq{da}, the Yukawa nullcline is given by 
\beq\label{nully}
\alpha_y=\0FE\,\alpha\,.
\eeq
The virtue of using \eq{nully} to approximate the separatrix is twofold. Firstly,  close to the Gaussian fixed point, the UV-IR connecting separatrix and the nullcline coincide, meaning that \eq{nully} is a very good approximation if the gauge coupling is matched to the SM at scales where $\alpha\ll \alpha^*$.  Secondly, rewriting \eq{nully} as
\beq\label{nully2}
\alpha_y=
\0FE(\alpha-\alpha^*)+\alpha^*_y\,,
\eeq
we conclude that the nullcline also coincides with the hypercritical surface at the UV fixed point. 
Hence, \eq{nully} can be viewed as a ``global linear approximation'' for the UV-IR connecting separatrix. Comparing this approximation with the UV hypercritical surface\eq{hyper} in the limit $B/C'\ll 1$ (with $C'=C-DF/E)$, we observe that \eq{Cy} becomes $C_y=F/E+{\cal O}(B/C')$, establishing that  the hypercritical surface \eq{nully2} exactly coincides with the Yukawa nullcline. For finite $B/C'< 1$, however, the slopes of the hypercritical surface and the nullcline differ. 
Inserting \eq{nully} into the running of the gauge coupling \eq{da} we find 
\beq\label{nullg}
\partial_t\alpha=\alpha^2(-B+C'\,\alpha)\,.
\eeq
The analytical solution to \eq{nullg} with initial condition $\alpha(\mu=M)=\alpha_M$ is given by \eq{almu} with \eq{z}, the sole difference being the value for the parameter $\vartheta$ which now reads
\beq\label{relevant2}
\vartheta={B^2/C'}<0
\eeq
instead of \eq{relevant}. 

Quantitatively, our results are illustrated in Fig.~\ref{fig:separatrix} at the example of a partially interacting fixed point \fp3 with $R_2= \bm 1, R_3=\bm{10}, N_F=30$. We compare the exact numerical solution for $\alpha_3(\mu)$ (full red line) with the hypercritical surface approximation (magenta) and with the Yukawa nullcline approximation (blue). We observe that the UV region (IR region) is well-approximated by the UV critical surface (Yukawa nullcline), respectively. We also observe that the exact separatrix is globally well approximated by the Yukawa nullcline, corresponding to \eq{almu} together with \eq{z} and \eq{relevant2}. 
This approximation offers good quantitative estimates for the matching scale $M$. 

Throughout the main body, we have  used the exact numerical separatrix for our results. We have also confirmed that the critical surface and the nullcline approximations offer very good accuracy in their respective domains of applicability.

\bibliographystyle{JHEP}
\bibliography{AS_pheno}

\end{document}